\documentclass{article}
\usepackage{jcappub}
\usepackage[utf8]{inputenc}
\usepackage[dvipsnames]{xcolor}
\usepackage{aas_macros}
\newcommand{\Mpc}{\mathrm{Mpc}}

\newcommand{\GeV}{\mathrm{GeV}}
\newcommand{\MeV}{\mathrm{MeV}}
\newcommand{\eV}{\mathrm{eV}}

\newcommand{\TRH}{T_\mathrm{RH}}
\newcommand{\Tkd}{T_\mathrm{kd}}
\newcommand{\arh}{a_\mathrm{RH}}
\newcommand{\krh}{k_\mathrm{RH}}
\newcommand{\ahor}{a_{\mathrm{hor}}}

\newcommand{\beq}{\begin{equation}}
\newcommand{\eeq}{\end{equation}}

\newcommand{\Mtot}{M_{\mathrm{tot}}}
\newcommand{\davg}{\delta_{\mathrm{avg}}}

\newcommand{\lcdm}{\Lambda\mathrm{CDM}}

\newcommand{\EM}{\mathrm{EM}}
\newcommand{\LM}{\mathrm{LM}}
\newcommand{\rad}{\mathrm{rad}}

\newcommand{\rhoem}{\bar{\rho}_{\mathrm{EM}}}
\newcommand{\rholm}{\bar{\rho}_{\mathrm{LM}}}
\newcommand{\rhoer}{\bar{\rho}_{\mathrm{ER}}}
\newcommand{\rholr}{\bar{\rho}_{\mathrm{LR}}}
\newcommand{\rhorad}{\bar{\rho}_{\mathrm{rad}}}
\newcommand{\rhomat}{\bar{\rho}_{\mathrm{mat}}}
\newcommand{\drhoemdt}{\dot{\bar{\rho}}_{\mathrm{EM}}}
\newcommand{\drholmdt}{\dot{\bar{\rho}}_{\mathrm{LM}}}

\newcommand{\drhoraddt}{\dot{\bar{\rho}}_{\mathrm{rad}}}

\newcommand{\aeeq}{a_{\mathrm{eeq}}}
\newcommand{\keq}{k_\mathrm{eq}}
\newcommand{\zeq}{z_\mathrm{eq}}
\newcommand{\tdyn}{\tau_\mathrm{dyn}}
\newcommand{\afree}{a_{\mathrm{free}}}
\newcommand{\tfree}{t_{\mathrm{free}}}

\newcommand{\vfree}{v_{\mathrm{free}}}
\newcommand{\Vfree}{V_{\mathrm{free}}}
\newcommand{\Rfree}{R_{\mathrm{free}}}
\newcommand{\Tfree}{T_{\mathrm{free}}}
\newcommand{\Efree}{E_{\mathrm{free}}}
\newcommand{\rvir}{r_{\mathrm{vir}}}
\newcommand{\rs}{r_{\mathrm{s}}}
\newcommand{\lad}{l_\mathrm{ad}}
\newcommand{\lfree}{l_\mathrm{free}}

\newcommand{\OmegaM}{\Omega_{\rm m}}
\newcommand{\OmegaR}{\Omega_{\rm r}}

\newcommand{\Mem}{M_{\mathrm{EM}}}
\newcommand{\Mlm}{M_{\mathrm{LM}}}
\newcommand{\MRH}{M_{\mathrm{RH}}}
\newcommand{\zrh}{z_{\mathrm{RH}}}
\newcommand{\rco}{r_{\mathrm{co}}}
\newcommand{\aeq}{a_{\mathrm{eq}}}
\newcommand{\teq}{t_{\mathrm{eq}}}
\newcommand{\akd}{a_{\mathrm{kd}}}

\newcommand{\Deltaeq}{\bar{\Delta}_{\text{eq}}}
\newcommand{\Deltatau}{\bar{\Delta}_\tau}
\newcommand{\Deltacrit}{\bar{\Delta}_\text{crit}}
\newcommand{\EH}{\mathrm{EH}}
\newcommand{\deltam}{\delta_\mathrm{m}}
\newcommand{\bfx}{\mathbf{x}}

\newcommand{\MEarth}{M_{\oplus}}
\newcommand{\SM}{M_\odot}

\title{Smallest Remnants of Early Matter Domination}
\author[a]{Gabriela Barenboim,}
\author[b,c]{Nikita Blinov,}
\author[b]{Albert Stebbins}
\newcommand\KICP{Kavli Institute for Cosmological Physics, University of Chicago, Chicago, IL 60637, USA}
\newcommand\FNAL{Fermi National Accelerator Laboratory, Batavia, IL 60510, USA}
\affiliation[a]{Departament de F\'{\i}sica Te\`orica and IFIC, Universitat de 
Val\`encia-CSIC, E-46100, Burjassot, Spain}
\affiliation[b]{\FNAL}
\affiliation[c]{\KICP}

\date{\today}
\proceeding{\hfill IFIC 21/29, FERMILAB-PUB-21-321-AE-T
}

\abstract{
The evolution of the universe prior to Big Bang Nucleosynthesis could have gone through a phase of early matter domination which enhanced the growth of small-scale dark matter structure. If this period was long enough, self-gravitating objects formed prior to reheating. We study the evolution of these dense early halos through reheating. At the end of early matter domination, the early halos undergo rapid expansion and eventually eject their matter. We find that this process washes out structure on scales much larger than naively expected from the size of the original halos. We compute the density profiles of the early halo remnants and use them to construct late-time power spectra that include these non-linear effects. 
We evolve the resulting power spectrum to estimate the properties of microhalos that would form after matter-radiation equality. 
Surprisingly, cosmologies with a short period of early matter domination lead to an earlier onset of microhalo formation compared to those with a long period. In either case, dark matter structure formation begins much earlier than in the standard cosmology, with most dark matter bound in microhalos in the late universe.
}
\begin{document}

\maketitle

\section{Introduction}

The standard model of cosmology ($\Lambda $CDM) has proven to be very successful in explaining essentially all observations of the universe. It also includes extrapolations of assumptions to domains which have not been observationally tested. Observable relics from the early universe such as light element abundances~\cite{Fields:2019pfx}, the cosmic microwave background (CMB) and its anisotropies (see, e.g., Refs.~\cite{Aghanim:2018eyx,Aiola:2020azj}) and the large scale distribution of matter (both luminous and dark; see, e.g., Refs.~\cite{Abbott:2018wzc,Abbott:2020knk,Costanzi:2020dgw,Troster:2020kai}) give strong support for a hot big bang including an early stage of accelerated expansion (inflation). In order to accommodate both Big Bang Nucleosynthesis (BBN)~\cite{Kawasaki:2000en,Hannestad:2004px,deSalas:2015glj,Hasegawa:2019jsa} and the CMB the model requires an extended period of radiation domination (RD) where most of the mass/energy density is in the form of a nearly thermal distribution of photons and neutrinos as the universe cools by expansion from $\sim 5\,\MeV$ to $\sim 1\,\eV$ ($1\,\sec$ to $10^5\,\text{yr}$). While it is usually assumed that radiation domination persisted throughout the period after inflation until just before BBN this is not required by observations. Given that the universe apparently has proceeded through several very different evolutionary phases (inflation, radiation domination, matter domination and currently dark energy domination) it is reasonable to suppose that there were additional intermediate phases which we are not currently aware of. 

Recently much effort has been given to understanding the observational implications of an early matter domination (EMD) phase between inflation and BBN. Such a cosmology can be realized in a multitude of settings. For example, inflation can be followed by a long period where inflaton oscillations (which can have an equation of state identical to matter) dominate the energy content of the universe~\cite{Jedamzik:2010dq,Easther:2010mr}. An analogous situation can arise in ultra-violet completions of the Standard Model (SM) like string and supersymmetric theories where long-lived, non-relativistic particles (or scalar field oscillations) can drive the expansion for an extended period~\cite{Kalashnikov:1983qv,Coughlan:1983ci,Khlopov:1985jw,Banks:1993en,deCarlos:1993wie,Banks:2002sd}; alternatively, a phase of EMD can result from the freeze-out of quasi-stable states in a dark sector~\cite{Zhang:2015era,Berlin:2016vnh,Berlin:2016gtr,Dror:2016rxc}. Such a cosmology can have a profound impact of the production of dark matter (DM), favouring regions of theory space that are radically different from naive expectations~\cite{Berlin:2016vnh,Berlin:2016gtr,Banks:1996ea,Moroi:1999zb,Giudice:2000ex,Acharya:2008bk,Visinelli:2009kt,Bose:2013fqa,Fan:2013faa,Blinov:2014nla,Blinov:2019rhb}.
An EMD phase would also produce potentially observable relics in the spatial distribution of dark matter (DM) on the very smallest scales which would be easily distinguishable from that expected in $\Lambda$CDM if one observes them~\cite{Erickcek:2011us,Barenboim:2013gya,Fan:2014zua,Dror:2017gjq,Nelson:2018via,Visinelli:2018wza,Blinov:2019jqc,Erickcek:2020wzd,Erickcek:2021fsu}. These smallest structures are gravitationally bound clumps of DM (microhalos). Microhalos have shallow gravitational potential wells and are much smaller than the Jeans' length of gas, so very little baryonic matter is bound to them. This would make them difficult to detect in any way except gravitationally. While microhalos are difficult to see we expect they will be detected eventually.

Matter domination allows for relatively rapid growth of inhomogeneities due to gravitational instability. An EMD epoch, with sufficient duration, will result in virialized (collapsed) structures in the very early universe which we call early halos (EHs). This paper explores the imprint of EHs on the smallest scale structure and how this is reflected in the microhalos present in the late universe we observe today. We find that, in contrast to the growth of small (linear) inhomogeneities during EMD, larger amplitude inhomogeneities which produce EHs suppress rather than enhance the late time inhomogeneities on the smallest scales. Such a suppression in the context of EMD was first pointed out in Ref.~\cite{Blanco:2019eij} though the suppression derived here is quantitatively different. This phenomenon which we call {\it explosive evaporation} has a simple intuitive explanation: the decay of early matter (EM) erases the gravitational potentials that bind EHs as the radiation decay products rapidly stream outward. The remaining stable particles, which constitute the DM of the current epoch, are also expelled outwards, but more slowly.  This outward motion eventually dilutes the large initially overdensity to a density below the cosmological mean at which point EH remnants overlap. If the decay were instantaneous the outward velocity would be comparable to the initial EH virial velocity, however a slower exponential decay leads to a much smaller outward velocity.  The EHs become unbound well into the radiation, long after the decay time.

This evaporation and dilution of dense halos is generic in the sense that it is independent of how the EHs form. In the scenario considered here EMD occurs when a non-relativistic species (EM) comes to dominate the cosmic density and ends when this species decays into relativistic SM particles. We consider the case where nearly all the matter coalesces into gravitationally bound structures. We also require that DM be present and non-relativistic during EMD so that it will gravitationally cluster with the decaying species. In this scenario we find the remnant inhomogeneities after evaporation depend crucially on the EH mass function (distribution of masses) but is otherwise insensitive to their origin. We give analytic formula for the dark matter power spectrum in terms of this mass function.

The rest of this paper is organized as follows. In \S\ref{sec:background_cosmo} we describe the background cosmology and briefly discuss the evolution of density perturbations during EMD. 
Enhanced growth during EMD ensures that most of the matter is bound up in EHs.
We then summarize the evolution of EHs in \S\ref{sec:halo_history}. 
The evolution of a halo progresses through periods of 1) stable clustering during matter domination, 2) adiabatic expansion which describes their evolution well into the radiation era and 3) free expansion as successive outer layers become unbound during the radiation era. We use a simple spherical model of an isolated halo in \S\ref{sec:isolated_halos} to study each of these steps quantitatively. 
Combining a population of EH remnants we construct the power spectrum of DM inhomogeneities which will eventually form microhalos in the late universe in \S\ref{sec:halo_superposition}. 
In \S\ref{sec:linear_evolution} we specialize to primordial Gaussian adiabatic inhomogeneities which are an extrapolation to small scales of what we observe on large scales today. In this model we can estimate the mass distribution of EHs using linear perturbation theory and the Press-Schechter formalism. This distribution depends on the duration of the EMD era.
Applying the previous results we predict the late time power spectrum for long and short durations of EMD in \S\ref{sec:late_time_ps_and_microhalos}. 
From these spectra we deduce the mass spectra of microhalos which first collapse and discuss various implications of their existence. Our general conclusion is that non-linear collapse during EMD suppresses inhomogeneities on small scales; the suppression length scale, however, is much larger than the scales of sizes of the original EHs. Thus, ``explosive evaporation'' gives a power spectrum suppression that is independent of any DM microphysics.
We outline avenues for further study and conclude in \S\ref{sec:conclusion}.

Our approach which starts with (nonlinear) bound halos complements linear theory estimates of the effect of EMD on the small scale power spectrum though the results are identical on much larger scales which do not collapse during EMD~\cite{Erickcek:2011us,Barenboim:2013gya,Blinov:2019jqc,Blanco:2019eij}. Our results do not apply in cases where structures do not collapse during EMD but we show that linear theory predictions are qualitatively incorrect in the highly nonlinear case where they do. Somewhat surprisingly the maximal enhancement of small scale inhomogeneity at late times from EMD growth lies somewhere in between the linear and highly non-linear cases.

\section{Cosmological Evolution with Early Matter Domination}
\label{sec:background_cosmo}
We model the universe model as containing four components: ``early matter'' (EM - a particle species which dominates the density when non-relativistic during EMD and then decays with lifetime $\tau$), ``late matter'' (LM - a stable species which is non-relativistic during EMD and is the dark matter (DM) today), late radiation (LR - the Standard Model decay products of EM assumed to rapidly thermalize into a relativistic gas of light species, e.g. photons, neutrinos, electrons, \dots) and (optionally) early radiation (ER - remnant relativistic matter from the epoch which precedes EMD). The combination of LR and ER is denoted ``rad'' (radiation) and of LM and EM is denoted ``mat'' (matter). The energy continuity equations for these homogeneous expanding fluids are
\begin{subequations}
\begin{align}
  \drhoemdt  + 3 H \rhoem  &= - \frac{1}{\tau} \rhoem \\
  \drhoraddt + 4 H \rhorad &= + \frac{1}{\tau} \rhoem           \\
  \drholmdt  + 3 H \rholm  &= 0\ .                
\end{align}
\label{eq:background_sys}%
\end{subequations}
where $\dot{}\equiv\partial/\partial t$, $H=\dot{a}/a$ gives the cosmological expansion in terms of the scale factor $a(t)$. The $\tau^{-1}$ terms transfer energy of decaying EM into LR. Note that we specialize to the case where LM is \emph{not} produced in the decays of EM.
The effect of gravity is given by the Friedmann equation for a spatially flat FLRW cosmology
\beq
H^2 = \frac{8\pi G}{3}\,\left(\rhoem + \rhorad + \rholm \right)\ .
\label{eq:background_flat}%
\eeq
which closes this system of equations. Solutions of Eq.~\ref{eq:background_sys} have the universe pass through four epochs which in chronological sequence are: early radiation domination (ERD), early matter domination (EMD), late radiation domination (LRD) and late matter domination (LMD). ERD is only present if $\rhoer\ne0$. The transition between EMD and LRD occurs at $t\sim\tau$ (the {\it decay time}); we will refer to this epoch as reheating. LRD and LMD are identified with radiation and matter domination in our universe (later dark energy domination is not relevant to our analysis). Given $\tau$ and defining $t=0$ by $H(0)=\infty$ the cosmological solution is fully specified by two parameters which set the duration of the EMD and LRD epochs. The LRD duration is determined by the LM to EM+LM density ratio at early times, $f\equiv\text{lim}_{t\rightarrow0}\,\rholm/(\rhoem+\rholm)$: the universe expands by a factor $\sim f^{-1}$ during LRD. Given CMB determination of the end of radiation domination and BBN constraints on the minimum decay time (giving the radiation ``reheating temperature'' $\TRH$ when $\rhorad=\rhomat$) one requires $f\lesssim10^{-8}$. $f$ could be much smaller. The duration of EMD is relevant here as it limits the amount of time for EHs to form.

\begin{figure}
    \centering
    \includegraphics[width=0.47\textwidth]{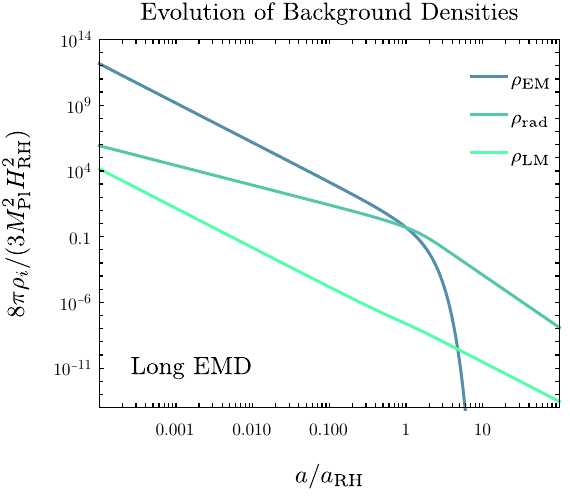}\;\;\;
    \includegraphics[width=0.47\textwidth]{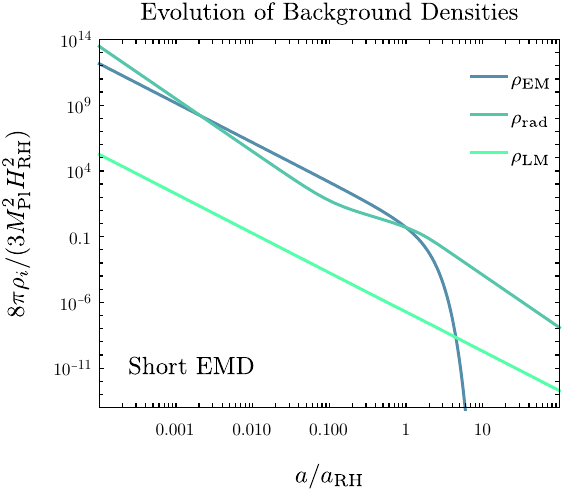}
    \caption{Evolution of background densities in a cosmology 
    with a period of early matter domination as function of the scale 
    factor normalized to its value at reheating. 
    In the left panel the initial radiation density is small compared to 
    EM, so all density fluctuations of interest enter the horizon during EMD. In contrast, in the right panel, the period of EMD is brief and is preceded by radiation domination. As a result, some modes enter the horizon before EMD starts. In both panels the densities are normalized to the total energy density at reheating.}
    \label{fig:bg_density_evolution}
\end{figure}

One can solve equations~\ref{eq:background_sys}--\ref{eq:background_flat} numerically. Two examples of such solutions are shown in Fig.~\ref{fig:bg_density_evolution}. The left panel shows the cosmological evolution of the various fluids for the case where EMD is very long (i.e., the initial radiation density is negligible); we will refer to this scenario as ``Long EMD''. An alternative possibility is that the period of EMD is brief and preceded by radiation domination; we will call this model ``Short EMD''.
We will use these two scenarios as benchmarks to study the evolution of density fluctuations. We will show that a longer period of EMD allows for the formation of more massive halos at early times. The initial densities of EM and ER and EM lifetime in the ``Long EMD'' (``Short EMD'') are chosen to give $\aeeq/\arh = 0$ ($1/2000$), where $\aeeq$ is the scale factor at ``early equality'' (i.e. when ERD ends and EMD begins) and $\arh$ is the scale factor at reheating. 

\begin{figure}
    \centering
    \includegraphics[width=0.47\textwidth]{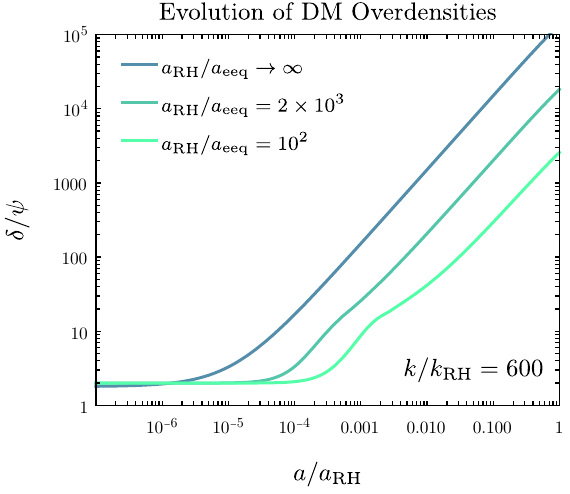}
    \caption{Evolution of the DM density contrast (normalized to the primordial value of the gravitational potential $\psi$) with $k/\krh = 600$  
    in several cosmologies with early matter domination as a
    function of the scale factor normalized to its value at reheating. 
    The upper line corresponds to a cosmology where the mode enters the horizon when the universe is already in the EMD phase. In the lower 
    two lines, the background cosmology experiences only a brief period of EMD that lasts a factor of $\sim 2000$ ($\sim 100$) in scale factor 
    for the middle (lowest) line ($a_\mathrm{eeq}$ refers to the scale 
    factor of the ``early matter-radiation equality'' when EMD began).
    The upper two lines correspond to the two background cosmologies shown in Fig.~\ref{fig:bg_density_evolution}.
    }
    \label{fig:density_contrast_evolution}
\end{figure}

It will be useful to have approximate analytic solutions to~\ref{eq:background_sys}--\ref{eq:background_flat} in certain limits. When only one species dominates simple solutions exist: during EMD ($a\gg\aeeq$ and $t\ll\tau$, where $\aeeq$ is the scale factor at ``early equality'' when the densities of ER and EM equal)
\beq
    \rholm(t) \simeq\frac{f}{6\pi\,G\,t^2}         \qquad 
    \rhoem(t) \simeq \frac{1-f}{6\pi\,G\,t^2}      \qquad
    \rholr(t) \simeq \frac{1}{10\pi\,G\,\tau\,t} \ ,
\label{eq:densities_emd}
\eeq
during LRD ($\tau\ll t\ll\tau/\sqrt{f}$)
\beq
    \rholm(t) \simeq\frac{f}{8.023\,\pi\,G\,\sqrt{\tau\,t^3}}         \qquad 
    \rhoem(t) \simeq \frac{(1-f)\,e^{-t/\tau}}{8.023\,\pi\,G\,\sqrt{\tau\,t^3}} \qquad
    \rholr(t) \simeq \frac{3}{32\pi\,G\,t^2} \ ,
\label{eq:densities_lrd}
\eeq
and during LMD ($t\gg\tau/\sqrt{f})$
\beq
    \rholm(t) \simeq\frac{1}{6\pi\,G\,t^2}         \qquad
    \rholr(t) \simeq \frac{3}{32\pi\,G\,t^2} \ .
\label{eq:densities_lmd}
\eeq

It is also useful to express the densities in terms of the cosmological scale factor, $a$. Defining $\arh$ such that during EMD $a=\arh (t/\tau)^{2/3}$ then\footnote{There are several possibilities for defining $\arh$, including via $t(\arh) = \tau$, $\rhomat(\arh) = \rhorad(\arh)$, 
$H(\arh) = 1/\tau$, which give similar numerical values for $\arh$ up to $\mathcal{O}(1)$ factors. We will use the first definition for the semi-analytic results and the second in our numerics without introducing different notation for this characteristic scale factor.}
\beq
\rholm(a)=\frac{f}{6\pi\,G\,\tau^2}\,\left(\frac{\arh}{a}\right)^3
\label{eq:density_lm}
\eeq
which is valid at all times. During LRD we find
\beq
a(t)\simeq1.1\,\arh\,\sqrt{\frac{t}{\tau}}
\label{eq:lrd_expansion}
\eeq
so after nearly all the EM has decayed, i.e. during LRD and LMD 
\beq
\rholr(a) \simeq \frac{1}{7.2\,\pi\,G\,\tau^2}\,\left(\frac{\arh}{a}\right)^4
\label{eq:density_lr_late}
\eeq
and the Hubble parameter is
\beq
H(a) \simeq \frac{1}{\tau}\,\left(\frac{\arh}{a}\right)^2\,
\sqrt{0.37+\frac{4}{9}\,f\,\frac{a}{\arh}}
=\frac{0.88\,f^2}{\tau}\,
\sqrt{\left(\frac{\aeq}{a}\right)^4+\left(\frac{\aeq}{a}\right)^3}
\label{eq:hubble_lr_late}
\eeq
where $a_{\rm eq}=0.83\,\arh/f$ is the scale factor when the LM and LR density are equal. This we identify with matter-radiation equality ($z_{\rm eq}\simeq3400$,
$t_{\rm eq}\simeq51\,\text{kyr}$) determined from late time observational cosmology~\cite{Aghanim:2018eyx}. Note that 
$f$ and $\TRH$ are not independent parameters, but are related by fixing the temperature and redshift of equality to the observed values, leading to $\TRH \sim \MeV \;(10^{-8}/f)$.

Density perturbations experience enhanced growth during the EMD phase of cosmological evolution. 
Perturbations that enter the horizon at $\ahor(k)$ during EMD (i.e., $\ahor < \arh$) are boosted by a factor of $\arh/\ahor$. This effect was first studied in detail in Ref.~\cite{Erickcek:2011us}. 
The growth of density perturbations for several durations of EMD (i.e., various values 
of $\arh/\aeeq$) is shown in Fig.~\ref{fig:density_contrast_evolution}. This evolution is obtained by 
numerically solving first order perturbation equations for EM, LM and radiation fluids coupled 
to gravity as in, e.g., Refs.~\cite{Erickcek:2011us,Blinov:2019jqc}. We will present partial 
analytic results for the scaling of the density contrast with wavenumber in \S\ref{sec:linear_evolution} and numerical details in Appendix~\ref{sec:linear_perturbations}. For now it is sufficient to observe that 
small-scale modes can be enhanced by orders of magnitude compared to the standard assumption 
of radiation-dominated growth of nearly-Harrison-Zeldovich perturbations. In particular, 
it is clear that certain perturbations can reach non-linearity and collapse before 
reheating, leading to the formation of EHs. The wavenumber in Fig.~\ref{fig:density_contrast_evolution} is chosen to roughly correspond to the largest scales that can collapse -- that is 
$\delta/\psi \sim 10^5$ at $a/\arh \sim 1$ means that $\delta\sim 1$ for $\psi \sim 10^{-5}$, where 
$\psi$ is the amplitude of the primordial gravitational potential. Smaller scales will collapse earlier, 
so if EMD lasts long enough most of the EM and LM will be bound in EHs of some size. The fraction of matter in EHs of a given mass and their distribution can be estimated by using linear perturbation theory in conjunction with the Press-Schechter formalism~\cite{Press:1973iz}, as we describe in \S\ref{sec:linear_evolution}.
In the following two sections we first focus on the evolution of individual EHs through 
reheating and beyond. Since these are non-linear structures, perturbation theory 
does not capture their dynamics, and we will instead study them using Newtonian equations of motion.

\section{A Brief History of Early Halos}
\label{sec:halo_history}

In what follows we show that in a broad range of scenarios the matter in the EHs progress through the following stages of evolution from early matter domination (EMD) to late matter domination (LMD):
\begin{itemize}
    \item{\bf stable clustering:} individual halo retain constant physical size during EMD, $a\lesssim0.3\,\arh$. Self gravity dominates and EM has not yet undergone significant decay.
    \item{\bf adiabatic expansion:} individual halos grow exponentially in physical size retaining their profile during the EMD/LRD transition, $0.3\,\arh\lesssim a\lesssim 10\,\arh$, while a significant fraction of EM has decayed but self gravity still dominates.
    \item{\bf peeling:} successive outer layers of the halo end adiabatic expansion and begin free expansion soon after LRD begins, $2\,\arh\lesssim a\lesssim10\,\arh$. The longer orbital timescale of outer layers become comparable to the expansion time invalidating the adiabatic approximation. 
    \item{\bf free expansion:} individual halos grow logarithmically in comoving size during LRD, $2\,\arh\lesssim a\lesssim\aeq$. Nearly all EMD has decayed and self-gravity is unimportant. The remnant LM moves ballistically in a radiation dominated universe. 
    \item{\bf halo overlap:} the ballistically expanding halo remnants will overlap with neighboring halos during LRD, $2\,\arh\lesssim a\lesssim\aeq$. Rapid expansion evacuates the LM from the initial halo position but this is mostly filled in by LM from neighboring expanding halos.
    \item{\bf recollapse:} the inhomogeneous LM distribution created by the   superposition of overlapping halo remnants will recollapse to form new structures during LMD,  $a\gtrsim\aeq$, when self gravity of the LM dominates over the LR.
\end{itemize}
Many of these stages are very short in duration and some are concurrent. Different parts of the halo may exhibit different behaviour at the same time. The intervals during which different stages occur depend weakly on the overdensity of the EHs and may vary somewhat from those quoted. This qualitative scenario is valid for EH overdensities $\lesssim10^{17}$. The first stage of ``stable clustering'' depends on the cosmology. In common scenarios 
where the early halos are formed from small initial density perturbations, this early structure formation proceeds via hierarchical assembly, so EHs undergo mergers with similarly-sized EHs or accrete smaller EHs. While these sub-halos undergo tidal stripping, they may survive for many dynamical times, maintaining roughly a constant physical size. Therefore in these models ``stable clustering'' is at best a coarse approximation. Despite this, we will assume initial stable clustering as it enables a simple analytical model; detailed $N$-body simulations are required to test the error made in this assumption.

In the next section we describe the evolution stages of an individual EH quantitatively, while the remainder of the paper is dedicated to studying the overlap of EH remnants at late times and the impact on the small-scale power spectrum.

\section{Evolution of an Isolated Halo}
\label{sec:isolated_halos}

Here we describe the evolution of matter, starting with an early halo (EH) which formed during EMD. We make three simplifying assumptions: 1) that EHs have collapsed sufficiently in advance of $t=\tau$ so that it has reached a quasi-equilibrium state of stable clustering before significant decay, 2) the EHs are sufficiently separated that tidal interactions and mergers are rare and 3) the EM and LM occupy the same phase space distribution in the halos. The first two assumptions are closely tied to each other, requiring mergers of halos to be episodic rather than continuous. One can imagine scenarios where 3) would not be valid: e.g. if one of the two components have sufficiently different velocity dispersions allowing one to collapse on scales where the other does not. These assumptions allow us to model the initial conditions purely in term of the density and velocity profiles of EHs along with their spatial correlations. As alluded to above, the first two assumptions may not hold 
in the strict sense in certain cosmologies. For example, if the primordial power spectrum is nearly flat, resulting continuous and hierarchical formation of EHs, 
it is not clear that a given EH is ever in the isolated or stable clustering regime. In such a situation, estimates of the merger halo merger rate indicate 
that most of the mass gain of a halo is through minor mergers which are less likely to disrupt the ``core'' density profile of the halo~\cite{Neistein:2008ht}. Moreover, 
$N$-body simulations of small $\Lambda$CDM halos show stable-clustering-like evolution of the concentration parameter~\cite{Diemer:2018vmz}. These qualitative observations 
suggest that the stable clustering and isolated ansatz might be a reasonable starting point even for these initial power spectra. In models with a 
different initial condition, such as a narrow power spectrum spike, the formation of isolated EHs may be explicitly realized and the first two assumptions would be clearly satisfied. Ultimately, the assumptions of stable clustering and rarity of disruptive tidal events enable a simple analytical study of the fate of EHs, but it is will be important to thoroughly validate these in $N$-body simulations.

Upon collapse during EMD the matter overdensity, $\delta\equiv\rho_\text{mat}/\rhomat-1$, increases to $\gtrsim100$ and then increases as $\sim a^3$, becoming extremely large in a short number of expansion times. The halo dynamical timescale, $\tdyn\sim(H\,\sqrt{1+\delta})^{-1}$, becomes much less than the expansion timescale $\sim 1/H$, allowing a halo to rapidly approach stable clustering. So long as $\tau_\text{dyn}\ll\tau$ a halo will respond adiabatically to its decaying matter content, a process described in Appendix~\ref{sec:adiabatic} and which we refer to as ``adiabatic expansion''. During this phase a halo undergoes homologous expansion, with its physical size growing as $r\propto e^{+t/\tau}$ and its internal velocities shrinking as $v\propto e^{-t/\tau}$. During adiabatic expansion $\tdyn\sim r/v\propto e^{2t/\tau}$.

Adiabatic expansion ends when $\tdyn\gtrsim\tau$, i.e. when there is significant EM decay during a single orbit. As halos start with such a large overdensity this will take many decay times. The gravitation of the radiation also works to disrupt halos but most the radiation is between the halos and not within them; as a result, these forces are not significant before or during disruption. When disruption does occur the gravitational attraction binding the halos becomes ineffectual and the LM particles free-stream away from the initial halo center, acting as test particles in a radiation dominated universe. We call this stage ``free expansion'' since the gravitation of EM and LM are unimportant. Free expansion is slower than the exponential growth during adiabatic expansion. The total free streaming length during the radiation era is larger than the inter-halo separation erasing structures on and above the mass scale of the EHs. Individual halos freely expand into a much larger volume such that their density is below the cosmological mean. The final LM distribution is the superposition of these underdense halos, so that the locations of the EH remnants are not necessarily physically underdense once their overlap is taken into account.

Individual EHs start as non-linear overdensities, $\delta\gg1$, and evolve into non-linear underdensities, $1+\delta\ll1$, so none of this evolution can be accurately described by linear theory. One can however use linear theory to track the evolution of the central position of halo remnants.

\subsection{Ejection Velocity and Free-Streaming Length}
\label{sec:Veject}

First, we develop some intuition for the expected behaviour of EHs based on simple scaling arguments. The size of the free-streamed remnant will depend mostly on the velocity at which the constituent particles are ejected. If one characterizes a halo by a single mass and size, $M$ and $R$, then $\tdyn\sim\sqrt{R^3/(G\,M)}$ and by the Virial theorem the internal velocities have magnitude $V\sim\sqrt{G M/R}$. During adiabatic expansion $M\simeq M_0\,e^{-t/\tau}$, $R\simeq R_0\,e^{t/\tau}$ and $V\simeq V_0\,e^{-t/\tau}$ but this ends when $\tdyn\sim\tau$ or $t\equiv\tfree\simeq \frac{1}{4}\tau\,\ln\delta_\tau$ where $\delta_\tau\sim G\,M_0\,\tau^2/R_0^3$ is roughly the overdensity of the halo when $t=\tau$. The characteristic size and internal velocity during the transition to free expansion is
\beq
\Rfree = R_0 e^{\tfree/\tau} \sim R_0 \delta_\tau^{1/4},\qquad
\Vfree\sim \frac{\Rfree}{\tdyn(\tfree)} \sim \frac{R_0}{\tau} \delta_\tau^{1/4}.
\label{sec:ejection_scaling}
\eeq
Collapsed halos have $\delta_\tau\gg1$ so $\tfree$ is significantly larger than $\tau$ and this transition occurs during the radiation era. After this transition the LM particles will free-stream outwards in a radiation dominated universe, traversing a physical distance
\beq
\Delta r\sim\Vfree\,\sqrt{t\,\tfree}\,\ln\left(\frac{t}{\tfree}\right)
    \sim\frac{1}{2}\,R_0\,
    \,\sqrt{\frac{t}{\tau}\,\sqrt{\delta_\tau}\,\ln\delta_\tau}
    \,\ln\left(\frac{4t}{\tau\,\ln\delta_\tau}\right)
\label{eq:free_streaming_length}
\eeq
at time $t$. $\Delta r\gg\Rfree$ soon after free expansion starts so $\Delta r$ which excludes an initial offset gives an accurate estimate of the distance from the halo center. 

If EHs are formed by hierarchical clustering during EMD they will not be well characterized by a single mass and size, $M$ and $R$, but instead span a large range of densities $\propto\delta_\tau$, the central regions having much larger $\delta_\tau$  while containing a small fraction of the mass. The time of transition from adiabatic to free expansion, $\tfree\propto\ln\delta_\tau$ will happen somewhat later for orbits near the center of the halo than at the edge. More significantly $\Vfree\propto\sqrt[4]{M_0\,R_0}$ will increase from the center to the edge for any halo profile. Thus the outer regions of the halo will be ejected first and at higher velocity than the inner regions and therefore free-stream to larger distances. Thus the inner parts of the halo will remain interior to the outer parts. %This ``peeling'' off of successive layers of the halo from the outside inward is similar to tidal stripping. 
The peeling dynamic determines the density profile of LM ejecta which in turn determines the remnants structure on the smallest scales. A quantitative description of peeling using a spherical halo model is given next.

\subsection{Spherical Halos}
\label{sec:shell_model_evol}

In almost any formation scenario one expects collapsed structures (EHs) to be roughly spherical, similar to dark matter halos present today. We therefore use a spherical approximation for EHs. For spherical structures one need only follow the dynamics of spherical shells of dark matter rather than point particles; the shells contain all particles with the same radius, radial velocity and absolute value of angular momentum about the center. 
One also expects EHs to be non-relativistic and far smaller than the horizon, so we use the Newtonian equations of motion
\beq
\ddot{r}(t) = - \frac{G\,M_<(r(t),t)}{r(t)^2} +\frac{L^2}{r(t)^3} 
- \frac{8\pi G}{3}\rhorad(t)\,r(t)
\label{eq:shell_eom_simple}
\eeq
where $\dot{}\equiv\partial/\partial t$, $r$ is the physical (not comoving) radius of a particular shell, $M_<$ is the total mass of non-relativistic matter within radius $r$ and $L=r\,v_\perp$ is the specific angular momentum which is conserved in spherically-symmetric potentials. The last term allows for the halo to be immersed in an expanding uniformly distributed bath of cosmological radiation of density $\rhorad$. Since non-relativistic halos have shallow gravitational potential wells ($-\Phi\ll c^2$) their presence does not alter the radiation distribution significantly. Recall that radiation contributes twice its mass density to its gravitationally attractive force since generally $\nabla^2\Phi=4\pi G\,(\rho+3\,p/c^2)$ using $\bar{p}_\mathrm{rad}=\rhorad\,c^2/3$ one has $\nabla^2\bar{\Phi}_\mathrm{rad}=8\pi G\,\rhorad$.

The quantity $M_<$ contains both EM and LM which we assume have identical phase space distribution, so
\begin{subequations}
\begin{eqnarray}
M_<(r,t) &=& \Mem(r,t) + \Mlm(r,t) \\
\Mem(r,t) &=& \frac{1-f}{f}\,\Mlm(r,t)\,e^{-t/\tau} \ . 
\end{eqnarray}
\label{eq:mass_interior_equal_phasespace}
\end{subequations}
$M_<(r,t)$ will vary if different shells cross. We next consider a special case where they do not cross and one can evolve each shell independently.

\subsection{Initially Circular Orbits}
\label{sec:circular}

\begin{figure}
    \centering
    \includegraphics[width=0.8\textwidth]{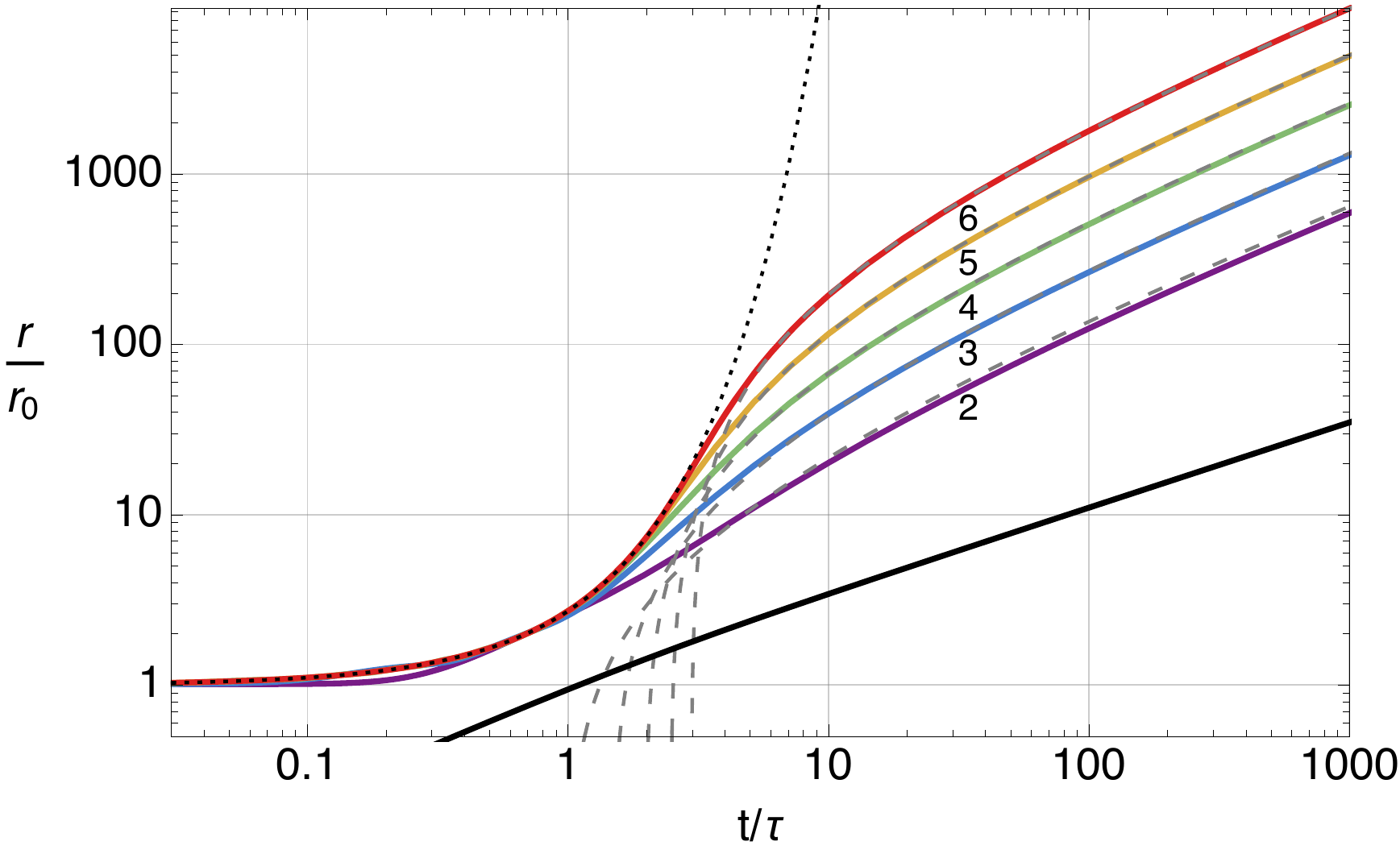}
    \caption{The colored curves give the evolution of the radius of shells in a spherical halo with initially circular orbits in units of the initial radius $r_0$. The curves are labeled by log${}_{10}\delta_\tau$ ($\delta_\tau\equiv G\,M_0\,\tau^2/r_0^3$ where $M_0$ is the mass initially contained within the shell and $\tau$ is the EM lifetime). The dotted curve gives the predicted initial adiabatic exponential expansion and the long-dashed curves the approximation of Eqs.~\ref{eq:r_scaling}-\ref{eq:expansion_coefficients} to the asymptotic free expansion behavior. Shells successively ``peel away'' from exponential expansion: outer shells with smaller $\delta_\tau$ first and inner shells with larger $\delta_\tau$ later. The solid black curve shows the cosmological scale factor $a(t)/a(\tau)$. During adiabatic expansion the shells briefly expand much faster than $a(t)$ so their comoving size grows rapidly. During free expansion the shells grow only logarithmically faster than $a(t)$, their comoving size increasing further.}
    \label{fig:r_evolution}
\end{figure}

A simple initial condition, where all the particles start in initially circular orbits illustrates the general behavior. In this case $r(0)=r_0$, $\dot{r}(0)=0$ and $L=\sqrt{G\,M_<(r_0,0)\,r_0}$. Throughout the evolution of multiple shells the shells never cross: during adiabatic expansion $r(t)\simeq r_0\,e^{t/\tau}$ while peeling retains the ordering of the shells. Thus $\Mlm$ is constant for each shell and one can replace $M_<(r(t),t)\rightarrow M(r_0)\,(f+(1-f)\,e^{-t/\tau})$. For initially circular orbits and $f\ll1$ Eq.~\ref{eq:shell_eom_simple} becomes
\beq
\tau^2\frac{\ddot{r}(t)}{r_0} =\delta_\tau(r_0)\,
\left(\frac{r_0}{r(t)}\right)^3\,
\left(1-e^{-t/\tau}\,\frac{r(t)}{r_0}\right)
- \frac{8\pi\,G\,\rhorad(t)\,\tau^2}{3}\,\frac{r(t)}{r_0}
\label{eq:shell_eom}
\eeq
where $\delta_\tau(r_0)\equiv G\,M(r_0)\,\tau^2/r_0^3$ as above. Assuming the mass density decreases from the center $\delta_\tau(r_0)$ will be a decreasing function of $r_0$. From the form of Eq.~\ref{eq:shell_eom} and initial conditions $E(0,\delta_\tau)=1$ and $\dot{E}(0,\delta_\tau)=0$ the solutions are of the form
\beq
r(t)=r_0\,E\left(\frac{t}{\tau},\delta_\tau(r_0)\right)
\label{eq:r_scaling}
\eeq
just as in Eq.~\ref{eq:free_streaming_length}. Initially ($t\lesssim\tau$) we expect adiabatic exponential expansion $E(T,\delta_\tau)\simeq e^T$. After adiabatic expansion ends in the radiation era at $T\sim T_\mathrm{free}$, $8\pi\,G\,\rhorad/3\rightarrow 1/(4t^2)$ and the first term on the right-hand side of Eq.~\ref{eq:shell_eom} quickly becomes negligible so $\ddot{r}\simeq- r/(4t^2)$; the asymptotic solution is therefore
\beq
E(T,\delta_\tau)\rightarrow
\Efree(\delta_\tau)\,\sqrt{T}\,\ln\left(\frac{T}{\Tfree(\delta_\tau)}\right)\ .
\label{eq:expansion_factor}
\eeq
Only this asymptotic limit is relevant for late time inhomogeneities, and it is characterized by two functions:
\beq
\Efree(\delta_\tau)\simeq0.435\,\sqrt{\sqrt{\delta_\tau}\,\ln\delta_\tau} \qquad\qquad
\Tfree(\delta_\tau)\simeq0.215\,\ln\delta_\tau
\label{eq:expansion_coefficients}
\eeq
The functional form is taken from the scaling derived in \S\ref{sec:Veject}. $\Tfree$ and $\Efree$ are determined by the cosmic time at the beginning of free expansion of a shell and by the velocity at this time:
\begin{eqnarray}
\tfree(\delta_\tau)&\equiv&\Tfree(\delta_\tau)\,\tau
                    \simeq 0.215\,\tau\,\ln\delta_\tau
                    =0.215\,\tau\,\ln\left(\frac{G\,M(r_0)\,\tau^2}{r_0^3}\right)
                                                                      \nonumber \\
\vfree(\delta_\tau)&\equiv&\frac{r_0}{\tau}
                           \frac{\Efree(\delta_\tau)}{\sqrt{\Tfree(\delta_\tau)}}
                    \simeq 0.938\,\frac{r_0}{\tau}\,\delta_\tau^{1/4}
                         = 0.938\,\left(\frac{G\,M(r_0)\,r_0}{\tau^2}\right)^{1/4}
                         \ .
\label{eq:EVfree}
\end{eqnarray}
For spherical halos with initially circular orbits individual shells accurately follow the $\delta_\tau$ scaling derived in \S\ref{sec:Veject} for entire halos with small corrections to numerical factors: $1/4\rightarrow0.215$ and $1\rightarrow0.938$.

Numerical solutions of Eq.~\ref{eq:expansion_factor} and asymptotic fitting functions are shown in Fig.~\ref{fig:r_evolution}. Adiabatic exponential expansion very accurately describes the early evolution for $\delta_\tau\gtrsim10^{3}$ as does the asymptotic fitting function. The accuracy of these  approximations rapidly become worse for $\delta_\tau<10^{2}$ because adiabaticity during EM decay is less closely realized.

\subsection{Spherical Halos in a Cosmological Context}
\label{sec:shell_model_evol_cosmo}

\begin{figure}
    \centering
    \includegraphics[width=0.8\textwidth]{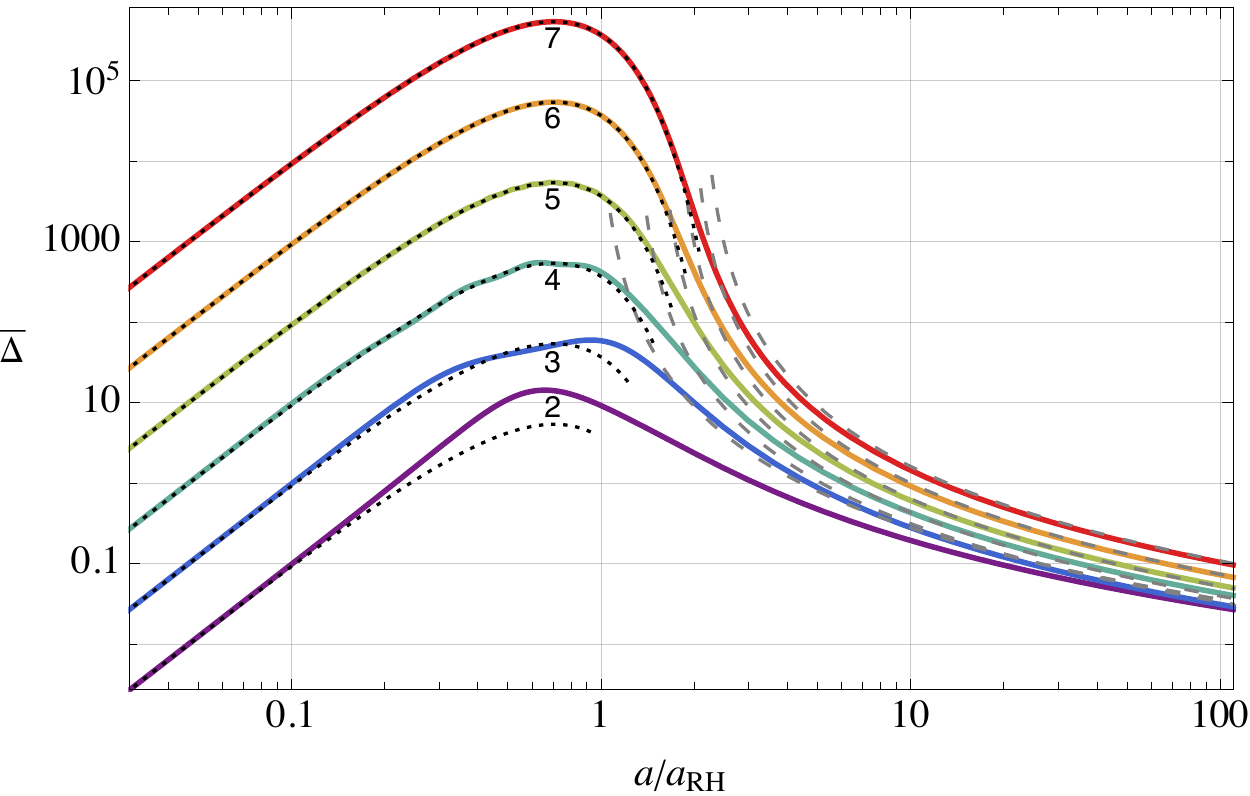}
    \caption{The colored curves show, as a function of the cosmological scale factor $a$, the ratio of the mean matter density within a shell to the mean cosmological density ($\bar{\Delta}$) for a halo with initially circular orbits. The curves are labeled by log${}_{10}\Deltatau$ (see Eq.~\ref{eq:delta_circular_emd}). The dotted curves give the predicted initial adiabatic exponential expansion and the long-dashed curves the asymptotic approximation of Eq.~\ref{eq:delta_circular_asymptotic}. $\bar{\Delta}$ does not include matter from overlapping neighboring halos.}
    \label{fig:Delta_evolution}
\end{figure}

One can translate the physical radius of shells into a quantity more closely related to the background cosmology by using the cosmological scale factor $a$ in place of $t$
and in place of $r$ using the ``interior density ratio'' defined by
\beq
\bar{\Delta}(a(t)) \equiv \frac{3 M_<(r(t),t)}{4\pi\,r^3\,\rhomat(t)} = 
\frac{9}{2}\,\frac{G\,\Mlm(r(t))\,\tau^2}{f\,r(t)^3}\,\left(\frac{a(t)}{\arh}\right)^3 \ ,
\label{eq:delta_defA}
\eeq
where Eqs.~\ref{eq:density_lm} and~\ref{eq:mass_interior_equal_phasespace} were used to obtain the second expression.
This is the ratio of the matter mass within a spherical shell to the mass within the same sphere at the cosmological average density. The last form emphasizes that the decay of EM does not in itself cause $\bar{\Delta}$ to change since $\Mlm(a)$ only changes due to shell crossing. To transform the Newtonian Eq.~\ref{eq:shell_eom_simple} into an equation for $\bar{\Delta}$ define density parameters $\OmegaM(a)\equiv8\pi\,G\,\rhomat/(3\,H^2)$ and $\OmegaR(a)\equiv8\pi\,G\,\rhorad/(3\,H^2)$ and use Eqs.~\ref{eq:background_sys} and~\ref{eq:background_flat} to obtain the flatness condition $\OmegaM+\OmegaR=1$, the deceleration parameter
$q(a)\equiv-a\,\ddot{a}/\dot{a}^2=1-\frac{1}{2}\,\OmegaM$ and find
\beq
a^2\,\bar{\Delta}'' = \frac{3}{2}\,\OmegaM\,\bar{\Delta}\,(\bar{\Delta}-1)
                 -\left(1+\frac{1}{2}\,\OmegaM\right)\,a\,\bar{\Delta}'
                 +\frac{4}{3}\,\frac{(a\,\bar{\Delta}')^2}{\bar{\Delta}}
-3\,\left(\frac{L}{H}\right)^2\,\left(\frac{2\,f}{9\,G\,\Mlm\,\tau^2}\right)^{4/3}
\,\left(\frac{\arh}{a}\right)^4\,\bar{\Delta}^{7/3}\ .
\label{eq:Delta_eomA}
\eeq
where ${}'\equiv \partial/\partial a$ and $\Mlm$ is the LM mass within the shell which may vary due to shell crossing but not due to EM decay.

Overdensity, $\deltam\equiv\rho_{\rm mat}/\rhomat-1$, is the quantity usually followed in cosmological perturbation analysis. For a spherical halo the mean matter overdensity within a spherical shell is $\bar{\delta}_{\rm m}=\bar{\Delta}-1$. One recovers linear perturbation theory of scalar (non-vortical) inhomogeneities by setting $L=0$ and taking the limits $|\bar{\delta}_{\rm m}|\ll1$ and $(\bar{\delta}_{\rm m}')^2\ll|\bar{\delta}_{\rm m}''|$ obtaining
\beq
\bar{\delta}_{\rm m}''
+\left(1+\frac{1}{2}\,\OmegaM\right)\,\frac{1}{a}\,\bar{\delta}_{\rm m}'
-\frac{3}{2}\,\frac{1}{a^2}\,\OmegaM\,\bar{\delta}_{\rm m}=0 \ .
\label{eq:delta_linear}
\eeq
For standard evolution of the background energy densities (i.e., no EMD) this reduces to the Meszaros equation~\cite{Meszaros:1974tb,Hu:1995en}.
Unfortunately, the conditions for the validity of this linear equation are not met for EHs and their subsequent evolution. During a brief period when $|\bar{\delta}_\mathrm{m}|\ll1$ the second condition on $\bar{\delta}_{\rm m}''$ is not satisfied.

During EMD where $\OmegaM\rightarrow1$ an exact circular orbit solution of Eq.~\ref{eq:Delta_eomA} is
\beq
\bar{\Delta}(a)=\Deltatau\,\left(\frac{a}{\arh}\right)^3 \qquad 
\Deltatau\equiv\frac{9}{2}\frac{(G\,M_0)^4\,\tau^2}{L^6}
=\frac{9}{2}\frac{G\,M_0\,\tau^2}{r_0^3}=\frac{9}{2}\,\delta_\tau
\label{eq:delta_circular_emd}
\eeq
where $\Mlm=M_0/f$ is constant since shells do not cross and $r_0=L^2/(G\,M_0)$ is the constant physical radius of the orbiting shell. Note that $\Deltatau$ here is a convenient short-hand, but the overdensity never reaches this value, since the decays of EM are already significant at $a=\arh$; the quantity $\delta_\tau$ was used in \S\ref{sec:circular}. Using Eq.~\ref{eq:lrd_expansion} the approximate asymptotic LRD solution for initially circular orbits of Eqs.~\ref{eq:r_scaling}--\ref{eq:expansion_coefficients} \& \ref{eq:lrd_expansion} in terms of $a$ is
\begin{eqnarray}
r(a) &\rightarrow&0.450\,\left(\frac{a}{f\,\aeq}\,r_0\right)
\,\left(\Deltatau^{1/4}\,\sqrt{\lad(\Deltatau)}\right)\,\lfree(\Deltatau,a) 
\label{eq:rofa_circular_asymptotic}\\
&&\lad(\Deltatau)\equiv\ln\frac{2}{9}\Deltatau \qquad
\lfree(\Deltatau,a)
=\ln\frac{1.62\,a}{f\,\aeq\,\sqrt{\ln\frac{2}{9}\Deltatau}}\ \nonumber.
\end{eqnarray}
The first non-numerical term gives the cosmological expansion while the second and third give additional growth during adiabatic expansion and free expansion, respectively. Here $\lfree\equiv\ln(a/\afree)$ where $\afree\equiv a(\tfree)$; it is interesting to note that free expansion is the same physical effect that gives rise to the logarithmic growth of small-scale density fluctuations in $\Lambda$CDM - it generates a decay of density perturbations in our case due to the initial period of adiabatic expansion.
Eq.~\ref{eq:rofa_circular_asymptotic} is valid for $a\lesssim\aeq$ so $\lfree$ will grow to be large since $f$ is very small. Substituting Eq.~\ref{eq:rofa_circular_asymptotic} into Eq.~\ref{eq:delta_defA} yields
\beq
\bar{\Delta}(a) \rightarrow
\frac{5.33\,\Deltatau^{1/4}}
{\lad(\Deltatau)^{3/2}\,\lfree(\Deltatau,a)^3}\qquad .\\
\label{eq:delta_circular_asymptotic}
\eeq
From $\Deltatau$ one can compute the density ratio
$\Delta\equiv\rho_\mathrm{LM}/\rholm
=\bar{\Delta}\,\left(1+\frac{1}{3}\,\frac{d\,\ln\bar{\Delta}}{d\,\ln\,r}\right)$
which is asymptotically given by
\beq
\frac{\Delta(a)}{\bar{\Delta}(a)} \rightarrow
\frac{\Delta_\tau}{\Deltatau-\frac{3}{4}\,
\left(1+\frac{2}{\lad(\Deltatau)}\,\left(1-\frac{1}{\lfree(\Deltatau,a)}\right)\right)
\,(\Deltatau-\Delta_\tau)} .
\label{eq:densityratio_circular_asymptotic}
\eeq
For a given shell the temporal dependence of $\bar{\Delta}$ and $\Delta$ is solely through $\lfree(\Deltatau,a)$ which is a large logarithm (since $f$ is small) whose $a$ dependence is slow. Thus $\bar{\Delta}(a)$ and $\Delta(a)$ decrease slowly during free expansion. If $\Deltatau\gg1$ then $\lad$ is large ($\lfree$ is usually even larger). In the ``large logarithm approximation'' $\Delta(a)/\bar{\Delta}(a)\approx 4\,\Delta_\tau/(\Deltatau+3\,\Delta_\tau)$.

In Fig.~\ref{fig:Delta_evolution} numerical solutions for $\bar{\Delta}$ are plotted for initially circular orbit model of \S\ref{sec:circular}. Initial adiabatic exponential expansion and the asymptotic approximation are very accurate for $\Deltatau\gtrsim10^4$ and become more accurate for larger $\Deltatau$. For $a\le\aeq$ the errors are $<10\%$ for $\Deltatau\ge10^4$ but $\le45\%$ and $\le63\%$ for $\Deltatau=10^3$ and $10^2$. The adiabatic approximation is poor for $\Deltatau\lesssim10^2$ which is not surprising since one sees that $\Deltatau=10^2$ only reaches $\mathrm{max}_a\bar{\Delta}(a)\simeq10$ so it does not correspond to a large overdensity. For $\Deltatau\gg10^3$ one finds $\mathrm{max}_a\bar{\Delta}(a)\simeq0.053\,\Deltatau$ at $a\simeq0.7\,\arh$. Thus while $\delta_\tau$ and $\Deltatau$ are algebraically convenient quantities they overestimate the maximum overdensity by a factors $\approx4$ and $19$, respectively. For $\Deltatau\lesssim10^2$ the adiabatic approximation becomes inaccurate and at $\Deltatau\le 9/2$ ($\delta_\tau\le1$) it yields unphysical results since $\lad<0$ and $\lfree$ is complex. One should not use these formulae for small $\Deltatau$; this approximation does not reduce to linear theory for $\Deltatau$ near unity.

So far we have focused on spherical EHs with circular, non-crossing orbits. It is important to verify whether our results are robust to more physical initial conditions. While an $N$-body numerical study of EHs is beyond our scope, in Appendix~\ref{sec:shell_code} we used an $N$-shell code to confirm that the qualitative and quantitative evolution described here holds even for more generic choices of shell angular momenta, initial density profiles of EHs and allowing for shell crossing. The shell model does not account for the 
distribution of velocities/angular momenta at each radius which will serve to smear out the remnant in position space. We will argue below that these effects do not qualitatively change our findings.

\subsection{Recollapse?}
\label{sec:recollapse}

\begin{figure}
    \centering
    \includegraphics[width=0.9\textwidth]{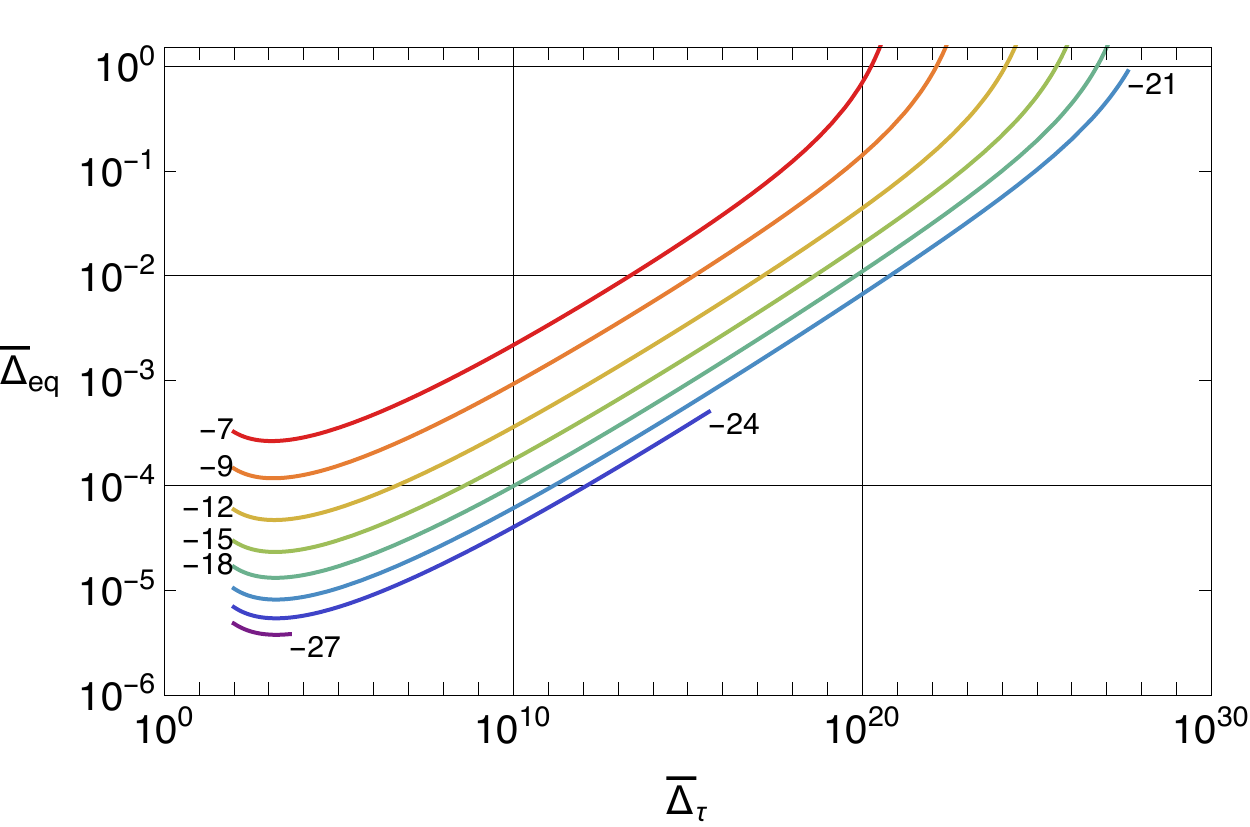}
    \caption{Density ratio of DM at equality $\Deltaeq$ as a function of $\Deltatau$ for different values $f$, the stable DM fraction which determines the duration of LRD.
    $\Deltatau$ determines the EH density before reheating via Eq.~\ref{eq:delta_circular_emd}. The numerical labels are values of $\log_{10}f$.
    These curves are computed for the model of \S\ref{sec:circular} over the region of physical validity delineated in \S\ref{sec:physical_parameters}. Where visible the right endpoints correspond to shells with $\tau_\text{dyn}=t_\text{pl}$. Objects with $\Deltaeq \gtrsim 1$ are expected to recollapse after matter-radiation equality. However, this requires enormous initial density contrasts before reheating.}
    \label{fig:Deltaeq}
\end{figure}

\begin{figure}
    \centering
    \includegraphics[width=0.9\textwidth]{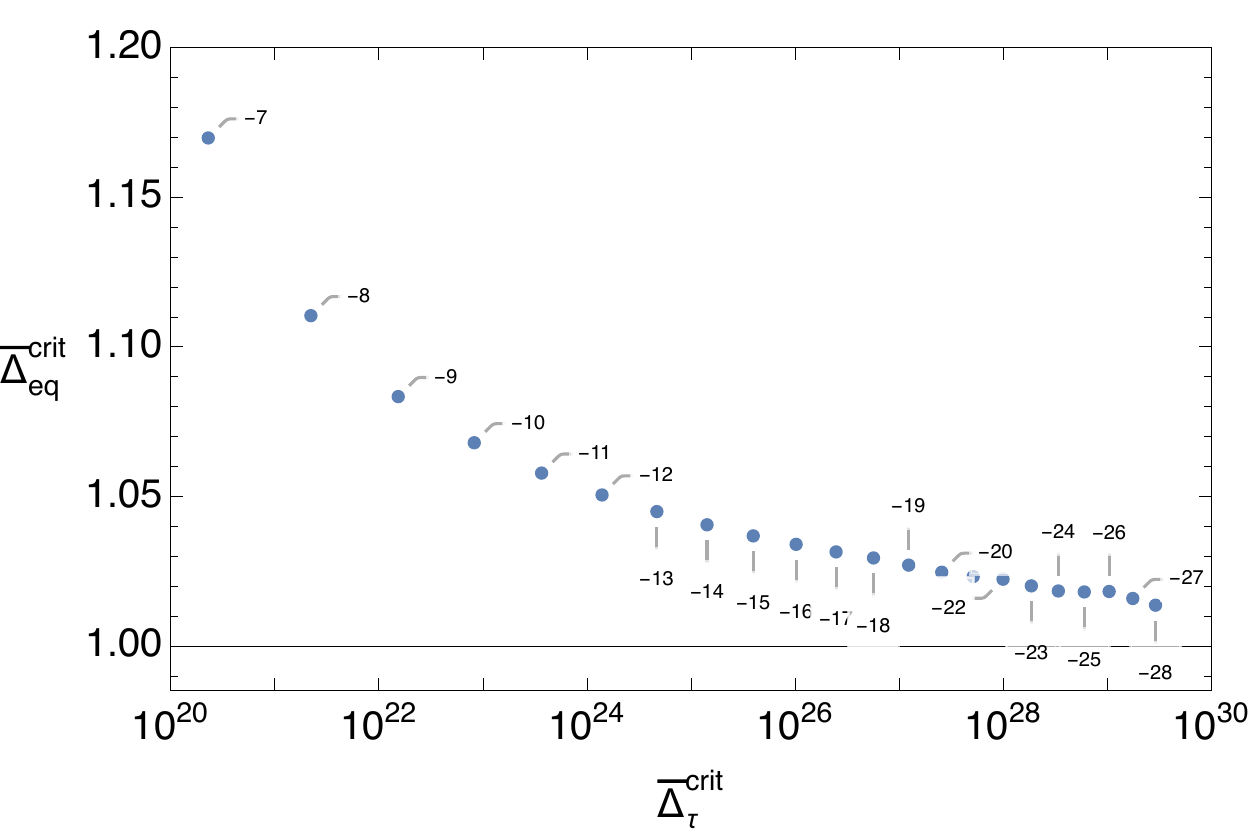}
    \caption{Each point gives $\bar{\Delta}$ for critical shells at two different epochs: $\Deltatau^\text{crit}=18.7\,\bar{\Delta}(\arh)$ and $\Deltaeq^\text{crit}=\bar\Delta(\aeq)$. Critical shells evolve to $\bar{\Delta}(\infty)\rightarrow1$. Shells with $\bar{\Delta}$ larger than these critical values at the corresponding time will recollapse while shells with smaller $\bar{\Delta}$ will evolve to $\bar{\Delta}\rightarrow0$. The numerical labels give $\log_{10}f$ where $f$ is the initial fraction of stable DM and determines the duration of late radiation domination (LRD). The values of $f$ plotted are limited to the region of physical validity delineated in \S\ref{sec:physical_parameters}.}
    \label{fig:DeltaCritical}
\end{figure}

During EMD stable clustering the halo maintains constant physical density so the density ratio increases rapidly: $\bar{\Delta}\propto a^3$. During LRD the LM halo experiences adiabatic and free expansion, both of which are faster than the cosmological expansion rate, causing decreases in the density ratio: $\bar{\Delta}\propto a^3\,e^{-3\,t/\tau}$ (adiabatic) and $\bar{\Delta}\propto(\ln a/\afree)^{-3}$ (free). During LMD a shell may either recollapse or continue expanding depending on whether the shell is gravitationally bound or free. A bound shell has specific gravitational binding energy, $U=G\,\Mlm/r$ larger than specific kinetic energy $K=(\dot{r}^2+(L/r)^2)/2$, while free shells have $K>U$. During adiabatic expansion $U$ and $K$ retain their Virial ratio, $U\approx2\,K$. During free expansion the trajectories become nearly radial so $K\rightarrow\dot{r}^2/2$ so we find 
using Eq.~\ref{eq:delta_defA} to relate shell radius to the density ratio $\bar{\Delta}$
\beq
   \frac{U}{K} = \frac{\rhomat}{(\rhomat + \rhorad)} \frac{\bar\Delta^3}{(\bar\Delta - \frac{a}{3}\bar\Delta')^2}\;\;\;
\genfrac{}{}{0pt}{1}{\longrightarrow}{a_\tau\ll a\ll\aeq}\;\;\;  \frac{a}{(\aeq + a)}
\left(\frac{\lfree}{1+\lfree}\right)^2\bar\Delta
\eeq
The last expression, derived from Eq.~\ref{eq:delta_circular_asymptotic} and valid during LRD free expansion, gives $\lim_{a\rightarrow\aeq}U/K\approx\Deltaeq/2$ where $\Deltaeq\equiv \bar{\Delta}(\aeq)$. Thus if $\Deltaeq\ll1$ a shell is unbound and will continue to expand and never recollapse; while if $\Deltaeq\gg1$ the shell is bound entering the matter era and will recollapse. 
We show $\Deltaeq$ as a function of $\Deltatau$ in Fig.~\ref{fig:Deltaeq}.
One sees that the sign of the overdensity $\Deltaeq-1$ is a rough indicator of collapse just as with linear theory growing modes. The separatrix between these different behaviors we denote by $\Deltacrit(a)$ which is the solution of Eq.~\ref{eq:Delta_eomA} for which $\Deltacrit(\infty)=1$. This specifies a value $\Deltatau=\Deltatau^\text{crit}$. In Fig.~\ref{fig:DeltaCritical} we plot $\Deltatau^\text{crit}$ and $\Deltaeq^\text{crit}\equiv\Deltacrit(\aeq)$ for various values of $f$.   One sees that for allowed values of $f$ that $\Deltatau^\text{crit}\gtrsim10^{20}$. One can understand this from Eq.~\ref{eq:delta_circular_asymptotic} which estimates
$\Deltaeq\approx5.33\,\Deltatau^{1/4}
\left(\ln\frac{2}{9}\Deltatau\right)^{-3/2}\,\left(\ln f^{-1}\right)^{-3}$
so the condition for recollapse, $\Deltaeq\gtrsim1$, is $\Deltatau\gg(\ln f^{-1})^{12}$ which is extremely large since $f\lesssim10^{-8}$. Such large overdensities are not totally implausible: since $\bar{\Delta}\propto a^3$ during stable clustering and $\Deltatau\sim10^{21}$ could be attained when the universe expands by a factor $\gtrsim10^7$ during EMD. Note that this means that these objects must have formed extremely early and therefore must be very light compared to the mass contained within the horizon at reheating.

For $\Deltatau<\Deltatau^\text{crit}$ an isolated halo becomes unbound and will not, by itself, recollapse to form a bound structure during LMD. In this case individual EHs which started as large overdensities during EMD are diluted below the cosmological mean density during LMD. It is only a superposition of these underdense remnants which will allow bound structures to form at late times.

\subsection{Halo Remnant Density Profile}
\label{sec:universal_profile}

\begin{figure}
    \centering
    \includegraphics[width=0.9\textwidth]{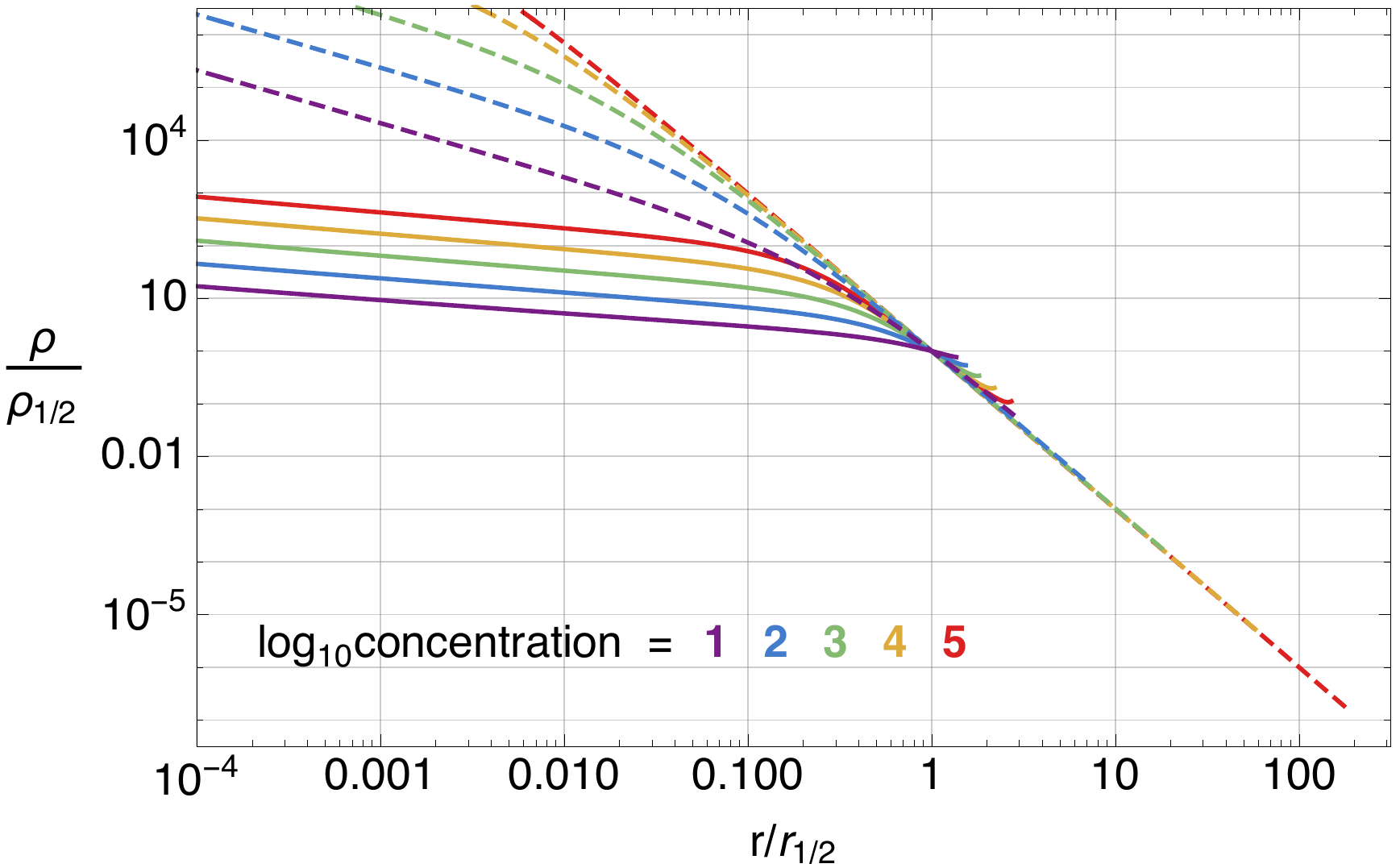}
    \caption{Density profiles of early halos relative to their half mass radius, $r_{1/2}$, and the density at this radius $\rho_{1/2}$. The dashed curves are the initial (NFW) density profile and the solid curve is the density profile of the asymptotic remnant from Eq.~\ref{eq:delta_circular_asymptotic} for $a=10^8\,\arh$ (the profile shape is insensitive to $a\gg \arh$). The different curves have different concentrations $=\ r_\text{vir}/r_\text{s}$.
    }
    \label{fig:NFWprofiles}.
\end{figure}

\begin{figure}
    \centering
    \includegraphics[width=0.9\textwidth]{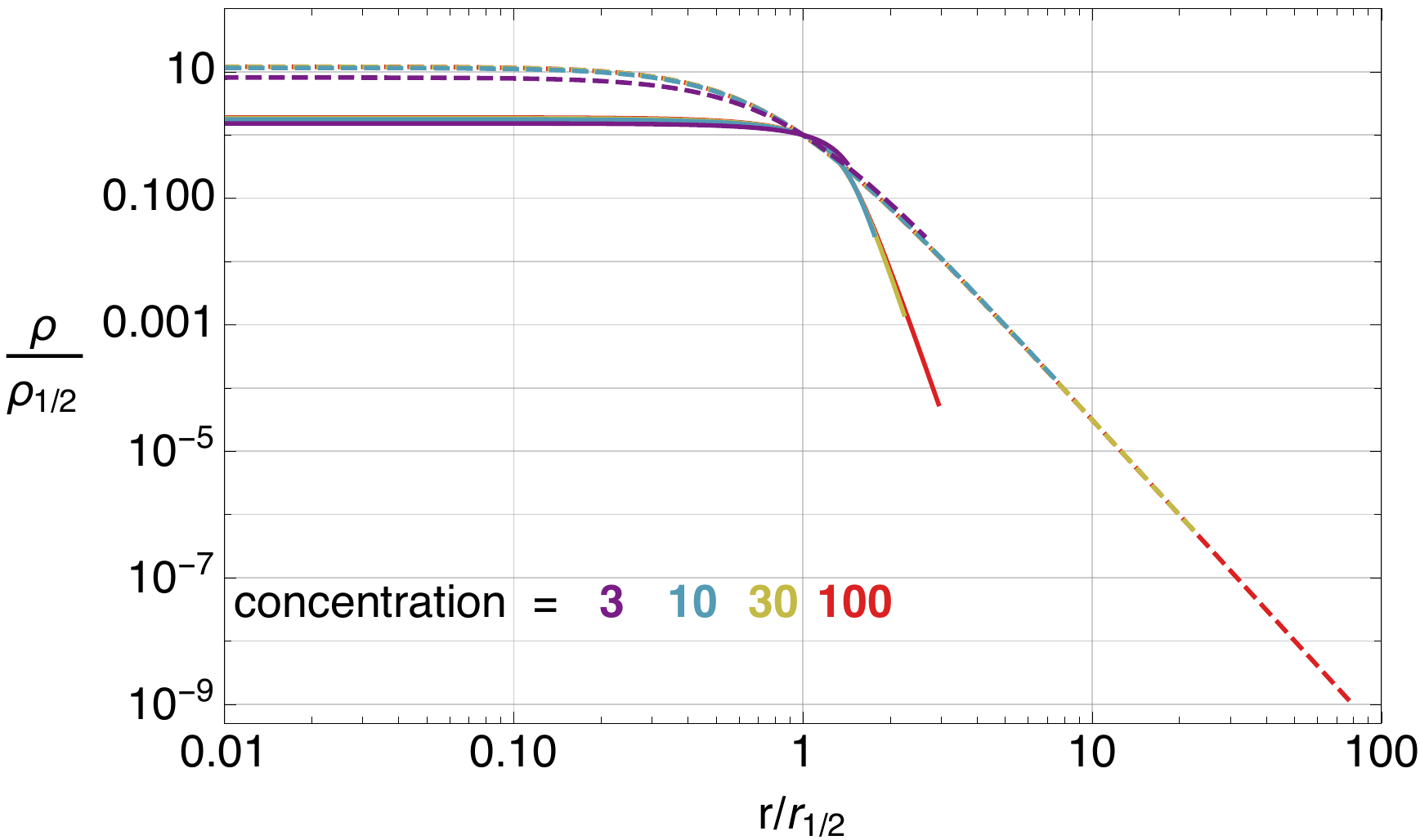}
    \caption{Same as Fig.~\ref{fig:NFWprofiles} except for initial Plummer density profiles.
    }
    \label{fig:PlummerProfiles}.
\end{figure}

The nature of the bound structures which collapse during LMD will depend on the spatial distribution of LM given by the density structure of individual halo remnants and how these remnants are clustered. Here we again consider only individual halos treating them as isolated objects. As shown above the density profile of a halo remnant depends on the initial density profile of the EH. As we have not specified how EHs form there is a large degree of uncertainty in what these initial profiles are, though they clearly must satisfy the rules of gravitational clustering such as stability. Since we have already assumed sphericity this generally requires the density decreases with radius. Here we consider a variety of models of static stable spherical non-relativistic halos of gravitationally bound non-interacting particles using Eqs.~\ref{eq:r_scaling}-\ref{eq:expansion_factor}, \ref{eq:delta_circular_asymptotic} to determine the asymptotic remnant profile. While this is formally correct for initially circular orbits Appendix~\ref{sec:shell_code} suggests it is a reasonable approximation more generally. Note that the density profile does not specify the phase space distribution and any density profile is consistent with purely circular orbits.

One characterizes a density profile by the radially-dependent logarithmic slope $\gamma(r)\equiv-\frac{\partial\ln\rho}{\partial\ln r}\ge0$ (so $\rho\genfrac{}{}{0pt}{1}{\propto}{\sim}r^{-\gamma}$). Steepness/shallowness increases/decreases with $\gamma(r)$. Near the center  
EHs may have diverging (cusped $0<\gamma(0)<3$) or finite (cored $\gamma(0)=0$) density. From Eq.~\ref{eq:delta_circular_asymptotic} one finds the central slope of the asymptotic remnant is related to the central slope of the EH by
\beq
\gamma_\text{rem}(0)=\frac{\gamma_\EH(0)}{4-\gamma_\EH(0)}
\label{eq:central_slope}
\eeq 
due to  $\bar{\Delta}\sim\Deltatau^{1/4}$ and $r\sim r_0 \Deltatau^{1/4}$. Thus a cored EH evolves to a cored remnant and a cusped EH evolves to cusped remnant but with a shallower central slope. This indicates that the density is more uniform in the remnant than in the EH. The difference can be quite dramatic, e.g.
$\rho_\EH\propto r^{-1}\rightarrow\,\rho_\text{rem}\propto r^{-1/3}$.

Cosmological halos usually form by gravitational coalescence from a uniform medium and will grow in size and mass as more matter is accreted and becomes virialized. The outer edge of accretion, known as the virial radius, $\rvir$, is usually characterized by 
$\Delta_\text{vir}\equiv\frac{\rho_\text{vir}}{\rhomat}
=\frac{d(r^3\,\bar{\Delta})}{d r^3}\vert_{r=\rvir}\approx200$.\footnote{Note that here we distinguish the density ratio $\Delta = \rho/\rhomat$ from the averaged density ratio $\bar\Delta$ defined in Eq.~\ref{eq:delta_defA}.} In contrast circular orbit halos have fixed physical size and do not allow for accretion since this involves radial not circular motion. The scenario we consider assumes most matter is virialized (described by stable clustering) by the end of EMD with only slow ongoing accretion / hierarchical clustering. To avoid small $\Deltatau$ we need to define an edge to EHs. We do this by choosing an appropriate $\rvir$ for the end of stable clustering and the beginning of adiabatic expansion at $t\sim\tau$. We choose this transition as the time when $\bar{\Delta}$ reaches its maximum value. As can be seen in Fig.~\ref{fig:Delta_evolution} $\max\bar{\Delta}\approx\frac{1}{20}\Deltatau$ so we define the edge of a halo, i.e. virial radius or virial shell, by $\Deltatau^\text{vir}=20\times200$.
This model may not be accurate for the small fraction of matter with $r\gtrsim\rvir$.

Interior to a virial shell some but not all halos have an ``extended atmosphere'' which we define as an outer region which has large overdensity $\Delta_\tau\gg1$ but contains a tiny fraction of the mass and $\Delta_\tau\ll\Deltatau$. If the EH has a power law atmosphere $\Delta_\tau\genfrac{}{}{0pt}{1}{\propto}{\sim}r_0^{-\gamma_\mathrm{EH}^\mathrm{atm}}$ then using the large logarithm approximation and Eqs.~\ref{eq:delta_circular_asymptotic}-\ref{eq:densityratio_circular_asymptotic} the remnant also has an atmosphere with
\beq
\gamma_\mathrm{rem}^\mathrm{atm}=4\,\gamma_\mathrm{EH}^\mathrm{atm}-9\ .
\eeq
Since an extended atmosphere only exists for $\gamma_\mathrm{EH}^\mathrm{atm}>3$ (otherwise the mass does not converge for large $r_0$) the slope of the remnant atmosphere is steeper than that of the EH: $\gamma_\mathrm{rem}^\mathrm{atm}>\gamma_\mathrm{EH}^\mathrm{atm}$.

Here we consider two contrasting models for EH density profile:
\beq
\rho_\text{NFW}(r)
=\rho_\text{vir}\,\frac{\rvir\,(\rvir+\rs)^2}{r\,(\rs+r)^2}
\qquad\qquad
\rho_\text{P}(r)=\rho_\text{c}\,\left(1+\left(\frac{r}{\rs}\right)^2\right)^{-5/2}\ .
\eeq
$\rho_\text{NFW}$ is the Navarro-Frenk-White profile~\cite{Navarro1996} which is currently the most commonly used model for dark matter halos in the late-time universe is and $\rho_\text{P}$ is the Plummer profile developed to model globular clusters \cite{Plummer1911}. $\rho_\text{vir}$ is the virial density, $\rho_\text{c}$ the central density and $\rs$ a characteristic radius. We define the concentration $\mathcal{C}\equiv\rvir/\rs$ which is usually a large number. The contrasting properties are 1) NFW has a cusp and Plummer a core, 2) for Plummer most mass resides at $r\lesssim\rs\ll\rvir$ the total mass rapidly converging outside of this while for NFW the mass increases $\propto\ln r$ for $r\gg\rs$ and most resides at $\rs\ll r\ll\rvir$.

Given the EH profiles above the asymptotic solution in Eq.~\ref{eq:delta_circular_asymptotic} can be used to derive the density profile of the remnant:
\beq
\rho_\mathrm{rem}(r) = \frac{3\Mlm(r)}{4\pi r^3} \left(1 + \frac{1}{3}\frac{d\ln\bar\Delta}{d\ln r}\right),
\label{eq:transformed_density_profile}
\eeq
where the $r$-dependence of $\bar\Delta$ is through $\Deltatau$ (see Eq.~\ref{eq:delta_circular_emd}) and the original radius of a mass shell, $r_0$ (related to $r$ via the asymptotic solutions in Eqs.~\ref{eq:r_scaling} and~\ref{eq:expansion_factor}). Note that since 
adiabatic expansion and peeling preserve shell order, the mass interior to $r$ at late times, $\Mlm(r, a\gg \arh)$ is equal to the mass 
interior to $r_0(r)$ at early times, $\Mlm(r_0,a<\arh)$.

In Fig.~\ref{fig:NFWprofiles} and  \ref{fig:PlummerProfiles} we show the density profiles of EHs and their remnants for various concentrations assuming the former are of the NFW and Plummer forms, respectively. The curves are normalized to the half mass radius and density. In both cases the remnant density profile is very shallow near the center and most of the mass resides at radii near the outer radius in the remnant which is a manifestation of the shallower density slope. 

In summary remnants of EHs are not only larger in physical and comoving size relative to the original halo but the their density profile significantly differs form the original halo. The inner profile is more shallow and the outer profile is steeper. In other words mass from an inner cusp is drawn outward being distributed more uniformly while the mass from any outer extended atmosphere is drawn inward leading to a sharper boundary to the remnant. The remnant mass is distributed among a smaller range of scales ($\ln r$)  than its progenitor, closer to that of a Gaussian or top hat profile.

Our calculation of the remnant 
density profile relied on a shell model that does not fully account for the distribution of DM velocities in a bound halo. Particles that are in some common volume element in the halo at an initial instant in time will generally have different angular momenta and will therefore be on very different orbits that are not described by the same dynamical time (and for highly elliptical orbits the dynamical time is different in different phases of the orbit); as a result, these particles will exit adiabatic expansion at different times and in different phases of their orbits, which intuitively leads to a further flattening of the remnant density profile. In the following section we will show that treating the remnant as having a top hat profile gives a simple quantitative understanding of the power spectrum even when the original EH was NFW. Thus, a further flattening of the remnant profile due to the velocity dispersion should only serve to make the analogy with a top hat better.

\subsection{Physical Parameters}
\label{sec:physical_parameters}
In this section we collect various observational and theoretical 
constraints on physical quantities encountered so far. 

Observational constraints limit viable values of the cosmological parameters ($\rhoem$, $\rholm$, $\tau$) and the derived parameter $f\equiv\lim_{t\rightarrow0}\rholm/(\rhoem+\rholm)$. 
The successful $\Lambda$CDM model of late time cosmology sets matter-radiation equality (LRD-LMD transition) at $t=\teq\simeq51\,\text{kyr}$. In order for Big Bang Nucleosynthesis to yield the observed abundance of light elements requires LRD to begin with a reheating temperature, $\TRH\ge5\,$MeV (\cite{Kawasaki:2000en,Hannestad:2004px,deSalas:2015glj,Hasegawa:2019jsa}) corresponding to $\tau\lesssim0.03\,$sec. Comparing $0.03\,$sec to $51\,$kyr one finds $f\lesssim10^{-7}$.

Physical constraints limit the region of validity of the Newtonian analysis of this section, i.e. limit viable values of the EH parameters ($M_0$ and $r_0$) and derived quantities $\delta_\tau=\frac{2}{9}\Deltatau=G\,M_0\,\tau^2/r_0^3$. In order for the halo to have attained stable clustering before EM decay one requires $\tau\gg\tdyn=\sqrt{r_0^3/(G\,M_0)}$ or $\delta_\tau\gg1$.

A less certain constraint comes from shortest timescale on which our physical model is valid. If one takes the requirement that all timescales be much greater than the Planck time, $t_\text{pl}$, then $\tau\approx f^2\,\teq\gg t_\text{pl}$ and $\tdyn=\tau/\sqrt{\delta_\tau}\gg t_\text{pl}$ then one finds $10^{-28}\ll f\lesssim10^{-7}$ and $1\ll\delta_\tau=\frac{2}{9}\Deltatau\ll 10^{83}\,f_{-7}^4$ where $f_{-7}\equiv f/10^{-7}$. This does not make any allowance for a period of inflation which would further bound $f$ from below. 

Newtonian clustering requires non-relativistic EHs: $G\,M_0/r_0\ll c^2$ or $r_0\ll10^4\,\text{km}\,f_{-7}^2\,\delta_\tau^{-1/2}$. The halos will become more non-relativistic during their subsequent expanding evolution. If $\Deltatau<\Deltatau^\text{crit}$ the halo remnant will expand by a factor given by Eqs.~\ref{eq:expansion_factor} and~\ref{eq:expansion_coefficients} until $a\sim\aeq$. Translating this limit on $r_0$ to the final expanded remnant size one finds 
$r_{\rm eq}\ll10^{11}\text{km}\,f_{-7}^2\,
\sqrt{\frac{\ln\delta_\tau}{\sqrt{\delta_\tau}}}$. EHs with $\Deltatau\gg\Deltatau^\text{crit}$ will collapse earlier during LRD and leave even smaller remnants. 

If EHs coalesce from a uniform expanding cosmology then causality requires that their mass be much less than the horizon mass at $t=\tau$ or earlier. This requirement is $M_0\lesssim c^3\,\tau/G=10^4\,\SM\,f_{-7}^2$. The remnant LM mass must then satisfy $\Mlm=f\,M_0\lesssim10^{-3}\,\SM\,f_{-7}^3$. The largest halo remnant in size and in mass is smaller than any dark matter structure yet observed.

Two physical quantities which do not enter our analysis are the masses of EM and LM particles. In order to form gravitationally bound structures both of their de Broglie wavelengths should be less than the size of the halo structures. This requirement may be written
$m_\text{min}\equiv\text{min}(m_\text{EM},m_\text{LM})\gg\hbar/\sqrt{G\,M_0\,r_0}
=\hbar\,(\sqrt{\delta_\tau}/(G\,M_0)^2/\teq)^{1/3}\,f^{-2/3}$
or using the upper limits on $M_0$ one finds
$m_\text{min}\,c^2\gg10^{-14}\;\text{eV}\,\delta_\tau^{1/6}\,f_{-7}^{-2}$. 
This can be an interesting constraint in models of ultralight DM, such as axion-like particles or vector DM. 

Thus the early halo model considered in this paper can only accommodate ratios of late matter to early matter in the range $10^{-28}\ll f\lesssim10^{-7}$ producing remnant halo structures with mass $\ll10^{-3}\SM$. This gives the characteristic mass scale of the smallest structures which will recollapse at late times (microhalos). Lower bounds on $f$ and the microhalo mass scale are more uncertain, depending on assumptions for very early cosmology at the inflation or Planck scale and on the particle nature and phase space distributions of EM and LM.

\section{Superposition of Halo Remnants}
\label{sec:halo_superposition}

As halos remnants expand in comoving size they will begin to overlap. Overlapping spherical halos are necessarily aspherical, so the spherical halo model used so far cannot accommodate the gravitational field of neighboring remnants. Fortunately this additional force is probably not important as we argue below that overlap occurs when $\bar{\Delta}\lesssim1$ and we see from Fig.~\ref{fig:Deltaeq}, that $\bar{\Delta}\gg1$ at the start of free expansion. During free expansion the self-gravity of the matter becomes unimportant so when neighboring halo remnants do overlap there is no significant interaction between them. As a result, the matter distribution is simply the linear superposition of the spherical matter distribution of individual remnants:
\beq
\rho(\bfx)=\rhomat\,\sum_h\Delta_h(|\bfx-\bfx_h|)
\label{eq:eh_superposition}
\eeq
where $h$ labels individual remnants, $\bfx_h$ is the halo center, and $\Delta_h(r)$ gives its density profile.
We first use the results from previous sections to show that 
the halo remnants overlap in \S\ref{sec:overlap}. We then 
apply this (early) halo model description of the density field 
to express the matter power spectrum at late times in terms of 
the density profiles of the remnants in \S\ref{sec:one_and_two_halo_power}.

\subsection{Remnant Overlap}
\label{sec:overlap}

One can quantify the amount of overlap of remnant $h$ by the ratio\footnote{The formulae of this section generalize to non-spherical remnants by $|\bfx-\bfx_h|\rightarrow\bfx-\bfx_h$.}
\beq
\mathcal{N}_h\equiv
\frac{\int d^3\bfx\,\Delta_h(|\bfx-\bfx_h|)\,\sum_{h'\ne h}\Delta_{h'}(|\bfx-\bfx_{h'}|)}
     {\int d^3\bfx\,\Delta_h(|\bfx-\bfx_h|)^2}\ ,
\label{eq:overlap_number}
\eeq
where the denominator is the density weighted average of $\Delta_h(r)$ of remnant $h$ and the numerator the same weighted average of the sum of $\Delta_{h'}$'s for all {\it other} halos. Roughly speaking the ``overlap number'', $\mathcal{N}_h$, is the number of other remnants which overlap remnant $h$: iff there is no overlap then $\mathcal{N}_h=0$ while $\mathcal{N}_h\gg1$ indicates a great deal of overlap. One can place lower limits on the amount of overlap from the values $\Delta$ in halos though specific values depend on the clustering of the initial EHs: if EHs tend to be near to each other then they overlap more readily. 

Let us consider the matter overdensity
\beq
\deltam(\bfx)\equiv\frac{\rho(\bfx)}{\rhomat} - 1
=-1+\sum_h \Delta_h(|\bfx-\bfx_h|)\ .
\label{eq:overdensity}
\eeq
Denote spatial averages by 
$\langle f(\bfx)\rangle_\bfx\equiv V^{-1}\int d^3\bfx\,f(\bfx)$
where the (usually infinite) cosmic volume is $V\equiv\int d^3\bfx$. Since the halos are assumed to contain all the mass it follows that 
$\langle\sum_h \Delta_h(|\bfx-\bfx_h|)\rangle_\bfx=1$ or $\langle\deltam(\bfx)\rangle_\bfx=0$. One can quantify the amount of overlap by the decomposition
\begin{eqnarray}
1+\langle\deltam(\bfx)^2\rangle_\bfx&=&
\left\langle\sum_h\sum_{h'}\Delta_h(|\bfx-\bfx_h|)\,\Delta_{h'}(|\bfx-\bfx_{h'}|)
 \right\rangle_\bfx \nonumber \\
&=&\left\langle\sum_h\Delta_h(|\bfx-\bfx_h|)^2\right\rangle_\bfx
+\left\langle\sum_{h\ne h'}\Delta_h(|\bfx-\bfx_h|)\,\Delta_{h'}(|\bfx-\bfx_{h'}|)
\right\rangle_\bfx \label{eq:delta2_decompositionA}\\  \nonumber
&=&\sum_h (1+\mathcal{N}_h)\,\left\langle\Delta_h(|\bfx-\bfx_h|)^2\right\rangle_\bfx\ .
\end{eqnarray}
For each remnant define the density weighted average $\Delta$ by
\beq
\langle\Delta\rangle_h\equiv\frac{\int d^3\bfx\,\Delta_h(|\bfx-\bfx_h|)^2}
                                 {\int d^3\bfx\,\Delta_h(|\bfx-\bfx_h|)  }
=\frac{\rhomat}{M_h}\,\int d^3\bfx\,\Delta_h(|\bfx-\bfx_h|)^2 \ .
\eeq
where $M_h=\rhomat\,\int d^3\bfx\,\Delta_h(|\bfx-\bfx_h|)$ is the mass of the remnant. Thus
\beq
1+\langle\deltam(\bfx)^2\rangle_\bfx
=\sum_h \frac{M_h}{\rhomat\,V} (1+\mathcal{N}_h)\,\langle\Delta\rangle_h
=\langle (1+\mathcal{N}_h)\,\langle\Delta\rangle_h\rangle_\mathrm{rem}\ .
\eeq
where $\langle f_h\rangle_\mathrm{rem}\equiv\sum_h \frac{M_h}{\rhomat\,V}\,f_h$ is a mass weighted average over halos since $\sum_h M_h=\rhomat\,V$. Defining a mean halo density ratio and a mean overlap number 
\beq
\langle\Delta\rangle\equiv\langle\langle\Delta\rangle_h\rangle_\mathrm{rem} \qquad
\langle\mathcal{N}\rangle
\equiv\frac{\langle\langle\Delta\rangle_h\,\mathcal{N}_h\rangle_\mathrm{rem}}
           {\langle\langle\Delta\rangle_h\rangle_\mathrm{rem}}\ .
\eeq
leads to the simple expression
\beq
\langle\deltam(\bfx)^2\rangle_\bfx=(\langle\mathcal{N}\rangle+1)\,\langle\Delta\rangle-1
\ .
\label{eq:total_power}
\eeq
Since $\deltam^2\ge0$ it follows that
\beq
\langle\mathcal{N}\rangle\ge\frac{1}{\langle\Delta\rangle}-1 \ .
\label{eq:overlap_number_bound}
\eeq
If $\langle\Delta\rangle\ge1$ this is not a new constraint since $\langle\mathcal{N}\rangle\ge0$ by construction. If $\langle\Delta\rangle\ll1$ we see that $\langle\mathcal{N}\rangle\gg1$ and there is a great deal of overlap.

The bound of Eq.~\ref{eq:overlap_number_bound} depends only on the density profiles and distribution of masses, $M_h$, of remnants and not how they are spatially distributed (clustering). The actual value of $\langle\mathcal{N}\rangle$ does depend on clustering. $\langle\mathcal{N}\rangle$ could be much larger than the bound of Eq.~\ref{eq:overlap_number_bound} if the remnant centers are strongly clustered. Our expectation is that clustering is modest so a baseline value is given by assuming the EH centers are uncorrelated in position in which case\footnote{In the limit $V\rightarrow\infty$ assuming the rms $M_h$ is finite.}
\beq
\langle\deltam(\bfx)^2\rangle_\bfx\approx\langle\Delta\rangle \qquad
\langle\mathcal{N}\rangle\approx\frac{1}{\langle\Delta\rangle} \ ,
\label{eq:total_power_unclustered}
\eeq
which follows from Eqs.~\ref{eq:delta2_decompositionA} and~\ref{eq:total_power}.
This is only slightly larger than the bound if $\langle\Delta\rangle\ll1$. Modest clustering would lead to similarly small fractional changes in the value of $\langle\mathcal{N}\rangle$. 

One finds from Fig.~\ref{fig:Deltaeq} that $\Delta<\bar{\Delta}\lesssim10^{-3}$ when $a\approx\aeq$ unless $\Deltatau\gg10^{10}$. Hence we expect $\langle\Delta\rangle\lesssim10^{-3}$ and $\langle\mathcal{N}\rangle\gtrsim10^3$ at the time of matter-radiation equality, i.e. a great deal of overlap.

\subsection{One- and Two-Halo Power Spectra}
\label{sec:one_and_two_halo_power}

The linear superposition of EH remnants 
in Eq.~\ref{eq:eh_superposition} is identical to the starting point of the halo model of 
DM density field as described in, e.g., Refs.~\cite{1991ApJ...381..349S,Cooray:2002dia}. 
We review aspects of this formalism in Appendix~\ref{sec:stats}. We are ultimately interested 
in estimating the power spectrum at late times in 
order to infer structure formation history after matter-radiation equality. In the halo model, the power 
spectrum decomposes into one- and two-halo terms as
\beq
P(k) = P_{1\mathrm{h}}(k)+P_{2\mathrm{h}}(k),
\eeq
where
\begin{subequations}
\begin{align}
P_{1\mathrm{h}}(k) & = \frac{1}{\rholm a^3}\int dm \frac{df}{d\ln m} |\mathcal{F}(k|m)|^2 \label{eq:nl_ps_1h_main}\\
P_{2\mathrm{h}}(k) & = \left\vert\int d\ln m \frac{df}{d\ln m} b(m) \mathcal{F}(k|m)\right\vert^2 \tilde{P}_{hh}(k)\label{eq:nl_ps_2h_main},
\end{align}
\label{eq:p1h_and_p2h}
\end{subequations}
where 
\beq
\frac{dn}{d\ln m} = \frac{\rholm a^3}{m}\frac{df}{d\ln m}
\eeq
is the halo mass function;  $\tilde{P}_{hh}(k)$ is a 
halo-halo power spectrum; $b(m)$ is a bias function that relates halo and mass power spectra\footnote{We will set $b \approx 1$ for simplicity but in principle the bias can be estimated analytically or from $N$-body simulations~\cite{Cooray:2002dia}.}; $\mathcal{F}$ is the mass-normalized 
Fourier transform of the density profile in comoving coordinates
\beq
\mathcal{F}(k, t | m) =  \frac{4\pi  a^3}{m} \int_0^{\infty} d\rco \rco^2\, j_0 (k\rco) \rho_\EH(\rco,t| m),
\label{eq:fdef_text}
\eeq
where $\rho_\EH = \rhomat \Delta_h$ for some $h$ (assuming all EHs have similar profiles and just differ in mass $m$).
Note that $\mathcal{F}\to 1$ as $k\to 0$, which ensures that 
$P(k)$ reduces to the linear power spectrum on large enough scales (see below).
The expressions in Eq.~\ref{eq:p1h_and_p2h} are remarkable in that they nearly factorize effects from linear and non-linear scales; the integrals over the halo mass distributions 
depend on the halo profiles, while the halo-halo 
power spectrum is closely related to the linear power spectrum. The tilde in $\tilde{P}_{hh}(k)$, however, indicates that we must model the fact that on small scales prior to reheating halos did not overlap. This exclusion effect ensures that $\tilde{P}_{hh}(k) \approx P^{\mathrm{lin}}(k)$ for $kr_e\ll 1$, and $\tilde{P}_{hh}(k) \approx C$ for $kr_e \gg 1$ with $r_e$ the typical exclusion radius; the constant $C$ is fixed such that the probability to find two halos with separation $r \lesssim r_e$ vanishes before EHs explosively evaporate. This prescription is described in detail in Appendix~\ref{sec:halo_exclusion}.

In order to make use of Eq.~\ref{eq:p1h_and_p2h} we will need to compute $\mathcal{F}$, $P^{\mathrm{lin}}$ and $df/d\ln m$ in specific cosmological models. The normalized density profile $\mathcal{F}$ can be calculated from the remnant profile 
derived in the previous section, Eq.~\ref{eq:transformed_density_profile}, together with the 
definition in Eq.~\ref{eq:fdef_text}. The linear 
power spectrum $P^{\mathrm{lin}}$ and mass distribution $df/d\ln m$ can be obtained from linear perturbation theory and the Press-Schechter formalism which we discuss in the following section.

\section{Linear Evolution of Density Perturbations}
\label{sec:linear_evolution}

As shown in \S\ref{sec:isolated_halos} and \S\ref{sec:halo_superposition} even extremely overdense EHs disassociate  and their remnants overlap with neighboring remnants producing an LDM distribution with rms overdensity  $\langle\deltam(\bfx)^2\rangle_\bfx\lesssim 10^{-3}$ 
(discussion following Eq.~\ref{eq:total_power_unclustered}) by the beginning of LMD.  Thus a very inhomogeneous (non-linear) matter distribution evolves into nearly homogeneous (linear) distribution. In this section we study the linear evolution of DM density perturbations both during EMD and after reheating.  These results along with the Press-Schechter (PS) formalism~\cite{Press:1973iz} will enable us to estimate the properties of bound object of both early halos that eventually form during EMD and microhalos that form from density fluctuations that survive reheating or result from the overlap of EH remnants.

\subsection{Linear Evolution and Collapse Before Reheating}
The fluid equations describing the evolution of perturbations in 
LM, EM and radiation are given in, e.g., Refs.~\cite{Erickcek:2011us,Barenboim:2013gya,Fan:2014zua} and in Appendix~\ref{sec:linear_perturbations}.
During EMD these equations can be solved for the LM density 
contrast $\delta$~\cite{Erickcek:2011us}
\beq
\delta = \delta_i - \frac{2}{3} \left(\frac{k}{\krh}\right)^2 \left(\frac{a}{\arh}\right) \psi,
\label{eq:approx_sol_long_emd}
\eeq
where $\psi$ and $\delta_i$ are the primordial (super-horizon) values of the gravitational potential and of the EM and LM 
fluctuations, respectively; if these were born as adiabatic fluctuations 
then $\delta_i \propto \psi$. This result is only 
valid for modes that enter the horizon during EMD; if a mode enters 
during an early period of RD (before the universe transitions to EMD) 
the evolution is more complicated. Numerical solutions 
exemplifying these different regimes are shown in Fig.~\ref{fig:density_contrast_evolution} for a mode with 
$k/\krh = 600$, where $\krh = \arh H(\arh)$ is the value of the 
conformal Hubble parameter at reheating. Equation~\ref{eq:approx_sol_long_emd} corresponds to the 
upper line of Fig.~\ref{fig:density_contrast_evolution}. For the background 
cosmologies in the lower lines, this mode enters during radiation domination, before transitioning to EMD; the impact of the gravitational driving effect, a hallmark of radiation domination~\cite{Hu:1995en}, is evident as these modes enter the horizon. Note that in Fig.~\ref{fig:density_contrast_evolution} modes with the same physical size at late times (after reheating), enter the horizon at different times and 
experience different amounts of growth. In the cosmologies 
with only a brief period of EMD, the total energy density is 
greater at earlier times, implying that horizon entry of a mode 
with the same $k/\krh$ occurs later in scale factor evolution. As
a result, these modes have a smaller amplitude at reheating, compared to models with a long period of EMD.
Semi-analytic solutions for the density contrast evolution 
can be constructed in models with a brief period of EMD, e.g., by following the methods of Ref.~\cite{Hu:1995en}. For simplicity, we 
only present analytical expression for a period of long EMD; 
other models are treated numerically.

After a long period of EMD the initial condition in Eq.~\ref{eq:approx_sol_long_emd} becomes 
irrelevant and we have\footnote{We use $\langle\cdot\rangle$ as a 
shorthand power spectrum, or equivalently, the 
autocorrelation function with $(2\pi)^3 \delta^3 (0)$ factored out.}
\beq
\langle \delta^2 \rangle \approx \left[\frac{2}{3} \left(\frac{k}{\krh}\right)^2 \left(\frac{a}{\arh}\right)\right]^2 \langle \psi^2 \rangle\;\;\;(\ahor \ll a < \arh),
\label{eq:linear_evolution_during_emd}
\eeq
which is valid well after the horizon entry of mode $k$ but before the 
end of EMD.
The autocorrelation function of the gravitational potential is related to the curvature power spectrum as 
\beq
\langle \psi^2 \rangle = \left(\frac{3}{5}\right)^{2} \times \frac{2\pi^2}{k^3}A_s \left(\frac{k}{k_0}\right)^{n_s-1},
\label{eq:primordial_psi_ps}
\eeq
where the factor $3/5$ arises from conservation of the curvature perturbation for adiabatic fluctuations. While we do not know the spectrum of primordial curvature fluctuations on the 
small scales of interest, we will for simplicity 
take $A_s =2.1\times 10^{-9}$, $n_s = 0.965$ with $k_0 = 0.05/\Mpc$~\cite{Aghanim:2018eyx}. Note that this is an enormous extrapolation of the simple powerlaw ansatz from scales probed by the CMB, $\sim \keq$, to $k> 10^7 \keq$.

\begin{figure}
    \centering
    \includegraphics[width=0.47\textwidth]{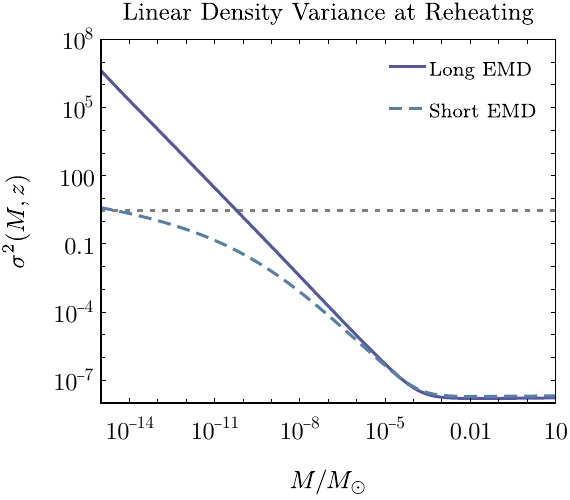}
    \includegraphics[width=0.47\textwidth]{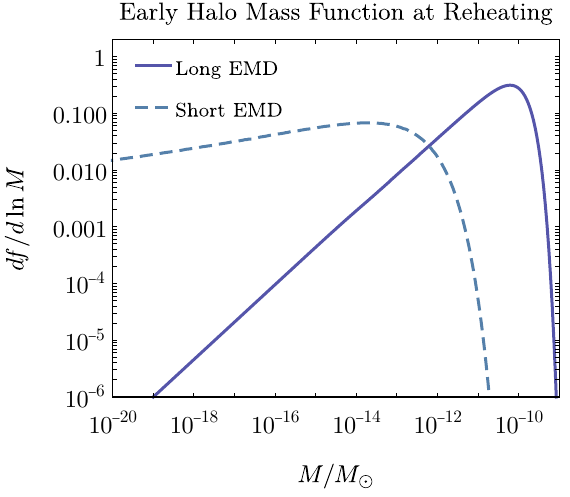}
    \caption{Linear theory density variance (left) and early halo mass function (right) computed using the PS formalism in Eq.~\ref{eq:PS_fraction} for the ``Long EMD'' and ``Short EMD'' cosmologies (solid and dashed lines in each panel, respectively). These quantities are evaluated at reheating (i.e., when the energy densities of EM and radiation are equal) and approximately describe the density field prior to adiabatic and free expansion of EHs, since these effects are not included in the linear theory. In the left panel, the gray dotted line 
    indicates the EMD spherical collapse threshold of $\delta_c^2 \approx 1.686^2$.}
    \label{fig:eh_density_variance_and_mass_function}
\end{figure}

The above results can be used to estimate various properties 
of collapsed objects that form \emph{during} EMD. 
First, it is useful to estimate the scale of the largest 
structures that can collapse during EMD. We can 
compute the mass and size of these non-linear 
structures using the Press-Schechter formalism 
by solving
\beq
\sigma(M_*, z_c) = \delta_c,
\label{eq:collapse_criterion}
\eeq
for the mass $M_*$ of one-sigma overdensities of the density field collapsing at $z_c$; in the above equation $\delta_c = 1.686$ is the collapse threshold and $\sigma^2$ is the density fluctuation variance
\beq
\sigma^2(M(R), z) = \int \frac{d^3 k}{(2\pi)^3} \langle \delta^2 \rangle W(kR)^2,
\label{eq:density_fluctuation_variance}
\eeq
where $W$ is a window function. The density variance for the ``Long EMD'' and ``Short EMD'' cosmologies  is shown in the left panel of Fig.~\ref{fig:eh_density_variance_and_mass_function} for $\TRH = 5\;\MeV$. Since this quantity is computed in linear perturbation theory, it does not include the effects of adiabatic and free expansion of collapsed objects near the end of EMD; however, we can use it to estimate the properties of the largest structures formed just before/at reheating, i.e., those with a collapse redshift $z_c = \zrh \sim 10^{10} (\TRH/5\;\MeV)$. 
The mass scale of these structures can be estimated from Eq.~\ref{eq:collapse_criterion} 
and corresponds to the intersection $\sigma^2$ with the gray dashed line equal to the collapse threshold $\delta_c^2$ in the left panel of Fig.~\ref{fig:eh_density_variance_and_mass_function}; numerically we find
\beq
\frac{M_* (\zrh)}{\MRH} \sim \begin{cases}
8\times 10^{-8} & \text{Long EMD} \\
5\times 10^{-12} & \text{Short EMD},
\end{cases}
\label{eq:largest_EH_mass}
\eeq
where $\MRH$ is the mass enclosed inside the horizon at reheating:
\beq
\MRH = 250\MEarth\left(\frac{5\;\MeV}{\TRH}\right)^3\left(\frac{10.75}{g_*(\TRH)}\right)^{1/2},
\label{eq:mrh_approx}
\eeq
where $\MEarth = 3\times 10^{-6}M_\odot$ is the Earth mass.
The tilde in the relation of Eq.~\ref{eq:largest_EH_mass} indicates that this is a window-function dependent quantity and therefore should be interpreted only as a order-of-magnitude characteristic mass of EHs forming near the end of EMD. 

We see that a shorter duration of EMD leads to much smaller EHs, since only the smallest scales have experienced enough growth to collapse. Using the Press-Schechter formalism~\cite{Press:1973iz} we can estimate the early halo distribution function from the density variance~Eq.~\ref{eq:density_fluctuation_variance}:
\beq
\frac{df}{d\ln m} = 
\sqrt{\frac{2}{\pi}} \frac{\delta_c}{\sigma} \left|\frac{d\ln \sigma}{d\ln m}\right| \exp\left(-\frac{\delta_c^2}{2\sigma^2}\right).
\label{eq:PS_fraction}
\eeq
This distribution, along with the linear power spectrum above and individual halo profiles are needed to evaluate the matter power spectrum \emph{after} reheating in Eq.~\ref{eq:p1h_and_p2h}. 
We show the distribution of EHs in the right panel of Fig.~\ref{fig:eh_density_variance_and_mass_function} for the ``Long EMD'' and ``Short EMD'' 
cosmologies. Note that in the latter example (dashed line) the distribution peaks at lower masses 
and has an extremely long tail as $M\to 0$; this is because modes that entered the horizon 
prior to the start of EMD grew only logarithmically with $a$ for $a < \aeeq$, resulting in a nearly scale-invariant fluctuation amplitude on as can be seen in the left panel of Fig.~\ref{fig:eh_density_variance_and_mass_function}.

To conclude the section we return to the evolution of linear perturbations described by Eq.~\ref{eq:linear_evolution_during_emd}.
After the transition to RD at $a = \arh$, linear perturbations 
continue to grow, albeit logarithmically with the scale factor. 
Matching the EMD and RD solutions at $\arh$ determines the 
power spectrum well after RH~\cite{Erickcek:2011us}
\beq
\langle \delta^2 \rangle \approx \left[\frac{2}{3} \left(\frac{k}{\krh}\right)^2 \ln\left(\frac{e a}{\arh}\right)\right]^2 \langle \psi^2 \rangle\;\;\;(\arh \ll a < \aeq).
\label{eq:long_emd_rd_correlation_func}
\eeq
This expression is the linear power spectrum of LM during radiation domination, $P^{\mathrm{lin}}$. More explicitly, using Eq.~\ref{eq:primordial_psi_ps}, gives
\beq
\frac{k^3}{2\pi^2} P^{\mathrm{lin}} =
\left[\frac{2}{5} \left(\frac{k}{\krh}\right)^2 \ln\left(\frac{e a}{\arh}\right)\right]^2 A_s \left(\frac{k}{k_0}\right)^{n_s -1}.
\label{eq:dimless_linear_ps_from_emd}
\eeq
In models with a finite duration of EMD, we evaluate $k^3 P/(2\pi^2)$ by 
using the numerical solution of $\langle \delta^2 \rangle$. 
Illustrative examples of these power spectra are presented in \S\ref{sec:late_time_ps_and_microhalos}.

\subsection{Linear Evolution and Collapse After Reheating}
\label{sec:evolution_after_mre}
In this section we briefly describe how the power spectrum after reheating evolves through matter-radiation equality and beyond. The power spectrum in Eq.~\ref{eq:dimless_linear_ps_from_emd} applies only to the ``Long EMD'' cosmology, and it is valid only for modes with $k>\krh$ during radiation domination. We therefore need a more general prescription to study the density field evolution for a wide range of scales and in different cosmologies. 
This evolution is captured by the Meszaros equation~\cite{Meszaros:1974tb,Hu:1995en}, whose general solution is 
\beq
\delta = \left[A_1 U_1(y)+  A_2 U_2(y)\right],
\label{eq:meszaros_sol}
\eeq
where $U_{1,2}(y)$ are functions of $y=a/\aeq$ (given in Ref.~\cite{Hu:1995en}), and $A_{1,2}$ are coefficients that are obtained by matching Eq.~\ref{eq:meszaros_sol} to numerics during radiation domination as described in Ref.~\cite{Blinov:2021axd}. The RD evolution can be written as (cf. Eq.~\ref{eq:long_emd_rd_correlation_func})
\beq
\delta = I_1 \psi \ln \left(\frac{I_2 a}{\ahor}\right)
\;\;\;(\arh \ll a < \aeq),
\label{eq:rd_form}
\eeq
where $\ahor(k)$ is the scale factor when the mode $k$ enters the horizon and $I_{1,2}$ are (typically) $k$-dependent coefficients that encode the pre-reheating evolution of small scales; $I_{1,2}$ are obtained by numerically solving for $\delta$ into RD and fitting the solution to Eq.~\ref{eq:rd_form}. Matching Eq.~\ref{eq:meszaros_sol} at $y\ll 1$ to Eq.~\ref{eq:rd_form}, yields expressions for $A_{1,2}$ in terms of $I_{1,2}$:
\begin{align}
  A_1 & \approx \frac{3}{2}\left[\ln \left(4I_2 e^{-3}\frac{\aeq}{\ahor}\right)\right]I_1 \psi.\\
  A_2 & \approx - \frac{4}{15}I_1 \psi.
\end{align}
For the modes of interest $k\gg\keq$, $A_1 \gg A_2$, and for $a\gtrsim \aeq$ Eq.~\ref{eq:meszaros_sol} simplifies to 
\beq
\delta \approx A_1 U_1(y),
\label{eq:meszaros_matched_sol_simplified}
\eeq
This is a useful result because the cosmology-dependent quantity $A_1$ factorizes from the scale factor evolution $U_1$, meaning that 
we can write the linear power spectrum in \emph{any} cosmology after MRE as 
\beq
P^{\mathrm{lin}}(k, a) = P_{\lcdm}^{\mathrm{lin}}(k,a) \left(\frac{\delta}{\delta_{\lcdm}}\right)^2,
\eeq
where $P_{\lcdm}^{\mathrm{lin}}$ is the $\lcdm$ power spectrum which can be computed using a Boltzmann code~\cite{Lewis:1999bs,2011JCAP...07..034B}, or using the transfer and growth functions of Ref.~\cite{Eisenstein:1997jh} (we use a combination of these methods as described in Appendix C of Ref.~\cite{Blinov:2019jqc}). The density contrast ratio 
$(\delta/\delta_{\lcdm})^2$ is independent of scale factor and therefore can be interpreted as a modification of the primordial power spectrum -- it is an explicit realization of the ``bump'' functions considered in Ref.~\cite{Blinov:2021axd}.

In summary, we have reduced the calculation of the linear power spectrum to the determination of $I_{1,2}$ which we obtain by numerically solving linear perturbation equations. In order to enable the evaluation of $P^{\mathrm{lin}}(k, a)$ across a wide range of scales, we construct fitting functions for $I_{1,2}$ which have 
the correct asymptotics for $k\ll \krh$ and $k\gg \krh$. For example, in the ``Long EMD'' model $I_1 \approx 2(k/\krh)^2 / 3$ and $I_2 \approx (\krh/k)^2$ for $k\gg \krh$; modes that enter the horizon during RD ($k\ll \krh$) modes that enter the horizon during RD have $I_1\approx 9.11$ and $I_2\approx 0.594$~\cite{Hu:1995en}. The details of the numerical solutions and the fitting functions are presented in Appendix~\ref{sec:linear_perturbations}. 
The resulting power spectra at matter-radiation equality are shown in Fig.~\ref{fig:ps_at_mre} for the ``Long EMD'' and ``Short EMD'' cosmologies as dashed lines. In the following section, we will include the non-linear effects of EH explosive evaporation at small scales.

\section{Power Spectrum and Formation of Microhalos}
\label{sec:late_time_ps_and_microhalos}
In this section we combine the results of 
Sections~\ref{sec:isolated_halos},~\ref{sec:halo_superposition} and~\ref{sec:linear_evolution} to compute the matter power spectra at small scales, and study early structure formation.

\begin{figure}
    \centering
    \includegraphics[width=0.47\textwidth]{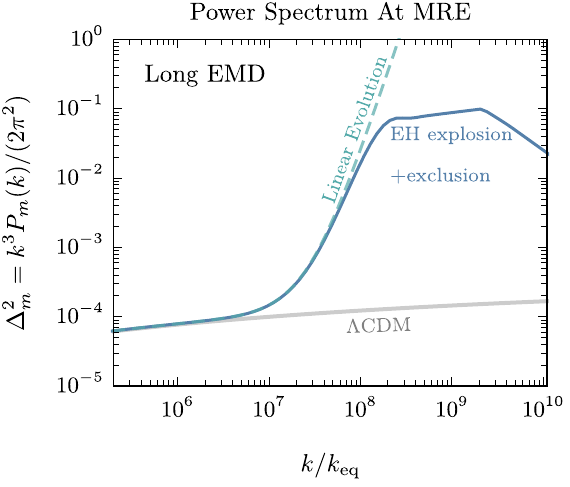}\;\;\;
    \includegraphics[width=0.47\textwidth]{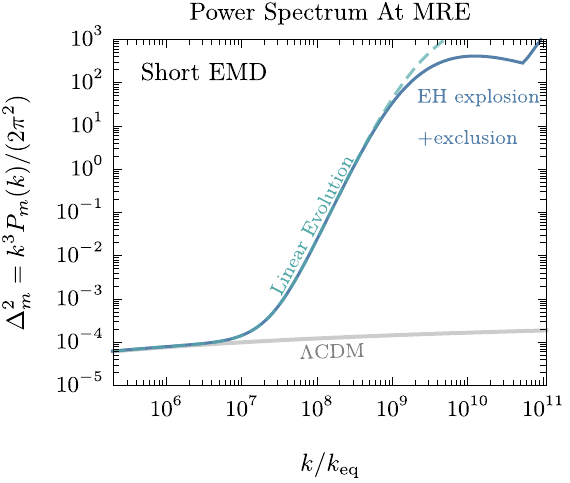}
    \caption{Matter power spectrum for a cosmology 
    with $\TRH = 5\;\MeV$ evaluated at matter-radiation equality. 
    In the left panel we show a variant where EMD lasted for a long time 
    and all modes with $k > \krh$ entered the horizon during EMD. 
    In the right panel EMD has a finite duration, and some modes entered the 
    horizon during a period of early radiation domination which was then followed by EMD. 
    In both panels the dashed lines are the linear power spectra without taking into account the disruption of halos that formed during EMD, while the solid line includes the effects of explosive evaporation of early halos at the end of EMD, which suppresses power at small scales. The gray band is the $\Lambda$CDM prediction. In the right panel the $\Delta_m^2 \gg 1$ region is shown for comparison with ``Long EMD'' only; linear perturbation theory used to evolve the PS to this point breaks down in this regime.
    \label{fig:ps_at_mre}}
\end{figure}

In Fig.~\ref{fig:ps_at_mre} we show two benchmark DM power spectra evaluated at matter-radiation 
equality, corresponding to the ``Long EMD'' and ``Short EMD'' background cosmologies in 
Fig.~\ref{fig:bg_density_evolution} and described in \S\ref{sec:background_cosmo}.
In each panel the gray line labeled $\Lambda$CDM indicates the expected spectrum in the absence of EMD based on the scale invariant primordial 
spectrum in Eq.~\ref{eq:primordial_psi_ps}; the slow growth with $k$ is a consequence of logarithmic 
growth of perturbations during radiation domination. 
The dashed lines indicate the linear power spectrum that does \emph{not} account for formation 
and eventual disruption of EHs during reheating. These power spectra monotonically increase at large $k$ unless there is some small-scale cutoff due to the nature of the DM particle. These spectra 
first diverge from the $\Lambda$CDM expectation for wavenumbers comparable to the comoving 
horizon size at reheating~\cite{Erickcek:2011us,Blinov:2019rhb}:
\beq
  \frac{\krh}{\keq} \approx 5.9\times 10^6 \left(\frac{\TRH}{5\;\MeV}\right)\left(\frac{g_*(\TRH)}{10.75}\right)^{1/6}.
\label{eq:krh_over_keq}
\eeq
In the left panel, the EMD enhancement scales as $(k/\krh)^4$ as expected from the analytic arguments around Eq.~\ref{eq:linear_evolution_during_emd}. In contrast, in the ``Short EMD'' example in the right panel, 
the enhancement becomes logarithmic in $k \gtrsim k_{\mathrm{eeq}} \sim \sqrt{\arh/\aeeq}\krh$ for modes that began their evolution before EMD in the early radiation era. 
The solid blue lines in both panels incorporate the effects of EH explosive evaporation as described in \S\ref{sec:one_and_two_halo_power}. We see that the impact of reheating on small scale power is significant: the maximum amplitude of the power spectrum can be significantly reduced compared to the naive linear theory 
expectation indicated by the dashed lines. The scale at which the full result begins to deviate from 
linear theory corresponds precisely to the size of EHs after they go through free expansion; this 
scale can be estimated by applying Eq.~\ref{eq:rofa_circular_asymptotic} to the virial radius of the largest EHs that have formed by the end of EMD; for EHs that formed at $a=\arh$ this virial radius
is 
\beq
\rvir \approx \arh \left(\frac{3\Mlm}{800 \pi \rho_\mathrm{cdm}}\right)^{1/3} = 9\times 10^{-4}\left(\frac{\arh}{\krh}\right)\left(\frac{\Mlm/\MRH}{10^{-7}}\right)^{1/3},
\label{eq:virial_radius_before_evap}
\eeq
where $\rho_{\mathrm{cdm}}$ is the present DM density and $\Mlm$ is the LM mass in the typical EHs. 
These typical masses are given in Eq.~\ref{eq:largest_EH_mass}. 
Using Eq.~\ref{eq:rofa_circular_asymptotic} we find the 
that the \emph{comoving} size of the remnant is approximately
\beq
\rvir^{\mathrm{co}} \sim 200\left(1 + 0.1\ln \frac{a/\arh}{10^8}\right)\rvir/\arh.
\label{eq:exploded_comoving_virial_radius}
\eeq
This radius maps onto a characteristic wavenumber beyond which the Fourier transform of density profiles, Eq.~\ref{eq:fdef_text}, begins to fall off and suppress the power spectra (see Eq.~\ref{eq:p1h_and_p2h}). $\rvir^{\mathrm{co}}$ is much larger than the comoving virial radius of the original halos due to the effects of adiabatic and free expansion. 
As as a result we expect an imprint on the power spectrum at scales larger than the typical size of the early halos.

Our numerically-derived profiles in \S\ref{sec:universal_profile} do not have an analytic form if they originate from NFW or Plummer EHs. We note however, that the adiabatic and free expansion tend to produce nearly-flat interiors and a sharp cutoff at $\rvir^{\mathrm{co}}$ (see, e.g.,  Fig.~\ref{fig:NFWprofiles}); as a result, the profiles resemble those of a spherical top hat, whose Fourier transform is just $3j_1(k \rvir^{\mathrm{co}})/(k \rvir^{\mathrm{co}})$. Using this approximate correspondence we see that 
$\mathcal{F}$ should significantly deviate from $1$ for $k \gtrsim  \sqrt{10}/\rvir^{\mathrm{co}}$; combining Eqs.~\ref{eq:exploded_comoving_virial_radius} and~\ref{eq:krh_over_keq} this gives an estimate of the cut-off scale\footnote{We dropped the logarithmic term from Eq.~\ref{eq:exploded_comoving_virial_radius} and the $g_*$ dependence in Eq.~\ref{eq:krh_over_keq} for simplicity.}
\beq
k_c/\keq \sim \sqrt{10}(\keq \rvir^{\mathrm{co}})^{-1} \approx 20 \left(\frac{\krh}{\keq}\right) \left(\frac{10^{-7}}{\Mlm/M_{\mathrm{RH}}}\right)^{1/3} =10^8 \left(\frac{\TRH}{5\;\MeV}\right)\left(\frac{10^{-7}}{\Mlm/M_{\mathrm{RH}}}\right)^{1/3},
\label{eq:approx_cutoff_scale}
\eeq
which is in excellent agreement with the two examples in Fig.~\ref{fig:ps_at_mre} which use the NFW profile for EHs. 
The comparison with a top hat profile is also useful in explaining 
the wavenumber scaling for $k \gg 1/\rvir^{\mathrm{co}}$:
we find that the Fourier transform of exploded NFW profiles is also similar to the Fourier transform of a spherical top hat, so $\mathcal{F}\sim 1/k^2$ at large $k$. This means that power spectra are suppressed by $|\mathcal{F}|^2 \sim 1/k^4$ for $k \gg (\rvir^{\mathrm{co}})^{-1}$. This suppression particularly evident in the left panel of Fig.~\ref{fig:ps_at_mre} where this scaling nearly flattens $k^4$ growth of the power spectrum in the ``Long EMD'' cosmology. The $k$ scaling for the ``Short EMD'' more difficult to derive because many different EH masses contribute to Eq.~\ref{eq:p1h_and_p2h}.

Equation~\ref{eq:approx_cutoff_scale} provides an approximate wavenumber beyond which modes are not enhanced by EMD. This model-independent cut-off allows for a very limited range of $k$ for the power spectrum to grow. It is also interesting that because this cutoff depends on the mass of the largest EHs, cosmologies where this mass is lower have boosted power over a broader range of scales. This somewhat counter-intuitive point is illustrated in the right panel of Fig.~\ref{fig:ps_at_mre}, where we see that the ``Short EMD'' cosmology has more power at smaller scales than the ``Long EMD'' case, precisely because the EHs were much smaller than in ``Long EMD''. Also, because the power is relatively flat at small scales in the ``Short EMD'' example (corresponding to the modes that entered the horizon in early radiation domination), the distribution of EHs has a long tail at small masses (see the right panel of Fig.~\ref{fig:eh_density_variance_and_mass_function}), ensuring that the expulsion of matter from EHs during adiabatic and free expansion has little effect on larger scales. Thus, the two features of ``Short EMD'' -- smaller characteristic mass of EHs and their relatively flat distribution -- ensure that it has significantly more power at small scales than ``Long EMD''.

\begin{figure}
    \centering
    \includegraphics[width=0.47\textwidth]{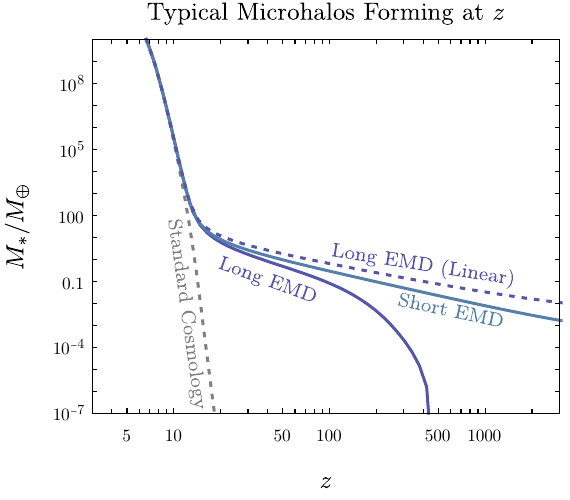}
    \caption{Mass of typical microhalos (normalized to Earth mass $\MEarth$) forming at redshift $z$ in cosmologies with Early Matter Domination. The solid lines 
    labelled ``Long EMD'' and ``Short EMD'' correspond to background cosmologies illustrated in Fig.~\ref{fig:bg_density_evolution} and include the effects of explosive evaporation 
    of EHs that formed during EMD. These effects wash out structure on small scales, resulting in smaller masses and later formation of microhalos compared to the naive expectation (blue dashed line) without accounting for the non-linear effects due to EHs.}
    \label{fig:mstar_vs_z}
\end{figure}

The power spectra at matter-radiation equality shown in Fig.~\ref{fig:ps_at_mre} can be evolved  into matter domination as described in \S\ref{sec:evolution_after_mre}. We use this  evolution to compute the density variance from Eq.~\ref{eq:density_fluctuation_variance}, which  enables us to estimate the formation history of microhalos of LM  as a function of redshift via Eq.~\ref{eq:collapse_criterion}. The typical microhalo mass forming at $z$, shown in Fig.~\ref{fig:mstar_vs_z}, highlights the physical importance of the explosive evaporation of EHs on late-time structure formation. The expulsion of matter from EHs at reheating erases small-scale power; in the ``Long EMD'' cosmology (solid blue line) this leads to a significant delay of the onset of microhalo formation compared to the naive expectation shown by the dashed blue line.  As we discussed above, for ``Short EMD'' these non-linear effects are less severe on the scales of interest since the EHs were much smaller. 

In \S\ref{sec:one_and_two_halo_power} we decomposed the power spectrum into one- and two-halo terms. The one-halo term measures the correlations within the EH remnants, whose evolution is encoded in the time-dependence of the remnant density profile. Since the DM particles in EH remnants are travelling outwards as a result of explosive evaporation, their correlations do not grow as described in \S\ref{sec:evolution_after_mre}. This is explicitly illustrated by the numerical solutions in Fig.~\ref{fig:Delta_evolution}. 
In contrast, the halo-halo power spectrum, which is related to the two-halo term via Eq.~\ref{eq:nl_ps_2h_main}, continues to grow since it is proportional to the linear spectrum (see Eq.~\ref{eq:hh_PS_from_lin_PS}). 
The physical interpretation here is that EH centers of mass continue to drift toward each other during RD even after the EHs have explosively evaporated. Well after reheating, the PS therefore becomes dominated by the two-halo terms. 

 The halo-halo power spectrum, however, must deviate from the linear matter power spectrum at small scales because of halo exclusion, as mentioned in \S\ref{sec:one_and_two_halo_power}: prior to evaporation EHs did not overlap by definition. A realistic implementation of exclusion effects is beyond the scope of the current work, so we opt for a simple prescription that is illustrated on a toy model in Appendix~\ref{sec:halo_exclusion}: we set the PS $P(k)$ to a constant for $k > 1/r_e$, where $r_e$ is the exclusion radius. $r_e$ is in principle different for different EHs, but we roughly estimate it as the typical separation of the EHs forming at the end of EMD (it is of the same order as the virial radius of EHs forming at this time) -- for more sophisticated models see, e.g., Ref.~\cite{Garcia:2020mxz}. The scales affected by halo exclusion effects are evident in both panels of Fig.~\ref{fig:ps_at_mre} at large $k$ as a sharp kink. In the absence of the density profile factor in the definition of $P_{2h}$, Eq.~\ref{eq:nl_ps_2h_main}, dimensionless power spectra for both cosmologies would scale as $k^3$. However, because the convolution of the EH mass function and the density profile $\mathcal{F}$ is different in the ``Long EMD'' and ``Short EMD'' examples, the behaviour of the PS also differs. For ``Long EMD'' the mass function integral 
gives a $1/k^4$ suppression, so $k^3 P_{2h}(k)  \sim 1/k$ decreases at large $k$; in ``Short EMD'' the mass function integral is instead logarithmically growing (owing to the very flat mass function shown in the right panel of Fig.~\ref{fig:eh_density_variance_and_mass_function}), so 
$k^3 P_{2h}(k)$ actually increases. 
It would be interesting to check whether this effect persists in other models of exclusion or in $N$-body simulations. However, it is likely irrelevant for the detectability of microhalos that form after equality since exclusion affects tiny length scales buried inside of microhalos.

In some cosmologies the linear theory logarithmic growth of density perturbations leads to overdensities $\gg 1$ at matter-radiation equality, as shown in the right panel of Fig.~\ref{fig:ps_at_mre} at large $k$. We only show this for comparison with ``Long EMD'', as linear perturbation theory is not valid in this regime; we do not use this highly-non-linear part of the power spectrum in Fig.~\ref{fig:mstar_vs_z}. In reality, as the EH centers drift in the RD universe, they can pass through each other, leading to either a saturation or even a decrease of the halo-halo PS. Large matter overdensities can collapse into microhalos during RD if matter-radiation equality is attained locally. Since our formalism does not account for these effects, we focus on microhalo formation for $z < \zeq$ in Fig.~\ref{fig:mstar_vs_z} which is sensitive to $k^3 P(k) /(2\pi^2) \lesssim \mathcal{O}(1)$ 
in Fig.~\ref{fig:ps_at_mre}. We leave the study of these issues to future work.

\subsection{Comparison with Other Cutoffs}
\label{sec:discussion}
The small-scale cut-off derived in the previous section is model-independent in the sense that it applies to any microphysical model of DM that is present during a period of EMD. In specific scenarios, however, other effects can lead to a more important suppression of power. 
For example, if DM is a thermal particle (that is, it is produced through 
contact with a plasma of relativistic particles), it can 
remain in kinetic equilibrium with the radiation bath long after chemical decoupling (freeze-out). This coupling to radiation provides pressure support to DM perturbations and prevents structure growth on scales that entered the horizon before kinetic decoupling. In fact, there 
are two different effects induced by DM-radiation 
coupling: free-streaming and acoustic oscillations, both of which 
suppress power at small scales~\cite{Loeb:2005pm,Green:2005fa}. 
The small scale cut-off is the smallest of the wavevectors
\beq
k_{\mathrm{cut}} \approx \min(k_{\mathrm{fs}}, k_{\mathrm{ao}}),
\eeq
where the free-streaming (fs) and acoustic oscillation scales are (ao)
\begin{subequations}
\begin{align}
    k_\mathrm{fs} & =\left(\int_{\akd}^{\aeq}\frac{da }{a H}\frac{v(\Tkd)}{a}\right)^{-1} \sim 300\xi^{-1/2} \left(\frac{m}{100\;\GeV}\right)^{1/2}\left(\frac{\Tkd}{50\;\MeV}\right)^{5/6}\left(\frac{5\;\MeV}{\TRH}\right)^{4/3}\krh\\
    k_\mathrm{ao} & = \akd H(\Tkd) \sim 20 \left(\frac{\Tkd}{50\;\MeV}\right)^{4/3}\left(\frac{5\;\MeV}{\TRH}\right)^{4/3}\krh
\end{align}
\label{eq:kfs_and_kao_during_emd}%
\end{subequations}
where $m$ is the DM mass, $\xi$ is the ratio of DM and SM temperature at the time of kinetic decoupling. 
In evaluating the above expressions we assumed that the kinetic decoupling 
temperature is much larger than the reheat temperature, $\Tkd \gg \TRH$, and we made use of the temperature scaling $T\sim a^{-3/8}$ during EMD. We see that for a large values of $\Tkd$ and $m$ that $k_{\mathrm{cut}}$ can be comparable or larger to the model-independent estimate in Eq.~\ref{eq:approx_cutoff_scale}. 

Small-scale cut-offs also exist in models where DM is an axion-like particle (ALP), where the wave-like nature of the particle suppresses power 
for wavenumbers larger than the comoving horizon at the start of 
ALP oscillations~\cite{Blinov:2019jqc}
\beq
k_\mathrm{osc} = \left(\frac{m_a}{H(\TRH)}\right)^{1/3} 
  \sim 10^3 \left(\frac{m_a}{10^{-5}\;\eV}\right)^{1/3} \left(\frac{5\;\MeV}{\TRH}\right)^{2/3} \left(\frac{10.75}{g_*(\TRH)}\right)^{1/6} \krh.
\eeq
Depending on the ALP mass, the model-independent cut-off can be 
easily the dominant source of suppression in these scenarios as well. 

Finally note that we have focused on the scenario where stable DM exists independently of the EM responsible for EMD. An alternative scenario is where the decays of EM also produce the majority of DM~\cite{Moroi:1999zb,Gelmini:2006pw,Arcadi:2011ev,Blinov:2014nla}. If the DM particles are weakly-interacting and do not thermalize, we expect them to free-stream a distance of order the horizon size at reheating, $k_\mathrm{fs} \sim \krh$, precluding any enhanced structure growth at all. If they do thermalize right after production, kinetic decoupling effects described above similarly wash out any potential enhancement from EMD~\cite{Fan:2014zua}. Therefore our model-independent small-scale cut-off is accurate for models with EMD-enhanced structure only if other suppressions due to the microphysics of the specific DM candidate are subdominant.

\section{Conclusion}
\label{sec:conclusion}

Despite its enormous success the Standard Model of Cosmology has experimental support over a relatively modest range of scales and includes an aggressive extrapolation to scales where no observational evidence is accessible. The  ``desert'', between the end of inflation and the BBN period is usually assumed populated by a long and uneventful epoch of adiabatically expanding thermal radiation. While this is certainly the simplest extrapolation it provides little challenge for observational cosmology.

A similar desert exists in the particle physics Standard Model (SM) from the electroweak scale to the GUT or Planck scale, where uncertain gravitational physics will likely dominate. In particle physics a plethora of Beyond the Standard Model (BSM) physics scenarios have been developed as possible UV extensions of the SM. Differing BSM models populate the desert with a wide variety of new particles and interactions, giving numerous targets for experiments on the energy, luminosity and precision frontiers.

This paper studies models with an early period of matter domination (EMD) in place of part of the radiation desert. EMD is particularly interesting, not only because it naturally arises in several BSM models but also due to its intimate connection with dark matter (DM). EMD changes the preferred parameter space for DM models and modifies the present day cosmological inhomogeneities on very small scales that have not yet been observationally probed.

While EMD enhances the growth of linear inhomogeneities, we have shown that the non-linear evolution instead suppresses power on small scales. This surprising and perhaps counter-intuitive result can be easily understood. If EMD is sufficiently long, most of the matter ends up in gravitationally-bound early halos (EHs). EMD ends when the matter that dominates the energy budget of the universe decays into SM radiation,  unbinding the EHs. As a result, the stable dark matter component present in the EHs free-streams and erases inhomogeneities on small scales.  In other works it is assumed that the decay was instantaneous to enable an estimate of the free-streaming cut off scale. We have shown here that the evolution of EHs is better characterized by several stages: adiabatic expansion of the gravitationally bound object; peeling of successive outer layers as EH self-gravity becomes less important; and finally free-streaming of DM particles. We referred to the entirety of this process as explosive evaporation. The resulting free-streaming velocity, eq.~\ref{eq:EVfree}, is significantly smaller than in the instantaneous decay approximation which would predict a velocity $\sim\sqrt{\frac{G\,M(r_0)}{r_0}\,\frac{\tau}{\tfree}}$ at time $\tfree$ which is a factor $\sim2\,\sqrt{\frac{\sqrt{\delta_\tau}}{\ln\,\delta_\tau}}$ too large.  The particle velocity distribution is also quite different.

The distribution of DM at late times is determined by a superposition of EH remnants. We use a halo model to construct the DM power spectrum that includes the non-linear effects described above, which cut off the EMD-driven enhancement.  This cutoff depends only on the mass distribution of the EHs and their density profiles. 

Our semi-analytic results for the evolution of individuals EHs are based on the assumptions of isolated spherical halos and circular orbits. We have shown that these simple simulations are similar to more general orbital distributions using a simple shell code. We expect that aspherical triaxial or rotationally-supported halos to yield similar remnant DM distributions. It would be interesting to verify these results for more realistic halos and to study the impact of halo overlap in $N$-body simulations.

We applied our results to two benchmark models containing either a long or short period of EMD, which enhance a nearly scale-invariant primordial spectrum of density fluctuations (as extrapolated from that observed on much larger scales).  In linear perturbation theory alone, one would expect the ``Long EMD'' cosmology to feature the greatest enhancement in small scale power. Interestingly, we find that the non-linear effects described above are more significant in this cosmology. As a result the ``Short EMD'' case tends to have larger inhomogeneity at small scales. As a final application of these results, we estimated the typical microhalo masses and formation times using the Press-Schechter approach. While both long and short EMD cosmologies lead to the formation of microhalos after matter-radiation equality, this occurs much earlier in ``Short EMD'' due to the relatively smaller impact of non-linear effects. We conclude that there is an ``optimal'' duration for EMD in order to generate the maximum enhancement of the power spectrum at late times, that minimizes the suppression due to explosive evaporation of EHs. This highlights the importance of the duration of EMD for the late-time matter distribution, an aspect of these cosmologies that has not yet been comprehensively studied.

While our calculation of explosive evaporation of EHs was agnostic to their formation history, we made several important assumptions about the nature of DM. We neglected any interactions between the dominant matter component responsible for EMD and the DM, and assumed that they have identical phase space distribution. It would be interesting to explore how relaxing some of these assumptions affects the late-time distribution of matter. We also specialized to EMD as a way to enhance density perturbations. Similar effects are generated in other cosmologies, such as an early period of kination. It is an open question as to whether non-linear physics plays a similarly important role here~\cite{Redmond:2018xty}.

Finally, we stopped short of examining the observational signatures of microhalos forming at late times, such as those in pulsar timing arrays~\cite{Lee:2021zqw} or microlensing~\cite{Dai:2019lud}. Given that enhancement of small-scale structure is so prevalent in many cosmological and particle physics models of DM, it is important and exciting to develop these and other techniques, as they can provide access to the earliest moments in the evolution of the universe.

\section*{Acknowledgements}
We thank Arka Banerjee, Carlos Blanco and Patrick Draper for helpful discussions. 
This manuscript has been authored by Fermi Research Alliance, LLC under Contract No. DE-AC02-07CH11359 with the U.S. Department of Energy, Office of Science, Office of High Energy Physics.
GB acknowledges support from the MEC Grant FPA2017-845438 and 
the Generalitat Valenciana under grant PROMETEOII/2017/033. 

\appendix
\section{Adiabatic Homologous Expansion of Halos}
\label{sec:adiabatic}

Here we model the evolution of early halos during EMD and the beginning of LRD. Assume that the early matter (EM) and late matter (LM) originally populate the orbits of a halo in the same way. When the EM decays we assume the decay products exit the halo on a timescale much smaller than the dynamical time of the halo so one may treat the decays as an instantaneous disappearance. This is accurate so long as the speed of the decay products is much larger than the escape velocity. The rate of disappearance is assumed uniform throughout the halo which would be accurate for particle decays if the halo is non-relativistic so the Lorentz factors are nearly unity.

Under these assumptions the phase space distribution of EM and LM are described by the same 1-particle distribution function $f[\mathbf{x},\mathbf{v},t]$ which for nonrelativistic halos obeys the Newton-Vlasov equation
\begin{equation}
 \frac{\mathbb{D}}{\mathbb{D}t}f\equiv \left(\frac{\partial }{\partial t}+\mathbf{v}\cdot \frac{\partial }{\partial \mathbf{x}}-\left(\frac{\partial
}{\partial \mathbf{x}}\Phi [\mathbf{x},t]\right)\cdot \frac{\partial }{\partial \mathbf{v}}\right)f[\mathbf{x},\mathbf{v},t]=0
\end{equation}
where $\mathbf{x}$ and $\mathbf{v}$  are the physical (not comoving) Cartesian coordinate and velocity. This equation describes the trajectories of the particles which do not decay and requires no modification for decaying particles if one normalizes the distribution function as:
\begin{eqnarray} 
  \rho _{\text{EM}}[\mathbf{x},t]&=&\left(1-f_{\text{LM}}\right)e^{-t/\tau }\int d^3\mathbf{v}\, f[\mathbf{x},\mathbf{v},t] \nonumber \\
  \rho _{\text{LM}}[\mathbf{x},t]&=&f_{\text{LM}}\int d^3\mathbf{v}\,f[\mathbf{x},\mathbf{v},t]
\label{NewtonVlasov}
\end{eqnarray}
where $\rho _{\text{EM}}$ and $\rho _{\text{LM}}$ are the EM and LM mass density; $\tau$ is the decay time; and $f_{\text{LM}}$ the fraction of mass in LM particles. The Newtonian gravitational potential is given by
\begin{eqnarray} 
  \Phi[\mathbf{x},t]&=&-\int d^3\mathbf{x'}\,
 \frac{G\,(\rho _{\text{EM}}[\mathbf{x'},t]+\rho _{\text{LM}}[\mathbf{x'},t])}{|\mathbf{x}-\mathbf{x'}|} \nonumber \\
  &=&-d[t]\int d^3\mathbf{x'}\,\int d^3\mathbf{v}\,\frac{G\,f\left[\mathbf{x'},\mathbf{v},t\right]}{|\mathbf{x}-\mathbf{x'}|} \ .
\end{eqnarray}
where $d[t]=\left(1-f_{\text{LM}}\right)e^{-t/\tau }+f_{\text{LM}}$ in the decaying particle EMD scenario.

During a prolonged EMD epoch we first expect gravitational collapse into bound structures (halos) after which halos successively merge into larger and larger halos (hierarchical clustering). Well into hierarchical clustering most matter is in a halo environment much denser that the cosmological mean where the dynamical timescale is much shorter than the expansion (Hubble) time: $H \tau _{\text{dyn}}\ll 1$. Mergers are assumed episodic after which halos quickly relax into a nearly stationary equilibrium which is represented by $\frac{\partial }{\partial t}f=0$ so
\begin{eqnarray} 
 0&=&\left(\mathbf{v}\cdot \frac{\partial }{\partial \mathbf{x}}
 -\left(\frac{\partial }{\partial \mathbf{x}}\Phi _{\text{eq}}[\mathbf{x}]\right)\cdot \frac{\partial
}{\partial \mathbf{v}}\right)f_{\text{eq}}[\mathbf{x},\mathbf{v}] \nonumber \\
\Phi _{\text{eq}}[\mathbf{x}]&=&-\int d^3\mathbf{x'}\,
\int d^3\mathbf{v}\frac{G\,f_{\text{eq}}\left[\mathbf{x'},\mathbf{v},t\right]}{| \mathbf{x}-\mathbf{x'}|}\ .
\label{StaticNewtonVlasov}
\end{eqnarray}
Such an equilibrium could persist indefinitely if not for the EM decay in the EMD scenario which causes $\Phi$ to vary introducing an unavoidable time dependence. So long as $\tau \gg \tau _{\text{dyn}}$ the rate of change is slow and one can follow the evolution of $f[\mathbf{x},\mathbf{v},t]$ in an adiabatic approximation. 

When $\tau \gg \tau _{\text{dyn}}$ each particle will undergo many orbits during a decay time, and hence $\tau$ introduces no new length scale into the dynamics. One therefore expects homologous expansion where the halo slowly passes through a scaling of the initial equilibrium. Arbitrary homologous scaling of position, velocity and phase space density is of the form:
\begin{equation}
 f[\mathbf{x},\mathbf{v},t]\cong\frac{1}{h_\text{f}[t]} f_{\text{eq}}\left[\frac{\mathbf{x}}{h[t]},\frac{\mathbf{v}}{h_\text{v}[t]}\right]\qquad
\Phi [\mathbf{x},t]\cong-\frac{h[t]^2h_\text{v}[t]^3}{h_\text{f}[t]}\,\Phi _{\text{eq}}\left[\frac{\mathbf{x}}{h[t]}\right]
\end{equation}
where $h[t]$, $h_\text{v}[t]$ and $h_\text{f}[t]$ scale the size, velocity and phase space density. In order to preserve the total LM mass one requires $h_\text{v}[t]=h[t]^{-1}$. The Vlasov equation preserves the mass weighted distribution of phase space density so one also requires $h_\text{f}[t]=1$. Thus adiabatic scaling must be of the form
\begin{equation}
 f[\mathbf{x},\mathbf{v},t]\cong f_{\text{eq}}\left[\mathbf{x}_\text{s},\mathbf{v}_\text{s}\right]    \qquad
\Phi [\mathbf{x},t]\cong\frac{1}{h[t]} \Phi _{\text{eq}}\left[\mathbf{x}_\text{s}\right] .
\end{equation}
where $\mathbf{x}_\text{s}\equiv\mathbf{x}/h[t]$ and $\mathbf{v}_\text{s}\equiv h[t]\mathbf{v}$.  Substituting this ansatz into the Eq.~\ref{NewtonVlasov} and using Eq.~\ref{StaticNewtonVlasov} one finds
\begin{eqnarray}
\frac{\mathbb{D}}{\mathbb{D}t}f=\left(d[t]-\frac{1}{h[t]}\right)
\left(\frac{\partial }{\partial\mathbf{x}_\text{s}} \Phi _{\text{eq}}\left[\mathbf{x}_\text{s}\right]\right)
\cdot
\frac{\partial }{\partial \mathbf{v}_\text{s}}f_{\text{eq}}\left[\mathbf{x}_\text{s},\mathbf{v}_\text{s}\right] && \nonumber\\
-\frac{\partial \ln h[t]}{\partial t}\,\left(\mathbf{x}_\text{s}\cdot \frac{\partial}{\partial\mathbf{x}_\text{s}}
     -\mathbf{v}_\text{s}\cdot \frac{\partial}{\partial\mathbf{v}_\text{s}}\right)
    \,f_{\text{eq}}\left[\mathbf{x}_\text{s},\mathbf{v}_\text{s}\right] &\cong&0\ .
\end{eqnarray}
In the adiabatic limit, $\left| \frac{\partial \ln h[t]}{\partial t}\right| \tau _{\text{dyn}}\ll 1$, so one neglects the last term and one finds that $h[t]=d[t]^{-1}$ gives the adiabatic scaling.

EMD halos will have differing mass, density profiles, angular momentum, etc. The distribution of halo properties will depend on the early universe scenario which gives rise to EMD. This adiabatic scaling is a general behavior and applies to each halo individually independent of halo properties. This scaling also works for any functional form $d[t]$. 

In the EMD scenario the LM contributes negligibly to the mass density before LMD, i.e. $f_{\text{LM}}\ll 1$, so during EMD and early LRD the LM particles act as test particles in the gravitational field. In this case $d[t]\cong e^{-t/\tau }$ and $h[t]\cong e^{t/\tau }$: the halo size grows while the particle velocities shrink, both exponentially, during the adiabatic phase. 

When a halo first collapses $\rho _{\text{halo}}\gg \bar{\rho }$ where $\rho _\text{halo}$ and $\bar{\rho}$ are the halo and cosmological mean  density. During EMD $\rho _\text{halo}$ remains nearly constant while $\bar{\rho}$ decreases rapidly. The dynamical to expansion timescale ratio, $H \tau _{\text{dyn}}\sim \left(\bar{\rho }/\rho _{\text{halo}}\right){}^{1/2}$, decreases rapidly during EMD but adiabatic expansion gives $\rho _{\text{halo}}\propto e^{-4t/\tau }$ so $H \tau _{\text{dyn}}$ rises rapidly during LRD. Since for any bound object $\rho _{\text{halo}}\gg \bar{\rho }$ adiabatic expansion persists for a time significantly greater than $\tau$ and bound EMD halos exist well into the radiation era but grow larger and less bound. Adiabatic expansion eventually ends when $\rho _{\text{halo}}\lesssim\bar{\rho }$ giving rise to a period of free expansion (see text). As a result of the adiabatic expansion phase one finds free expansion rather than free contraction in the radiation era.

\section{Non-Circular Orbits and Shell Crossing}
\label{sec:shell_code}
In this section we relax some of the assumptions that were used in \S\ref{sec:isolated_halos} to describe the evolution of early halos through reheating. We use a simple shell code which retains spherical symmetry of an isolated halo but allows us to consider different angular momenta (i.e., non-circular orbits) and sensitivity to the initial density profile, as well as automatically including shell crossing. From the perspective of the late-time power spectrum, the most interesting quantity is the asymptotic velocity of (formerly) EH shells well after reheating, which determines the EH remnant profile. The physical velocity with respect to the comoving frame ($v=ad(r/a)/dt$) yields:
\beq
v(a) = -H \left(\frac{3M_<}{4\pi\rholm}\right)^{1/3}\mathcal{V},
\label{eq:velocity_from_shell_model}
\eeq
where we defined
\beq
\mathcal{V} = -\frac{a \bar\Delta'}{\bar\Delta^{4/3}}
\label{eq:velocity_factor_def}
\eeq
with $\bar\Delta$ defined in Eq.~\ref{eq:delta_defA}. 
Since $\bar\Delta' < 0$, $\mathcal{V}>0$; $\mathcal{V}$ is a useful 
quantity because it encodes the dynamical effects due to EH 
explosive evaporation (the usual redshifting 
of velocity is captured by the first two factors in Eq.~\ref{eq:velocity_from_shell_model}).
Moreover, $\mathcal{V}$ should approach a constant deep in the radiation era which can be seen from Eq.~\ref{eq:Delta_eomA}:
\beq
\Delta'' = -\frac{\Delta'}{a^2} + \frac{4}{3}\frac{(\Delta')^2}{\Delta}
\Rightarrow \mathcal{V}' = 0.
\eeq
In principle the constant value of $\mathcal{V}$ depends on the parameters 
of the EH, such as the initial mass profile and average overdensity. However, 
using numerical solutions of Eq.~\ref{eq:Delta_eomA} we have checked 
that $\mathcal{V} \approx 2$ over a wide range of ``initial'' conditions 
$\bar\Delta_\tau$. We will therefore use $\mathcal{V}$ as a simple diagnostic and check this result with an $N$-shell simulation allowing for non-circular orbits and shell crossings. We will find that $\mathcal{V}\approx 2$ even in this more general set-up, implying that the results in the main text are robust. We leave more detailed comparisons, the impact EH non-sphericity, substructure and halo overlap to future work. 

We follow the evolution of $N$ spherical, equal-mass shells, each with an equation of motion given by Eq.~\ref{eq:shell_eom_simple}. Their 
evolution is coupled through the mass interior term 
\beq
M_<(r_j,t) = \sum_{r_j < r_k} m \left[f + (1-f)e^{-t/\tau}\right],
\eeq
where $j$ and $k$ are shell labels, $m$ is the shell mass at early times 
and the two terms in the brackets represent LM and EM contributions to the mass.

We initialize the shell configuration similarly to Refs.~\cite{Nusser:2000xn,Lu:2005tu}, which we summarize below.
In order to specify the initial positions of the shells, we need postulate an initial mass profile for the overdensity:
\beq
1 + \delta(r_k,t) = \frac{3M_<(r_k,t)}{4\pi r_k^3\rhomat(t)},
\label{eq:overdensity_defn}
\eeq
where $\rhomat$ is the background matter density.
Following Ref.~\cite{Nusser:2000xn} we take at the initial time $t_i$
\beq
\delta(r_k,t_i) = \davg \left(\frac{M_<(r_k,t_i)}{\Mtot}\right)^{-\epsilon} = \davg \left(\frac{k+1}{N}\right)^{-\epsilon},
\label{eq:initial_overdensity_profile}
\eeq
where $\davg$ and $\Mtot$ represent the average density contrast and mass of the whole shell configuration at early times, $N$ is the number of shells and $\epsilon >0$. 
The initial position of shell $k$ is then~\cite{Lu:2005tu}
\beq
r_k(t_i) = \left[\frac{3M_<(r_k,t_i)}{4\pi \rhomat(t_i)(1 + \delta(r_k,t_i))}\right]^{1/3}.
\eeq
If the initial mass distribution corresponds to $\delta$ in the linear regime (i.e., $\delta\ll1$) then 
the initial velocity can be obtained from linear perturbation theory during matter domination. 
A standard result is that $\delta \propto a$, or equivalently $\dot\delta /\delta = H$. 
Differentiating Eq.~\ref{eq:overdensity_defn} with respect to time, using $\dot\delta/\delta$, $H = 2/(3t)$ during EMD and 
evaluating the result at $t=t_i$ gives
\beq
\frac{\dot r_k }{r_k} = H - \frac{\dot\delta_k}{3(1+\delta_k)} = \frac{2}{3 t_i} \left[1 - \frac{\delta_k}{3(1+\delta_k)}\right],
\label{eq:ic_velocity}
\eeq
where $\delta_k = \delta(r_k,t_i)$, the first term is the usual Hubble expansion (i.e., if we only kept it the overdensity would remain constant in time) 
and the second term is responsible for the linear growth of $\delta$. In practice we initialize our simulations with large overdensities $\delta_\mathrm{avg}  \gtrsim 1$; we find that initial density and velocity profiles do not impact the subsequent evolution since the shell configuration quickly virializes. Thus, as long as the initial time $t_i$ is well before reheating, $t_i \ll \tau$, our results are not sensitive to the precise choice of 
shell position and velocity. 

In order to assess the impact of shell angular momentum, we parametrize $L$ similarly to Ref.~\cite{Nusser:2000xn}\footnote{This differs slightly from Ref.~\cite{Nusser:2000xn} which took $L^2 = 2\alpha G \Delta M_<(r_k,t_i) r_k$, with $\Delta M$ the mass excess over background contained within $r_k$.}
\beq
L^2 = 2\alpha G M_<(r_k,t_i) r_k,
\label{eq:angular_momentum_alpha_def}
\eeq
where $\alpha$ is a dimensionless parameter.
The shells are gravitationally bound if
\beq
\alpha \leq 1 - \frac{(1-\xi)^2}{1+\delta},
\eeq
where $\xi = \delta/[3(1+\delta)]$ is the deviation
from the Hubble expansion in the initial velocity, Eq.~\ref{eq:ic_velocity}.
In this parametrization circular and radial orbits correspond to $\alpha = 1/2$ and $\alpha = 0$, respectively.

Following initialization, we evolve the shell ensemble forward in time using an adaptive leapfrog integration scheme well into radiation equality. In order to ensure stability of the solution we follow Ref.~\cite{Lu:2005tu} to endow the shells with a finite thickness and constant density.
This modification ensures that the inter-shell forces 
change continuously as the shells move past each other, leading to a more stable integration of the equations of motion.

\begin{figure}
    \centering
    \includegraphics[width=0.47\textwidth]{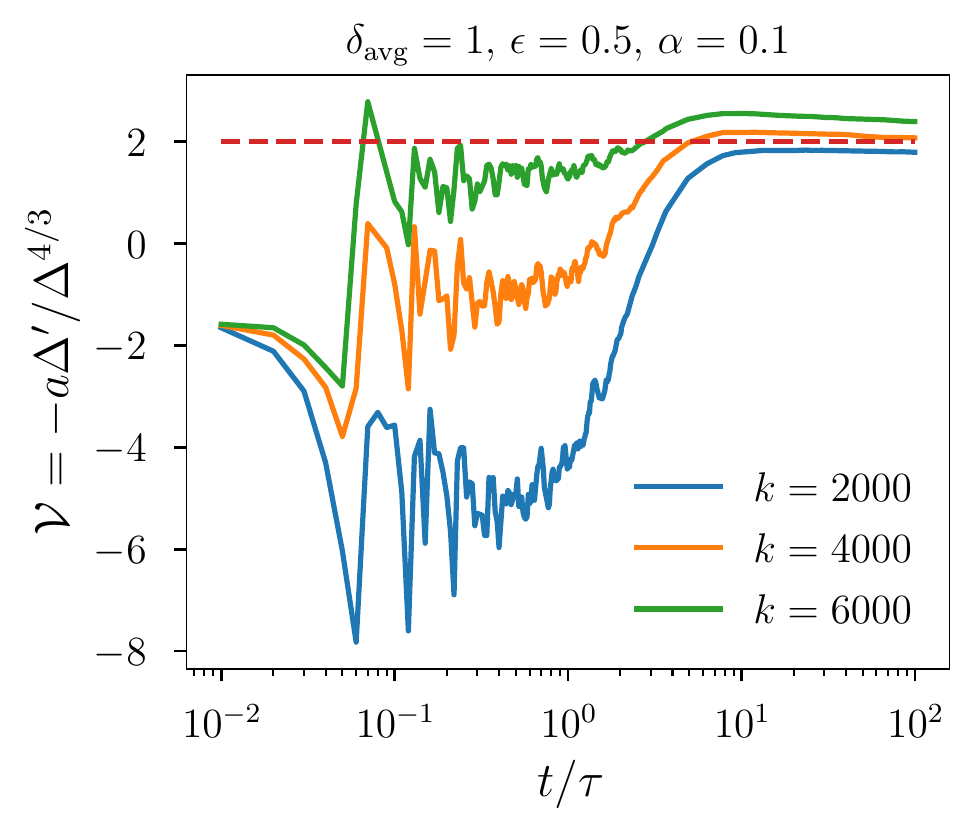}
    \includegraphics[width=0.47\textwidth]{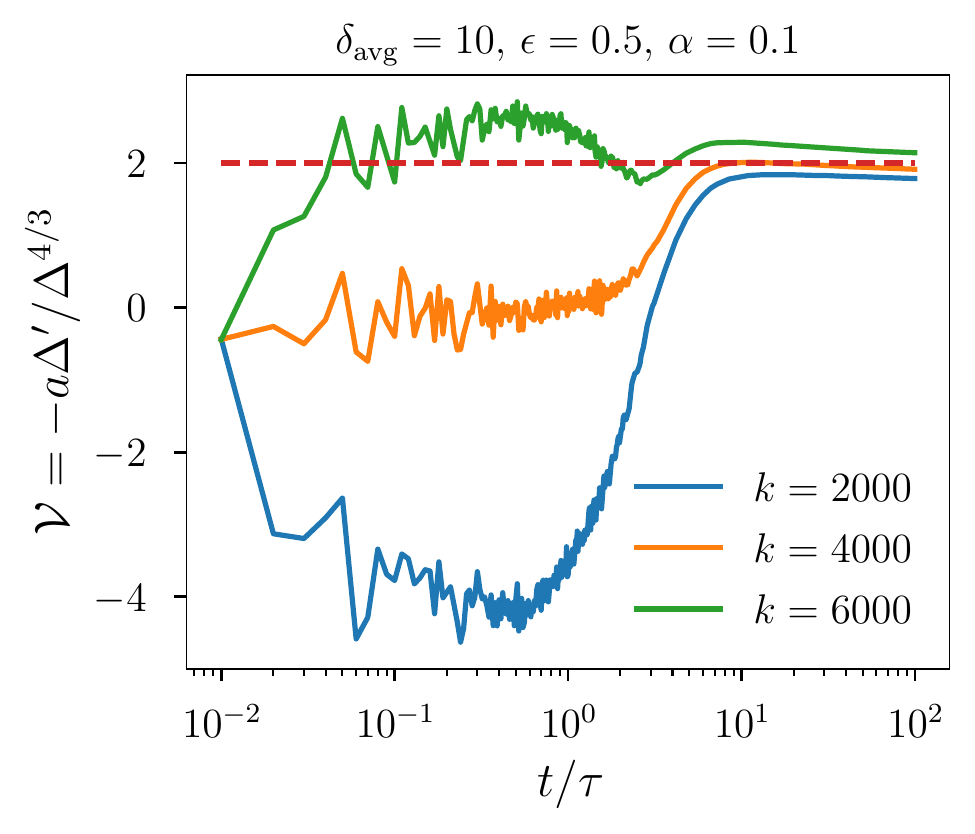}
    \\
    \includegraphics[width=0.47\textwidth]{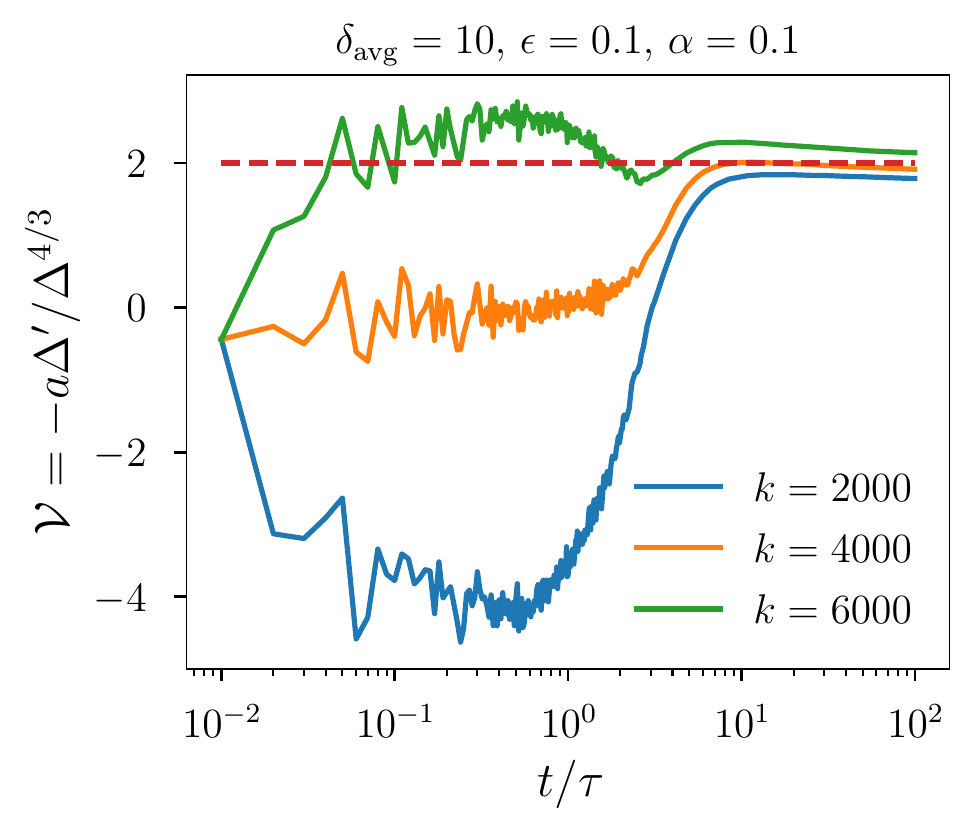}
    \includegraphics[width=0.47\textwidth]{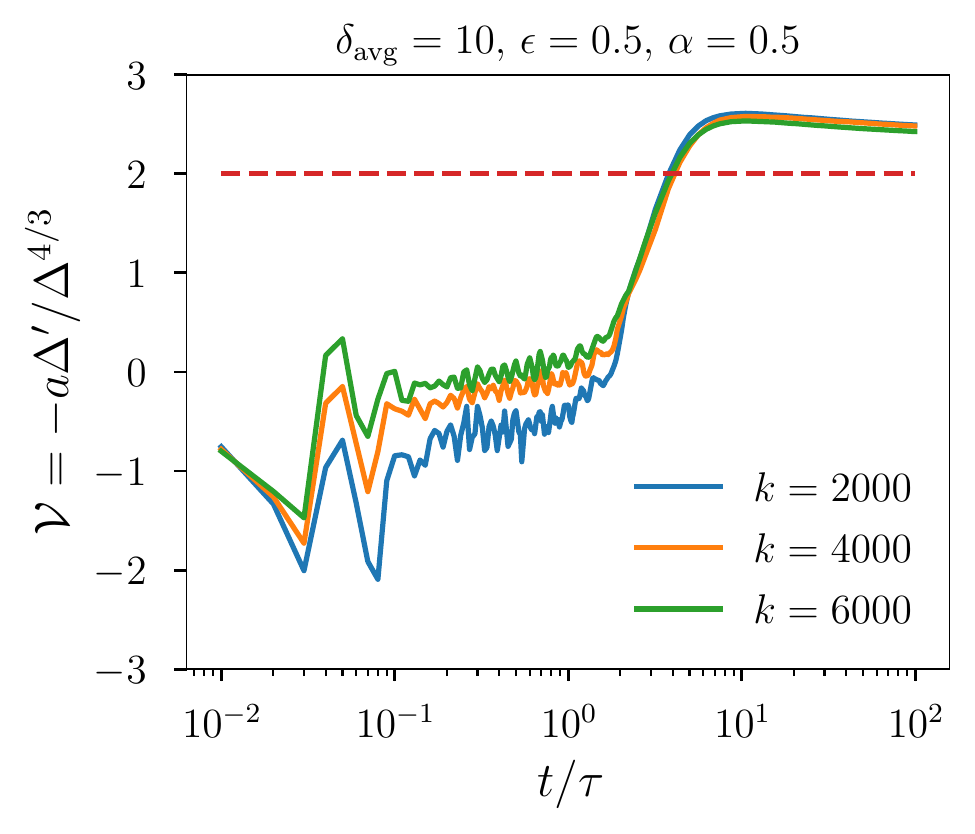}
    \caption{Evolution of velocity factor $\mathcal{V}$ (Eq.~\ref{eq:velocity_factor_def}) for three representative shells (out of a $N=10000$ shell simulation) as a function of $t/\tau$. Different panels correspond to different initial shell parameters, such as the initial average overdensity $\delta_{\mathrm{avg}}$, initial density profile slope $\epsilon$ (both defined in Eq.~\ref{eq:initial_overdensity_profile}) and the angular momentum parameter $\alpha$ (Eq.~\ref{eq:angular_momentum_alpha_def}).
    For a wide range of parameters $\mathcal{V}$ approaches $\sim 2$ (indicated by the dashed line) well after reheating, $t/\tau \gg 1$. The asymptotic value of $\mathcal{V}$ is mildly dependent on the initial angular momentum and overdensity. 
    }
    \label{fig:velocity_factor}
\end{figure}

In Fig.~\ref{fig:velocity_factor} we show 
the evolution of $\mathcal{V}$ for several choices of $\delta_\mathrm{avg}$, $\epsilon$ and $\alpha$. These results confirm our initial finding that $\mathcal{V} \sim 2$ asymptotically for $t/\tau\gg 1$, although this value does have some mild dependence on the angular momentum of the shells. 

\section{Linear Evolution}
\label{sec:linear_perturbations}

The Fourier-space density $\delta_i$ and velocity divergence $\theta_i = i \vec{k}\cdot \vec{v}_i$ perturbation equations for $i=\EM$ (the EMD field),
$\LM$ (the dark matter), and $\rad$ (SM radiation) are~\cite{Erickcek:2011us,Blinov:2019jqc}:
\begin{subequations} 
\begin{align}
a^2 E \delta_{\EM}^\prime + \theta_{\EM} + 3a^2E \Phi^\prime &= \frac{a}{\tau}\Phi, \label{eq:delta_EM}\\
a^2 E \theta_{\EM}^\prime + a E \theta_{\EM} + k^2\Phi &=0, \label{eq:theta_EM}\\
a^2 E \delta_{\rad}^\prime+\frac{4}{3}\theta_{\rad}+4a^2 E \Phi^\prime&= \frac{a \rhoem}{\tau \rhorad}\left[\delta_{\EM}-\delta_{\rad}-\Phi\right], \label{eq:delta_rad}\\
a^2 E \theta_{\rad}^\prime+k^2\Phi -k^2\frac{\delta_{\rad}}{4}&= \frac{a \rhoem}{\tau \rhorad}\left[\frac{3}{4}\theta_{\EM}-\theta_{\rad}\right], \label{eq:theta_rad}\\
  a^2 E \delta_{\LM}^\prime + \theta_{\LM} +3a^2E\Phi^\prime &= 0 \label{eq:delta_LM} \, ,\\
  a^2 E \theta_{\LM}^\prime+ aE\theta_{\LM}+k^2\Phi &=0 \label{eq:theta_LM} \, ,\\ 
k^2\Phi +3aE^2\left[a^2\Phi^\prime+a\Phi\right] &= \frac{3}{2}a^2\left[\rhoem\delta_{\EM}+\rhorad\delta_{\rad}+\rholm\delta_{\LM}\right], \label{eq:phi}
\end{align}
\label{eq:lin_pert_eq} 
\end{subequations}
where $\Phi$ is the gravitational potential ($\Phi = -\psi$ in the absence of anisotropic stress) and the prime denotes differentiation with respect to $a$, $E$ is the dimensionless Hubble parameter 
\beq
E^2 = \rhoem + \rhorad + \rholm. 
\eeq
The energy densities are normalized such that $E(a=1) = 1$, while 
$1/\tau$, $k$ and $\theta_i$ are in units of $H_1$, the Hubble parameter at $a=1$. These quantities can be obtained in physical units by, e.g., identifying the scale factor at reheating, $\arh$ with a physical temperature $\TRH$ (or, equivalently, using Eq.~\ref{eq:krh_over_keq}).

The initial conditions for the above system of equations depends on whether modes enter during matter or radiation domination. In the ``Long EMD'' cosmology we assume all modes of begin their evolution during EMD, such that the initial conditions are the same as in Ref.~\cite{Blinov:2019jqc}:
\begin{align}
\delta_\EM = \delta_\LM = 2\delta_\rad = 2\Phi_i \\
\theta_\EM = \theta_\LM = \theta_\rad = -\frac{2}{3}k^2\sqrt{a}\Phi_i, \label{eq:long_emd_ic}
\end{align}
where $\Phi_i$ is the superhorizon value of $\Phi$.
Since we specify $\Phi_i$ during EMD, its power spectrum is related to the primordial curvature power spectrum by
\beq
\langle \Phi^2_i \rangle =\left(\frac{3}{5}\right)^2 \langle \mathcal{R}^2\rangle.
\label{eq:phi_curve_emd}
\eeq

In the ``Short EMD'' cosmologies we assume that all modes begin their evolution during early RD, requiring a different set of initial conditions:
\begin{align}
(4/3)\delta_\EM = (4/3)\delta_\LM = \delta_\rad = 2\Phi_i \\
\theta_\EM = \theta_\LM = \theta_\rad = -\frac{1}{2\sqrt{\rhorad(a=1)}}k^2 a \Phi_i,
\label{eq:short_emd_ic}
\end{align}
which is in agreement with Newtonian gauge initial conditions in the standard cosmology~\cite{Ma:1995ey}. Now $\Phi_i$ is fixed during early radiation domination, so its relationship to the curvature perturbation is
\beq
\langle \Phi^2_i \rangle =\left(\frac{2}{3}\right)^2 \langle \mathcal{R}^2\rangle.
\label{eq:phi_curve_rd}
\eeq

We solve the system of equations~\ref{eq:lin_pert_eq} numerically with the initial conditions given in Eq.~\ref{eq:long_emd_ic} or Eq.~\ref{eq:short_emd_ic}. Well after reheating, the solutions have the form given in Eq.~\ref{eq:rd_form}, which we fit to numerics for a wide range of $k$. We present these fits below for the ``Long EMD'' and ``Short EMD'' cosmologies.

\paragraph{Long EMD}
 We find that
\beq
I_1 \approx \frac{2}{3}\left(\frac{k}{\krh}\right)^2\left(0.757 + \frac{10.97}{(k/\krh)^{2/\alpha}}\right)^\alpha,
\eeq
where $\alpha = 1.13$; 
\beq
I_2 \approx \frac{3.15}{(5.73 + (k/\krh)^{2/\beta})^\beta},
\eeq
where $\beta = 0.923$;
\beq
\ahor/\arh \approx 1.108\left(\frac{1}{(k/\krh)^{1/\gamma} + 0.426(k/\krh)^{2/\gamma}}\right)^{\gamma},
\eeq
where $\gamma \approx 0.302$. Note that for $k/\krh \ll 1$, $(9/10)I_1 \approx 9$, $I_2 \approx 0.63$, and $\ahor/\aeq \approx \keq/(\sqrt{2}k)$ reproduce the standard results for modes that enter the horizon during radiation domination well after reheating~\cite{Hu:1995en}.\footnote{The additional factor of $9/10$ is needed when comparing ``Long EMD'' to the standard cosmology because the relationship between the superhorizon gravitational potential (which sets the initial conditions for the density perturbations) and the curvature perturbation (whose power spectrum is usually specified) is different during EMD or RD -- see Eqs.~\ref{eq:phi_curve_emd} and~\ref{eq:phi_curve_rd}.} In the opposite limit $k\gg \krh$, these functions reproduce the approximate solution in Eq.~\ref{eq:long_emd_rd_correlation_func}.

\paragraph{Short EMD} 
This cosmology has a finite period of EMD that is preceded by early RD; EMD starts 
when $a/\arh = 4.5\times 10^{-4}$, so in terms of scale factor EMD lasts a factor of $\sim 2200$ -- see the right panel of Fig.~\ref{fig:bg_density_evolution}. 
The mode that enters at the time of early equality is $k/\krh \approx 90$. As before 
we solve the linear equations numerically for the DM density contrast and try to find simple analytic functions 
for $I_1$ and $I_2$ that fit the numerics well; these are more complicated because they need to capture 3 regimes now - early RD, EMD, and late RD.
We find the following convenient fitting functions: 
\beq
I_1 \approx \frac{2}{3}\left(\frac{k}{\krh}\right)^2 \left(\frac{11.7}{(k/\krh)^{2/\alpha}} + \frac{1}{(1.15 + 0.00512(k/\krh)^{2/(\alpha \beta)})^\beta} \right)^\alpha
\eeq
where $\alpha \approx 1.06$, $\beta \approx 2.07$;
\beq
I_2 \approx \left(\frac{1}{745 + 24.2 (k/\krh)^{1/\kappa}} + \frac{1}{1.68 + 0.289(k/\krh)^{2/\kappa}}\right)^{\kappa}, 
\eeq
where $\kappa\approx 0.902$;
\beq
\ahor/\arh \approx \left(\frac{0.005}{(k/\krh)^{1/\gamma}} + \frac{1}{(0.702 (k/\krh)^{1/(\delta \gamma)} + 0.304(k/\krh)^{2/(\delta\gamma)})^\delta}\right)^\gamma,
\eeq
where $\delta\approx 0.406$, $\gamma \approx 0.686$. 

The fitting functions given above match the numerical results 
to better than 5\% for $10^{-2} \leq k/\krh \leq 500$. They also 
have correct asymptotics where numerical solutions are impractical, enabling us to study the density field across an even wider range of scales.

\section{Early Halo Model}
\label{sec:stats}
In this section we describe the halo model of the matter power spectrum making use of the results of Refs.~\cite{1991ApJ...381..349S,Cooray:2002dia}. 
We start by writing the density field as a superposition of 
halos with different masses $m_i$:
\beq
\rho(\mathbf{r}) = \sum_{i} \rho_\EH(\mathbf{r}-\mathbf{r}_{i}| m_{i})
 = \sum_i \int d m d^3 r' \delta(m-m_i) \delta^3(\mathbf{r}'-\mathbf{r}_i) m u(\mathbf{r}-\mathbf{r}'| m),
\eeq
where we assume that all halos have the same functional form for  the density profile $\rho_\EH$ that only depends on the halo mass. In the second equality we have written the sum over halos in terms of the mass-normalized density profile $u$. A realization of the halo model consists of a set of $\{\mathbf{r}_i,m_i\}$; an ensemble average over possible realizations is then defined by 
\beq
\left\langle \sum_i  \delta(m-m_i) \delta^3(\mathbf{r}'-\mathbf{r}_i) \right\rangle = \frac{dn}{dm}\equiv n(m),
\eeq
where $n(m)$ is the density of halos per unit mass (i.e., the halo mass function) and we have performed an average over space. It is easy to check that this leads to 
\beq
\langle \rho(\mathbf{r}) \rangle = \int dm\,m\,n(m) = \rholm a^3, 
\eeq
using $\int d^3 r u(r|m) = 1$.
The two point function depends both on the  one point function above, as well as on the halo-halo correlation function $\xi_{hh}(r|m_1,m_2)$:
\begin{align}
\left\langle \rho(\mathbf{r}_1)\rho(\mathbf{r}_2) \right\rangle
&= \int dm\,m^2\, n(m) \int d^3 y u(\mathbf{r}_1-\mathbf{y}|m)u(\mathbf{r}_2-\mathbf{y}|m) +\int dm_1\,dm_2\,m_1\,n(m_1)\,m_2\, n(m_2) \times\nonumber\\
&\times
\int d^3 y_1 d^3 y_2 \left[1 + \xi_{hh}(|\mathbf{y}_1 - \mathbf{y}_2||m_1,m_2)\right]
u(\mathbf{r}_1-\mathbf{y}_1|m_1)u(\mathbf{r}_2-\mathbf{y}_2|m_2).
\end{align}
The first term is the one-halo contribution, encoding the correlation between elements of DM mass distribution inside a single halo; the second two-halo term describes the correlation between mass elements in different halos.

The above expression can be used to find the two-point correlation function of the density contrast. In Fourier space we have 
\beq
\langle \delta(k) \delta(k') \rangle 
= [P_{1h}(k)+P_{2h}(k)](2\pi)^3\delta^3 (\mathbf{k} + \mathbf{k}') \equiv P(k)\, (2\pi)^3\,\delta^3(\mathbf{k} + \mathbf{k}'),
\eeq
where the one- and two-halo terms are 
\begin{subequations}
\begin{align}
P_{1h}(k) & = \frac{1}{\rholm^2}\int dm\,m^2\,n(m)\,|\tilde{u}(k|m)|^2 \\
P_{2h}(k) & = \frac{1}{\rholm^2}\int dm_1\,dm_2\,m_1\,n(m_1)\,m_2\,n(m_2)\, 
P_{hh}(k|m_1,m_2)\,\tilde{u}(k|m_1)\,\tilde{u}(k|m_2)^*.
\end{align}
\label{eq:halo_model_ps_def}
\end{subequations}%
The tildes denote Fourier transform (in comoving coordinates) and 
\beq
P_{hh}(k|m_1,m_2) = \int d^3 r\,\xi_{hh}(r |m_1,m_2)\,e^{i \mathbf{k}\cdot \mathbf{r}}
\eeq 
is the halo-halo power spectrum.  This result relates the non-linear mass density contrast power spectrum to the halo mass function and power spectrum of the halo centers. Note that from Eq.~\ref{eq:fdef_text} we have
\beq
\tilde{u}(k|m) = \frac{1}{a^3} \mathcal{F}(k|m).
\label{eq:profile_from_dist}
\eeq

The Press-Schechter formalism and its extensions provide a model of the halo mass function in terms of the density variance. Following Refs.~\cite{Seljak:2000gq,Cooray:2002dia} we approximate
\beq
P_{hh}(k|m_1,m_2) \approx b(m_1)\,b(m_2)\,P(k)^{\mathrm{lin}},
\label{eq:hh_PS_from_lin_PS}
\eeq
where $b(m)$ is bias parameter that can be estimated for a given mass function~\cite{Cooray:2002dia}. Note that this form cannot be valid up to arbitrarily large $k$: the  halo-halo correlation function $\xi_{hh}$ must turn over  since halos are mutually exclusive - equivalently $P_{hh}$ must decrease at large $k$. We ignore this problem for now, but return to it in \S\ref{sec:halo_exclusion}.

Combining Eqs.~\ref{eq:halo_model_ps_def},~\ref{eq:profile_from_dist} and~\ref{eq:hh_PS_from_lin_PS} yields the halo model for the matter  power spectrum $P = P_{1h} + P_{2h}$, where
\begin{subequations}
\begin{align}
P_{1h}(k) & = \frac{1}{\rholm a^3}\int dm \frac{df}{d\ln m} |\mathcal{F}(k|m)|^2 \label{eq:nl_ps_1h}\\
P_{2h}(k) & = \left\vert\int d\ln m \frac{df}{d\ln m} b(m) \mathcal{F}(k|m)\right\vert^2 P(k)^{\mathrm{lin}}\label{eq:nl_ps_2h},
\end{align}
\end{subequations}
and we used $n(m) = (\rholm a^3/m^2)df/d\ln m$. In the following section we apply these results to an analytic model.

\subsection{An Analytic Example}

In this section we apply some of the results from the previous  subsection to a simple power law power spectrum of the form 
\beq
P(k)^\mathrm{lin} = A k^\alpha,\,\Delta^2(k) = \frac{A}{2\pi^2} k^{3+\alpha}
\label{eq:power_law_spectrum_no_exclusion}
\eeq
where $A$ and $\alpha$ are constants; for EMD-motivated  power spectra we expect $\alpha \approx 1$ (see Eq.~\ref{eq:dimless_linear_ps_from_emd}), while in $\Lambda$CDM  have $\alpha \approx 0$ for modes that enter the  horizon during radiation domination (neglecting the $k$-dependence  of the logarithmic growth during RD).  Using the definition in Eq.~\ref{eq:density_fluctuation_variance} we can compute the density fluctuation variance for this power spectrum:
\beq
\sigma^2(R(M)) = \frac{A}{2\pi^2(3+\alpha)} \frac{1}{R^{3+\alpha}} = 
\delta_c^2 \left(\frac{M_*}{M}\right)^{1 + \alpha/3},
\eeq
where in the last equality we used the typical collapse mass $M_*$ defined in Eq.~\ref{eq:collapse_criterion} to simplify the expression. Note that we are focusing here on a particular instant in time; redshift dependence can be easily restored by introducing a growth function into $\sigma^2$ (or alternatively into the collapse threshold $\delta_c$). The Press-Schechter halo mass function is naturally presented in terms of the variable 
\beq
\nu = \frac{\delta_c^2}{\sigma^2} = \left(\frac{M}{M_*}\right)^{1+ \alpha/3}, 
\label{eq:press_schechter_nu}
\eeq
such that 
\beq
\frac{m^2}{\rholm a^3} \frac{dn}{dm} d\ln m 
= \frac{df}{d\ln m} d\ln m = \sqrt{\frac{\nu}{2\pi}} e^{-\nu/2} d\ln\nu.
\label{eq:press_schechter}
\eeq

We can now estimate the non-linear power spectra of the halo model given in Eqs.~\ref{eq:nl_ps_1h} and~\ref{eq:nl_ps_2h}. We will do this in two cases: for the physically motivated late-time form of $\mathcal{F}$ computed in \S\ref{sec:isolated_halos} which encodes the explosion of EHs, and for the scenario where no explosion occurs (and therefore $\mathcal{F}$ is just the Fourier transform of the density profiles). These cases can be handled simultaneously because $\mathcal{F}$ has a universal behaviour: $\mathcal{F}\to 1$ for $k \ll k_s$, and it decreases as $\mathcal{F}\sim (k_s/k)^\beta$ for large $k$, where $k_s(m)$ is a characteristic scale that depends on the mass of the EH and $\beta$ is an $m$-independent constant. For example for the EMD case, $k_s = k_c$ given in Eq.~\ref{eq:approx_cutoff_scale} and $\beta = 2$; for unexploded NFW-like EHs, $k_s\propto r_s^{-1}$ and $\beta = 2$ follows from the Fourier transform of the NFW profile. In either case, the $m$-dependence of $k_s$ can be expressed as a function of $\nu$ using Eq.~\ref{eq:press_schechter_nu}. The one-halo term then has the general form 
\beq
P_{1h}(k) \sim \frac{m_*}{\rholm a^3} \begin{cases}
1 & k \leq k_s(m_*) \\
\left(k_s(m_*)/k\right)^{2\beta} & k > k_s(m_*),
\end{cases}
\eeq
where the integral over $\nu$ just gives $m_* \times \mathcal{O}(1)$, where the $\mathcal{O}(1)$ constant depends on $\alpha$ and $\beta$. Similarly we find for the two-halo power spectrum 
\beq
P_{2h}(k) \sim P(k)^\mathrm{lin}
\begin{cases}
1 & k \leq k_s(m_*) \\
\left(k_s(m_*)/k\right)^{2\beta} & k > k_s(m_*),
\end{cases}
\eeq
We can immediately see that both the one- and two-halo terms are expected to turn over at $k = k_c(m_*)$; for $\TRH = 5\,\MeV$ $m_* \approx 6\times 10^{-11}M_\odot$, which yields $k_c(m_*)/\keq \approx 7\times 10^7$ at $a=\aeq$. Note that for the EMD-inspired case with $\alpha = 1$ and $\beta=2$, and the dimensionless two-halo powerspectrum $k^3 P_{2h}(k)/(2\pi)^3 \sim \mathrm{const}$ for $k > k_c$.

It is well known that the simplest application of the halo model in Eqs.~\ref{eq:nl_ps_1h} and~\ref{eq:nl_ps_2h} results in some unphysical features of the matter power spectrum~\cite{Cooray:2002dia}. First, the one-halo term in Eq.~\ref{eq:nl_ps_1h} contributes a constant to the power spectrum as $k\to 0$, which implies that the halo model power spectrum does not reduce to the linear one on large scales (alternatively, the integral constraint on the matter correlation function $\int_0^\infty  r^2 \xi(r) dr = 0$ is not satisfied). Second, the above calculation ignores the fact that halos are non-overlapping objects by using the approximation in Eq.~\ref{eq:hh_PS_from_lin_PS}. Ref.~\cite{2011PhRvD..83d3526S} found that these issues are, in fact, related, and addressing exclusion in a physical way results in the correct behaviour of the matter power spectrum as $k\to 0$. We therefore consider a simple implementation of halo exclusion effect in the context of a simplified model like Eq.~\ref{eq:power_law_spectrum_no_exclusion} in the following subsection.

\subsection{Halo Exclusion}
\label{sec:halo_exclusion}
The approximate relationship between the halo-halo power spectrum and the linear matter power spectrum in Eq.~\ref{eq:hh_PS_from_lin_PS} is valid only on scales exceeding the typical halo size, i.e. on linear scales. We will attempt to understand the non-linear regime by starting with simple model of the PS, similar to the one in the previous section:
\beq
P(k)^\mathrm{lin} = A k^\alpha e^{-\lambda k},\,\Delta^2(k) = \frac{A}{2\pi^2} k^{3+\alpha}e^{-\lambda k},
\label{eq:cutoff_powerlaw}
\eeq
where we introduced an exponential cut-off to ensure the finiteness of the various integrals (the functional form of the cut-off is chosen to enable some analytic results); the cut-off wavenumber $\lambda^{-1}$ is distinct from the one computed in the main text ($k_c$) -- it is just a regulator that we will take to be much larger than other scales. We want to implement the fact that the probability of finding two halos separated by less than some exclusion radius $r_e$ is 0; since the probability of finding two halos at separation $r$ is $\propto 1 + \xi_{hh}(r)$, this requirement becomes 
\beq
\xi_{hh}(r\leq r_e) + 1 = 0.
\label{eq:hh_exclusion_condition}
\eeq
We can mock up this behavior by modifying $\xi^{\mathrm{lin}}(r)$, the Fourier transform of $P(k)^\mathrm{lin}$, as follows. The exclusion condition, Eq.~\ref{eq:hh_exclusion_condition}, is enforced if we define
\beq
\xi_{hh}(r) = \begin{cases}\xi^{\mathrm{lin}}(r) & r> r_e\\
-1 & r \leq r_e
\end{cases}.
\label{eq:hh_correlation_excised}
\eeq
This is a very coarse model for the halo-halo correlation because it uses the linear theory result in the mildly-non-linear regime $r\sim r_e$ and because it treats halos as hard spheres of equal sizes -- see, e.g., Ref.~\cite{Baldauf:2013hka} for possible improvements. 
The Fourier transform of this function is~\cite{Baldauf:2013hka}
\begin{align}
P_{hh}(k) & = \int_{r_e}^{\infty} d^3 r \xi^\mathrm{lin}(r) e^{ik\cdot r} - \int_0^{r_e}d^3 r e^{ik\cdot r} \\
& = P^{\mathrm{lin}}(k) - \int_0^{r_e} d^3 r \xi^\mathrm{lin}(r) e^{ik\cdot r}- \int_0^{r_e}d^3 r e^{ik\cdot r}.
\label{eq:hh_exclusion_explicit}
\end{align}

For the spectrum in Eq.~\ref{eq:cutoff_powerlaw} some of the integrals can be performed analytically. Let us first discuss the explicit form of the linear correlation function. For $\alpha=1$ (the EMD-like case) we have~\cite{1980lssu.book.....P}
\beq
\xi(r)^{\mathrm{lin}} = \frac{A}{\pi^2\lambda^4}\, \frac{\sin(3\arctan r/\lambda)}{ (r/\lambda) (1+  r^2/\lambda^2)^{3/2}}.
\eeq
This correlation function approaches zero from below at large distances $r/\lambda \gg 1$
\beq
\xi(r)^{\mathrm{lin}} \sim -\frac{A}{\pi^2 r^4},
\eeq
and is nearly constant and positive at short distances $r/\lambda \ll 1$
\beq
\xi(r)^{\mathrm{lin}} \sim \frac{3A}{\pi^2\lambda^4}.
\eeq
This means that matter is \emph{anti}-correlated at large distances and strongly correlated at small distances. The positive correlation, however, occurs for $r \lesssim \sqrt{3}\lambda$ which is sensitive to the precise nature of the cut-off. Note that this positive contribution is necessary to ensure that the integral constraint is satisfied~\cite{1976A&A....53..131P}
\beq
 \lim_{k\to 0}  P^{\mathrm{lin}}(k)= 0 = \int d^3 r \xi(r)^{\mathrm{lin}}.
\eeq
The excision of the halo overlap, Eq.~\ref{eq:hh_correlation_excised}, removes the region with $r<r_e$, including the region $r \lesssim \sqrt{3}\lambda$ where $\xi(r)^{\mathrm{lin}}$ is positive. This means that the two-halo term computed from the halo-halo correlation function in Eq.~\ref{eq:hh_correlation_excised} cannot satisfy the integral constraint by itself; this is fine since only the total power spectrum (the sum of one- and two-halo terms) is an observable. 

We therefore find that exclusion of overlap regions \emph{reduces} the power spectrum, even on large $k\to 0$ scales, as discussed in, e.g., Ref.~\cite{Smith:2006ne,Baldauf:2013hka}. This is just the statement that the last two terms in Eq.~\ref{eq:hh_exclusion_explicit} are negative. These two terms are constant in the $k\to 0$ limit and therefore represent a reduction in the halo shot noise (which is nominally part of the one-halo term). Ref.~\cite{2011PhRvD..83d3526S} argued that $k\to 0$ limit of these terms must nearly cancel against the one-halo shot-noise-like contribution ensuring the correct asymptotic behaviour of the entire power spectrum. This cancellation, however, is sensitive to the behaviour of $\xi(r)$ in the non-linear regime, which we do not model. We therefore opt for a simplified calculation of the power spectrum, focusing on the two-halo term from which we subtract off the shot-noise-like contribution (the $k\to 0$ limit); the small scale correlations encoded in the one-halo term (and the subtracted piece) become subdominant at late times.

The $k\to 0$ limit, $C$, of the exclusion terms (last two terms of Eq.~\ref{eq:hh_exclusion_explicit}) can be computed analytically
\beq
C \equiv \int_0^{r_e} d^3 r \xi^\mathrm{lin}(r) + \int_0^{r_e}d^3 r 
= \frac{4 A r_e^3}{\pi\lambda^4 (1 + r_e^2/\lambda^2)^2} + \frac{4\pi r_e^3}{3} \to \frac{4A}{\pi r_e} + \frac{4\pi r_e^3}{3},
\label{eq:shot_noise_from_2h}
\eeq
where the last step corresponds to $r_e/\lambda \gg 1$; note that the cut-off dependence drops out, so the result is insensitive to the precise nature of the regulator as long as $r_e$ is well above $\lambda$.

\begin{figure}
    \centering
    \includegraphics[width=0.47\textwidth]{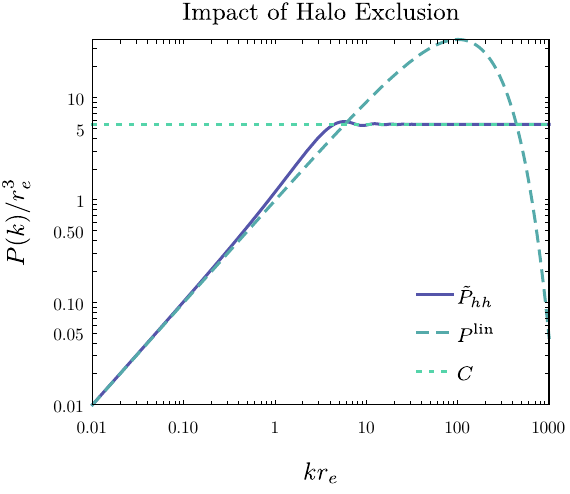}
    \caption{Impact of halo exclusion on the halo-halo power spectrum in a toy model with linear matter power spectrum in Eq.~\ref{eq:cutoff_powerlaw}. The wavenumber is given in units of inverse exclusion radius $1/r_e$.
    }
    \label{fig:halo_exclusion_impact}
\end{figure}

We now define a ``subtracted'' halo-halo PS $\tilde{P}_{hh}$ via
\beq
P_{hh}(k) = \tilde{P}_{hh}(k) - C.
\eeq
The discussion in the previous paragraph implies that $\tilde{P}_{hh}$ has two nice properties: $\tilde{P}_{hh}(k) \approx P^{\mathrm{lin}}(k)$ for $kr_e\ll 1$, and $\tilde{P}_{hh}(k) \approx C$ for $kr_e \gg 1$. The subtracted PS $\tilde{P}_{hh}(k)$ and the constant $C$ are compared to the linear power spectrum in Fig.~\ref{fig:halo_exclusion_impact} for $r_e/\lambda = 10^2$ and $A = r_e^4$. We see that the main impact of halo exclusion is the flattening of $\tilde{P}_{hh}(k)$ for $k r_e \gtrsim 1$ (in other words, halo exclusion leads to a $1/k$ suppression at these scales).Because the $r_e/\lambda \gg 1$, these features are not sensitive to the cut-off $\lambda$.

\begin{figure}
    \centering
    \includegraphics[width=0.47\textwidth]{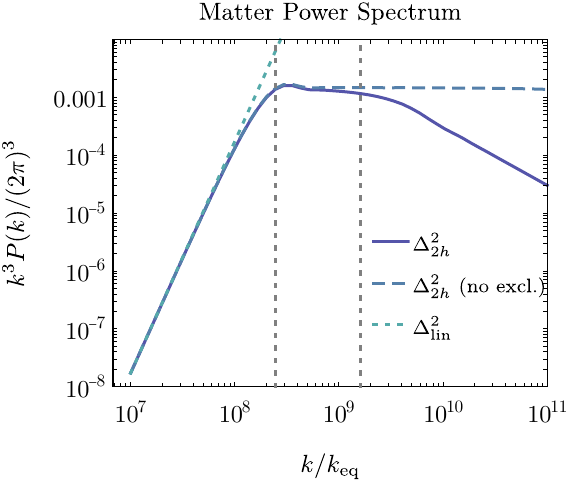}
    \caption{Matter power spectrum after the end of EMD in a toy model given by Eq.~\ref{eq:cutoff_powerlaw} with the normalization $A$ corresponding to $\TRH \approx 5\;\MeV$. The solid line shows the dimensionless power spectrum including both the explosive evaporation of early halos at the end of EMD, and the halo exclusion effect discussed in the text. The dashed line neglects the impact of halo exclusion, while the dotted line is the original linear power spectrum. The cutoff scale $\lambda$ is taken to be a factor $10^{-3}$ smaller than the exclusion scale $r_e$, indicated by the right-most vertical dotted line. The left-most vertical dotted line indicates the expulsion cutoff $k_c$ from Eq.~\ref{eq:approx_cutoff_scale}. Only the two-halo contribution to the matter power spectrum is shown since the one-halo term becomes negligible well after reheating.}
    \label{fig:halo_exclusion_impact_after_expulsion}
\end{figure}

The above results motivate the following definitions of the one- and two-halo power spectra
\begin{subequations}
\begin{align}
P_{1h}(k) & = \frac{1}{\rholm a^3}\int dm \frac{df}{d\ln m} |\mathcal{F}(k|m)|^2 - C\left\vert\int d\ln m \frac{df}{d\ln m} b(m) \mathcal{F}(k|m)\right\vert^2\label{eq:mod_1h_ps}\\
P_{2h}(k) & = \left\vert\int d\ln m \frac{df}{d\ln m} b(m) \mathcal{F}(k|m)\right\vert^2 \tilde{P}_{hh}(k).
\label{eq:mod_2h_ps}
\end{align}
\end{subequations}
As mentioned above, we focus on the two-halo contribution that dominates at late times, thereby avoiding the difficulties of enforcing the proper cancellations in Eq.~\ref{eq:mod_1h_ps} which is irrelevant at late times anyway. In Fig.~\ref{fig:halo_exclusion_impact_after_expulsion} we show the result of evaluating Eq.~\ref{eq:mod_2h_ps} using $\mathcal{F}$ for a spherical top hat density profile. This power spectrum roughly corresponds to $\TRH = 5\;\MeV$ (i.e., we chose $A = 2\pi^2 A_s/\krh^4$ with $\krh$ from Eq.~\ref{eq:krh_over_keq}), but does not include the logarithmic growth of perturbations on large scales. However, it illustrates the two main features of this result, which are apparent by comparing it with the original linear PS (dotted line) and the PS without halo exclusion (dashed line). First, we see that on scales smaller than $1/k_c$ (the left vertical dotted line) the expulsion of matter due to reheating flattens the power spectrum. On yet smaller scales, below the exclusion radius $r_e$ (the right vertical dotted line), the PS decreases as $1/k$. The plateau feature arises because $r_e k_c < 1$, i.e. EHs grow to be much larger than their original size $\sim r_e$ during adiabatic expansion and subsequent free expansion.

\bibliographystyle{JHEP}
\bibliography{biblio}
\end{document}